\documentclass[letterpaper,twocolumn,10pt]{article}
\usepackage{usenix}

\usepackage{graphicx}
\graphicspath{{figs/}}

\usepackage{amsmath}
\usepackage{amsfonts}
\usepackage{amsthm}

\usepackage[english]{babel}
\usepackage{blindtext}
\usepackage{lipsum}
\usepackage{soul}
\usepackage[ruled,vlined,linesnumbered]{algorithm2e}

\usepackage{filecontents}
\usepackage{xspace}
\usepackage{subcaption}
\usepackage{xcolor}
\usepackage{xurl}
\definecolor{newblue}{RGB}{100,140,177}
\definecolor{newgreen}{RGB}{90,176,80}

\usepackage{hhline}
\usepackage{multirow}
\usepackage{soul}
\usepackage{array}
\usepackage{booktabs}
\usepackage{mathtools}
\usepackage{comment}

\usepackage{pifont}
\usepackage{marvosym}

\usepackage{csquotes}
\usepackage[font=itshape]{quoting}
\usepackage{filecontents}

\usepackage{tikz}
\newcommand{\highlight}[2]{\colorbox{#1!17}{#2}}
\definecolor{ballblue}{rgb}{0.13, 0.67, 0.8}
\usepackage{todonotes} % for comments

\newcommand{\cameraold}[1]{{\color{red}{}}}

\newcommand{\passed}[1]{{}}
\newcommand{\jc}[1]{{\color{red}{\footnotesize [JC: #1]\xspace}}}

\newcommand{\amz}[1]{{\color{green!70!black}{\footnotesize [Amrita: #1]\xspace}}}
\newcommand{\yc}[1]{{\color{blue}{\footnotesize [YC: #1]\xspace}}}

\newcommand{\review}[1]{{\color{red}{#1}\xspace}}
  \renewcommand{\review}[1]{{\color{black}{#1}\xspace}}

\definecolor{editcolor}{RGB}{120,100,0}
\newcommand{\edit}[1]{{\color{editcolor}{#1}}}

\newcommand{\ignore}[1]{{\xspace}}
\newcommand{\REMOVE}[1]{}

\newcounter{packednmbr}

\newenvironment{packeditemize}{\begin{list}{$\bullet$}{
\setlength{\itemsep}{0.5pt}\addtolength{\labelwidth}{-4pt}\setlength{\leftmargin}{2.5ex}\setlength{\listparindent}{\parindent}\setlength{\parsep}{1pt}\setlength{\topsep}{2pt}}}{\end{list}}

\newcommand{\tightcaption}[1]{\vspace{-0.4cm}\caption{{\normalfont{\textit{{#1}}}}}\vspace{-0.4cm}}

\newcommand{\tightsection}[1]{\vspace{-0.35cm}\section{#1}\vspace{-0.25cm}}

\newcommand{\tightsubsection}[1]{\vspace{-0.35cm}\subsection{#1}}\vspace{-0.25cm}

\newcommand{\eg}{{e.g.,}\xspace}
\newcommand{\ie}{{i.e.,}\xspace}

\newcommand{\myparashort}[1]{\smallskip\noindent{\bf {#1}}~}
\newcommand{\mypara}[1]{\vspace{0.1cm}\noindent{\bf {#1}:}~}

\newcommand{\myparaq}[1]{\vspace{0.1cm}\noindent{\bf {#1}?}~}

\newcommand{\aecode}{{coded tensor}\xspace}

\newcommand{\name}{\textsc{Grace}\xspace}
\newcommand{\namelite}{\name-Lite\xspace}
\newcommand{\namepretrain}{\name-P\xspace}
\newcommand{\namedecoder}{\name-D\xspace}

\newcommand{\Coder}{\ensuremath{f_\phi}\xspace}
\newcommand{\Decoder}{\ensuremath{g_\theta}\xspace}
\newcommand{\Quality}{\ensuremath{L}\xspace}
\newcommand{\Distort}{\ensuremath{D}\xspace}
\newcommand{\Size}{\ensuremath{S}\xspace}

\newcommand{\x}{\ensuremath{\textbf{x}}\xspace}
\newcommand{\y}{\ensuremath{\textbf{y}}\xspace}

\newcommand{\loss}{\ensuremath{P}\xspace}

\newcommand{\cmedit}[2]{#1}
\newcommand{\tempstrike}[1]{}

\usepackage{titling}
\begin{document}
%-------------------------------------------------------------------------------

%don't want date printed
\date{}

%\title{\scalebox{0.9}{\Large \name: Loss-Resilient Real-Time Video Communication Using Data-Scalable Autoencoder}}
%\title{\scalebox{0.97}{\Large \name: Real-Time Video Communication Using Data-Scalable Autoencoders}}
%\title{\scalebox{0.87}{\Large \name: Balancing Real-Time Video Quality and Frame Delay via Data-Scalable Autoencoders}}
%\title{{\Large \name: Improving Real-Time Video Communication in High Network Latency}}
%\title{\scalebox{0.9}{\huge \name: Loss-Resilient Real-Time Video Communication under High Network Latency}}
%\title{\name: Loss-Resilient Neural Codec for Real-Time Video Communication}
%\title{{\Large \name: Taming Latency and Jitter for Real-Time Video Communication}}
\title{\Large \bf \name: Loss-Resilient Real-Time Video through Neural Codecs \vspace{-10pt}}
%\normalfont \large \em Paper \#1386
%\vspace{-3.5em}}

%for single author (just remove % characters)
% \author{
% Yihua Cheng$^1$, Anton Arapin$^2$, Ziyi Zhang$^1$, Qizheng Zhang$^1$, Hanchen Li$^1$, Yuhan Liu$^1$, Xu Zhang$^1$, Nick Feamster$^1$, Junchen Jiang$^1$ \\
% $^{1}$\{yihua98, ziyizhang, qizhengz, lihanc, yuhanl, zhangxu, feamster, junchenj\}@uchicago.edu \\
% $^{2}$anton.arapin@gmail.com \\
% The University of Chicago
% }

\author{
Yihua Cheng$^1$, Ziyi Zhang$^1$, Hanchen Li$^1$, Anton Arapin$^1$, Yue Zhang$^1$,  Qizheng Zhang$^2$, Yuhan Liu$^1$, \\ Kuntai Du$^1$, Xu Zhang$^1$, Francis Y. Yan$^3$, Amrita Mazumdar$^4$, Nick Feamster$^1$, Junchen Jiang$^1$ \\ \vspace{-10pt}
\textit{$^1$The University of Chicago, \xspace\xspace $^2$Stanford University, \xspace\xspace $^3$Microsoft, \xspace\xspace $^4$NVIDIA \vspace{-20pt}
}}

% \text{Yihua Cheng}^{1}, \text{Anton Arapin}^{2}, \text{Ziyi Zhang}^{1}, \text{Qizheng Zhang}^{1,3}, \text{Hanchen Li}^{1}}, \\ 
% \text{Nick Feamster}^{1},\text{Junchen Jiang}^{1} \\
% ^{1}\{yihua98,ziyizhang,qizhengz,lihanc,feamster,junchenj\}@uchicago.edu \xspace \xspace \\
% ^{2}anton.arapin@gmail.com \\
% ^{1,2}\text{The University of Chicago}, \xspace \xspace ^{3}\text{Stanford University}

% \author{
% Paper ID \fillme
% }
\pagestyle{empty}

\maketitle
\thispagestyle{empty} 

%\input{sections/abstract_v5}
% \input{sections/abstract_v6}
% \input{sections/abstract_v7}
%\input{sections/abstract_v8}
%!TEX root = ../main.tex
%!TEX spellcheck = en_US

\begin{abstract}
In real-time video communication, retransmitting lost packets over
high-latency networks is not viable due to strict latency requirements.
% FYY: high-latency networks might be Wi-Fi or cellular, not necessarily wide-area networks
To counter packet losses without retransmission, two primary strategies are 
employed---encoder-based forward error correction (FEC) and
decoder-based error concealment.
The former encodes data with redundancy before transmission, yet determining the
optimal redundancy level in advance proves challenging.
% The latter, on the other hand, enables video reconstruction from partially
% received packets at the expense of compression efficiency.
%the decoder to reconstruct video from partially received packets, but it hurts compression efficiency and without changing the encoder, not all lost data can be reconstructed by the decoder.
% The latter, on the other hand, enables the decoder to reconstruct video from partially received packets, but it inevitably hurts compression efficiency to make each packet independently decodable, and without changing the encoder, not all lost data can be reconstructed by the decoder.
The latter reconstructs video from partially received frames,
but dividing a frame into independently coded partitions inherently compromises compression efficiency, and the lost information cannot be effectively recovered by the decoder without adapting the encoder.

We present a loss-resilient real-time video system called \name,
which preserves the user's quality of experience (QoE) across a wide range of packet losses through a new neural video codec.
Central to \name's enhanced loss resilience is
its {\em joint training of the neural encoder and decoder under
a spectrum of simulated packet losses}.
% We introduce \name, a loss-resilient, learning-based video system designed to
% maintain video quality across a wide range of packet losses. Capitalizing on
% the recent advances in neural video codecs,
% \name markedly enhances loss resilience by jointly optimizing the
% encoder and decoder---both integrated with neural network
% components---under various simulated packet losses.
In lossless scenarios, \name achieves video quality on par with
conventional codecs (\eg H.265).
As the loss rate escalates, \name exhibits a more graceful, 
less pronounced decline in quality, consistently outperforming other loss-resilient schemes.
%consistently delivers higher quality than
%other loss-resilient schemes, exhibiting a more graceful decline in quality.
Through extensive evaluation on various videos and real network traces, we demonstrate that
\name reduces undecodable frames by 95\% and stall duration by 90\%
compared with FEC, while markedly boosting video quality over
error concealment methods.
% ; it also significantly boosts video quality over
% state-of-the-art error concealment---without notably affecting other metrics in both comparisons.
%\jc{Yihua, the previous sentence may seem a bit too good to be true. maybe we need to tune it down a bit}
In a user study with 240 crowdsourced participants and 960 subjective ratings, \name registers
a 38\% higher mean opinion score (MOS) than other baselines. %, further attesting to its effectiveness.
% \fyy{Shorten the last two sentences if needed; I have been trying to fit
% Figure 1 into the first page.}
We make the source codes and models of \name public at \url{https://uchi-jcl.github.io/grace.html}.
\end{abstract}

\tightsection{Introduction}

Real-time video communication has become an integral part of our daily
lives~\cite{blum2021webrtc}, spanning online
conferences~\cite{webrtc-vc-1,webrtc-vc-2},
cloud gaming~\cite{webrtc-gaming-1,webrtc-gaming-2},
interactive virtual reality~\cite{webrtc-vr-1,webrtc-vr-2},
and IoT applications~\cite{webrtc-iot-1,webrtc-iot-2}.
To ensure a high quality of experience (QoE) for users,
real-time video applications must protect against
packet losses\footnote{In this study, we use the term
``packet loss'' to refer to both packets
dropped in transit and those not received by the decoding deadline.
Under this definition, a video frame could experience a high packet
loss rate (e.g., 50\%) even if the actual network loss
rate is low~\cite{rudow2023tambur}.}.
%Thus, many real-time video frames can have a high (over 50\%) packet loss rate, even if the network has a low average packet drop rate, which corroborates with other studies~\cite{??,??}.}.
However, retransmitting lost packets across high-latency networks
is not feasible due to stringent real-time latency requirements~\cite{itu-g114}.

\begin{figure}[t]
  \centering
  \includegraphics[width=0.57\columnwidth]{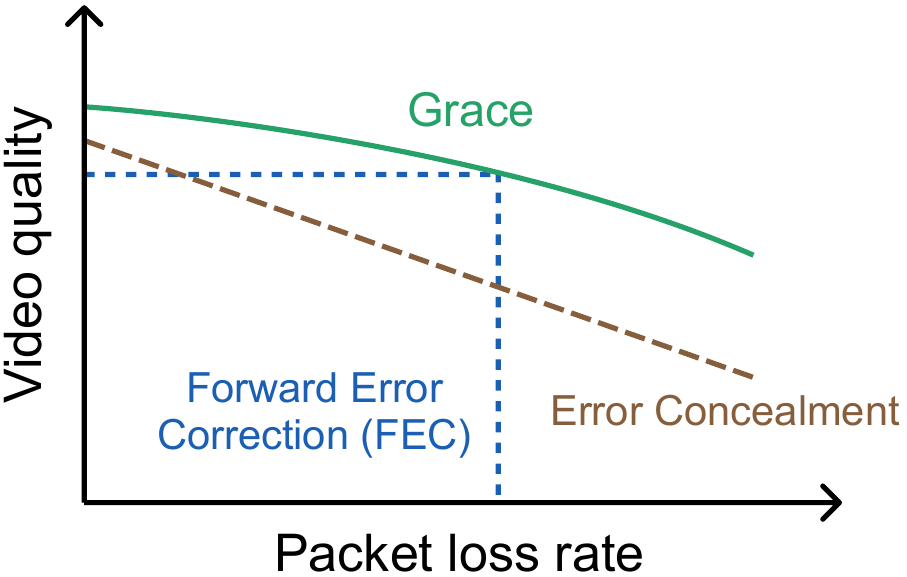}
  \vspace{8pt}
  \tightcaption{Illustration of the video quality achieved by
  different loss-resilient schemes, operating under the same bandwidth budget,
  across varying packet loss rates.
  Actual experimental results are shown in Figure~\ref{fig:loss_by_dataset}.}
  %(Actual figures from our evaluation appear later, \eg Figure~\ref{??}.)}
  \label{fig:intro-example}
  \vspace{-8pt}
\end{figure}

Loss-resilient techniques generally fall into two categories.
First is encoder-side forward error correction (FEC), such as Reed-Solomon codes~\cite{wicker1999reed}, fountain codes~\cite{mackay1997near, mackay2005fountain}, and more recently, streaming codes~\cite{rudow2023tambur, badr2017fec}.
%, are widely used.
% to recover packet losses without retransmission.
FEC incorporates redundancy into data prior to transmission.
With a redundancy rate of $R\%$---the percentage of redundant data relative to the total data size---up to $R\%$ of lost data can be recovered.
Beyond that, the video becomes undecodable,
rendering a sharp collapse in video quality (Figure~\ref{fig:intro-example}).
Increasing $R$ protects against higher losses but also entails a higher bandwidth overhead, which in turn reduces video quality.
Thus, determining the optimal $R$ in advance poses a practical challenge.
%As illustrated in Figure~\ref{fig:intro-example}, a larger $R$ can tolerate more packet losses at the expense of lower video quality, and the video is no longer decodable when the loss rate exceeds $R\%$, causing a sharp collapse in video quality.
%\passed{\fyy{This is not a good place to insert parentheses as they break the flow and sound informal. Generally refrain from overuse of parentheses.}} 
%Nevertheless, determining the optimal $R$ value involves forecasting the packet loss rate, posing a practical challenge.
%Under the same bandwidth, a larger $R$ is more resilient to packet losses but also entails lower video quality.

%Another popular solution to recover from packet loss is decoder-side error concealment,
The second category is decoder-side error concealment,
which reconstructs portions of a video frame affected by packet losses, through handcrafted heuristics~\cite{kolkeri2009error, zhou2010efficient,wang2013novel} or neural networks~\cite{xiang2019generative,kang2022error,sankisa2018video, mathieu2015deep,reparo}.
Nevertheless, implementing error concealment requires partitioning a video frame into independently decodable units (\eg slices~\cite{wenger2002scattered} or tiles~\cite{kumar2006error}) first, thus reducing compression efficiency.
Moreover, since the encoder is not optimized for loss resilience, the lost information cannot be effectively recovered by the decoder alone.
%it aggressively compresses each frame, resulting in each packet containing little redundancy to help reconstruct the missing information.
As a result, the video quality tends to deteriorate rapidly with increasing packet loss, as illustrated in Figure~\ref{fig:intro-example}.

In this paper, we present \name, a loss-resilient real-time video system designed to maintain the user's quality of experience (QoE) across a wide range of packet losses.
Our key insight is that {\em jointly optimizing the encoder and decoder under a spectrum of simulated packet losses} considerably strengthens loss resilience.
To facilitate this joint optimization, \name strategically employs a neural video codec (NVC)~\cite{dvc}, integrating neural networks into the core components of a conventional video encoder and decoder.
%which architecture also holds promise for robustness against data perturbation (such as packet loss)~\cite{??}.
%\edit{to make the encoder and decoder trainable.}\hc{"to train encoder and decoder for this specific purpose"}
%\yc{maybe we can say NVC is inherently more robust to XXX (think about the term here) + citation}
%\fyy{This sentence sounds incomplete}.
%\jc{i removed the text about replacing core components of conventional codecs, since it's not critical and it's already implicit in the term neural video codecs. feel free to revert if you want}
% and integrates neural networks into the core components 
%of a conventional encoder and decoder (rather than replacing them entirely).
In contrast to FEC, \name's NVC is trained to handle \textit{diverse} packet losses, eliminating the need to predict a loss rate beforehand and preventing the undecodable video under exceedingly high losses.
Unlike decoder-side error concealment, \name \textit{jointly} trains the encoder and decoder, so that the encoder 
learns to properly distribute each pixel's information across multiple output elements in anticipation of losses, facilitating the decoder's frame reconstruction when packets are actually lost.
%thereby  enhancing its output's resilience to packet loss and allowing the decoder to better reconstruct the original frame when packet loss happens.
% making the encoder's output more robust to packet loss. 
%As a result, the decoder can better reconstruct the original frame when packet loss happens.
%learns to add some redundancy \fyy{what do you mean?} to enable a better reconstruction of information critical to frame quality when packet loss happens.
%to the information critical to the frame quality, enabling a better reconstruction when packet loss happens.
%By jointly optimizing the encoder and decoder, rather than only the decoder, \name's encoder learns to add redundancy to the information that is critical to the frame quality, enabling a better reconstruction than error concealment methods when packet loss happens.
Consequently, \name displays a more graceful quality degradation amid varying losses, while consistently delivering higher video quality than previous solutions (Figure~\ref{fig:intro-example}).
To materialize the above benefits of \name's codec, 
our design of \name addresses three system challenges.
%however, three system challenges remain to be addressed.

First, to ensure loss tolerance, each packet must be independently decodable. 
Existing solutions achieve this by dividing the frame into independently decodable units. 
However, this introduces a size overhead because the data in each unit follows different distributions and thus cannot be compressed efficiently. 
In response to this challenge, we train \name's neural encoder to regularize the values in its output to conform to the same distribution, thereby reducing the partitioning overhead. 
We also utilize reversible random mapping~\cite{lcg} during such partitioning, making it more amenable to NVCs. 
While training \name under packet losses, simulating random partitioning and packet losses is inefficient and precludes differentiability.
Hence, we apply random zeroing to the encoder's output directly, simulating packet losses without actually dropping packets (\S\ref{sec:core}).

% packet loss simulation during training must be carefully designed to ensure that simulated losses mirror real-world effects.
% %the same effect as the real ones.
% %the effect of real packet losses is the same as during training.
% To this end, we randomly zero (``mask'') values in the encoder's output during training to simulate packet loss (\S\ref{sec:core}).
% %values in the encoder's output are randomly zeroed (masked) 
% %before feeding into the decoder (\S\ref{sec:core}).
% At runtime, \name employs a novel packetization scheme that partitions the encoder's output using a reversible random function~\cite{lcg} before entropy-encoding each partition into a separate packet.
% If a packet goes missing, its corresponding values in the encoded output are zeroed. 
% As a result, each packet is individually decodable, and the impact of a lost packet matches that of simulated packet loss (\S\ref{subsec:packetization}).
% %is mapped to the random erasure of values in the encoder's output.
%Section~\S\ref{sec:core} and \S\ref{subsec:packetization} provides more details about the training and packetization process of \name. 

Second, packet losses can lead to discrepancies between the reference frames at the encoder and decoder side, resulting in sustained quality degradation in the decoded video stream if synchronization is not maintained. 
Traditional remedies, such as retransmission or sending a new keyframe, fall short of seamlessly rectifying this inconsistency. 
\name introduces an innovative protocol to adeptly realign the encoder and decoder states without hindering video decoding. 
In the event of packet loss, the decoder leverages the loss resilience of \name to decode partially received packets. 
Simultaneously, the decoder communicates the loss details to the encoder. 
This feedback mechanism enables the encoder to swiftly adjust its recent reference frames to match those at the decoder end, eliminating the need for additional data transmission (\S\ref{subsec:protocol}).

%the video system should efficiently synchronize the encoder and decoder states to avoid video stream corruption after packet loss ends.
%To this end, we design a new protocol for \name to resynchronize the encoder and decoder states without blocking the decoder.
%retransmission.

%When the decoder decodes a frame based on partially received packets, 
%When getting the packet loss feedback from the decoder, the encoder quickly re-decodes the recent frames as seen by the decoder to align its state with the decoder.
% Such re-decoding only increases the encoding time by 7\% whenever it happens (\S\ref{subsec:protocol}).
%Such re-decoding only requires 7\% of the encoding time (\S\ref{subsec:protocol}).

Third, \name must be efficient to encode and decode video in real-time across various devices, from laptops to mobile phones.
Existing NVCs, however, often employ expensive neural networks, particularly for motion estimation and post-processing.
%Among the performance optimizations,
We show that by downscaling the image input for motion estimation and simplifying post-processing, \name accelerates the encoding and decoding by $4\times$ without a noticeable impact on loss resilience. 
Moreover, with hardware-specific runtimes such as OpenVINO and CoreML, \name attains over 25~fps on CPUs and iPhones (\S\ref{subsec:fast-coding}).

Comprehensive experiments (\S\ref{sec:eval}) on a diverse set of videos and real network traces show that 
with the same congestion control logic, 
\name reduces undecodable frames by 95\% and stall duration by 90\% compared with state-of-the-art FEC baselines. It also boosts 
the visual quality metric of SSIM by 3~dB
%video quality by 3~dB of SSIM (a visual quality metric)
over a recent neural error concealment scheme (\S\ref{subsec:eval:setup}).
%defined in \S\ref{subsec:eval:setup}. 
%(We run \name and baselines on the same congestion control logic.)
Our IRB-approved user study with 240 crowdsourced participants and
a total of 96 subjective ratings demonstrates 
a 38\% higher mean opinion score (MOS) for \name,
further attesting to its effectiveness.
Regarding computational efficiency, \name achieves more than 25~fps on popular mobile devices (e.g., iPhone 14 Pro), meeting the real-time requirement.
%resource-constrained devices such as , matching the requirement of real time videos.

Our contributions are summarized as follows. 
{\em (i)} We present \name, which, to the best of our knowledge, represents \textbf{the first effort to \textit{jointly} train a neural video encoder and decoder under a spectrum of packet losses, aiming to improve loss resilience in real-time video} (\S\ref{sec:core}).
Different from other recent ML-based real-time video systems~\cite{concerto, onrl, locki}
that use ML-based rate adaptation to minimize packet loss, \name uses ML to make the video codec itself resilient to packet loss.
{\em (ii)} We build the end-to-end video system to address the practical challenges associated with integrating a new NVC, developing optimization techniques related to packetization, encoder/decoder state synchronization, and runtime efficiency (\S\ref{sec:delivery}).

\tightsection{Background}
\label{sec:motivate}

%Video coding has a rich history of research and it is hard to fit a comprehensive overview in a limited space.
%Nonetheless, 

\vspace{0.3cm}

\tightsubsection{Real-time video coding}
\label{subsec:background}

To help explain \name's design, we first introduce some key concepts in
real-time video coding and streaming. 
%\fyy{The comparison between video conferencing and streaming
%is a waste of space to me. It's fine to focus on conferencing.}\jc{how about we keep it? most reviewers dont have the expertise to understand the difference}
%before explaining the need for a different approach (\S\ref{subsec:ds},\ref{subsec:prior}). 
%The video sender runs a rate adaptation logic to dynamically set the frame rate and bitrate of the encoded video.

\begin{figure}[t!]
         % \hspace{-0.2cm}
        \includegraphics[width=0.97\linewidth]{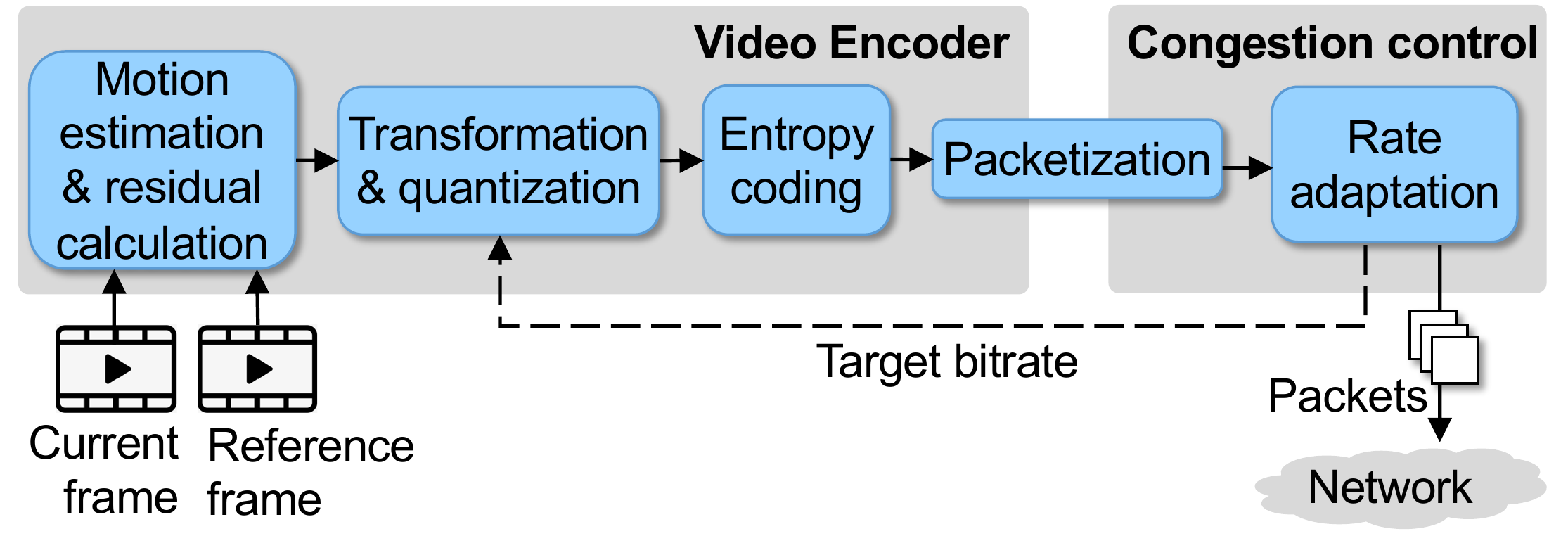}
         \label{fig:}
         % \vspace{-10pt}
         \tightcaption{A typical workflow of video frame encoding.
         % \fyy{I prefer not to use "tightcaption" or italics for the caption. Regular USENIX template looks much nicer. Leave ample vertical space after each figure.}
         }
    \label{fig:codec-workflow}
    \vspace{-8pt}
    % \vspace{-0.25cm}
\end{figure}

The sender encodes video at a specific frame rate and bitrate, \eg
%A video stream is encoded at certain frame rate and bitrate, both are controlled by an adaptation logic in the player.
with 25~fps (frames per second) and 3~Mbps, the encoder generates a 15~KB frame on average every 40~ms.
%, though the actual frame size varies.
%depends on the {\em type} of a frame
An encoded video is composed of groups of consecutive frames, called a group of pictures (GoP). Each GoP starts with an I-frame (or key-frame), followed by multiple P-frames (or inter-frames)\footnote{On-demand video also uses B-frames (bidirectional predicted frames), which 
% are smaller than P-frames since they 
refer to both past and future P-frames. However, real-time video rarely uses B-frames in order to render frames as soon as possible.
% However, to render frames as soon as possible, real-time video rarely uses B-frames~\cite{??}, since they refer to future frames to be decoded.
}.
% and B-frames (bidirectional predicted frames). 
I-frames are independently encoded without referencing other frames, while P-frames encode only the differences relative to previous reference frames.
In real-time video, the majority of frames are P-frames to minimize frame sizes, so our discussion here focuses on P-frames.
Figure~\ref{fig:codec-workflow} shows the workflow of P-frame encoding and decoding. 
Given a new frame and a reference frame, the encoder 
(1) calculates motion vectors (MVs) and residuals for each macroblock (MB), e.g., 16×16-pixel samples, 
(2) transforms and quantizes the MVs and residuals, 
(3) performs entropy encoding on the transformed data, 
(4) divides the entropy-encoded data into packets, and 
(5) transmits these packets with congestion control, such as GCC~\cite{carlucci2016analysis}. 
%To encode a frame, the encoder first calculates the differences (\eg motion vectors and residuals) among macroblocks (4×4,8×8,…4\times4,8\times8,\dots) in the current frame and reference frames (or within an I-frame),  then transforms and quantizes these differences to a more compact form, and finally losslessly encodes the data using arithmetic coding. 
%The entropy-encoded data is then packetized and sent out by the transport layer, which tries to deliver as many packets as possible 
%% (though not always all packets) 
%to the receiver by the time when the frame is decoded.
%algorithm decides how fast to send a frame's packets, 
%in order to ensure enough data be delivered to the receiver when the frame is decoded. 
Correspondingly, the receiver depacketizes and decodes the received data to 
% runs a reverse process (depacketization and decoding) to 
reconstruct each frame from the received packets.

% A key metric in real-time video is {\em frame delay}, the time between the encoding and the rendering of a frame.
%\yc{video conferencing has more strict realtime constraints -- short buffer, no time window for retransmission on high latency network}
To reduce frame delay, which denotes the time from frame encoding to rendering, real-time video clients (\eg WebRTC) commonly make two choices
%~\cite{10.1145/3410449,salsify}
that differentiate them from video streaming (\eg Netflix, ESPN Live): %\footnote{\jc{any exceptions?}}:
% that affect the requirements of a desirable video coding scheme.
\begin{packeditemize}
\item Real-time video employs notably {\em shorter} (tens of ms) buffers, as opposed to video streaming that utilizes several seconds of playback buffer for on-demand or live content.
%The buffer in real-time video is much {\em shorter} (tens of ms) than that of streaming video, which is at least several seconds for on-demand and live video. 
Thus, retransmission delay is difficult to conceal with such short buffers,
especially in high-latency networks.
%\amz{I swapped the "10s" in this section to "10's" to reduce confusion, please revert if undesired. }
% This short buffer often cannot conceal retransmission delay in long-latency networks.
%With a short buffer, any missing packet might delay a frame longer than the jitter buffer can conceal.
\item To maintain a short buffer, real-time video sends each frame in a burst and decodes it as soon as its packets are received. As a result, {\em any} lost packets, whether due to drops or queuing, can affect frame decoding.
% the packets of each frame are sent in a burst (without waiting for acks), and %without waiting for acknowledgement.
% {\em any} packets of a frame could be delayed or dropped when the receiver tries to decode the frame.
In contrast, streaming video is transmitted in chunks (each with hundreds of frames) over the HTTP/TCP protocol.
% \fyy{"100s of frames" without the apostrophe after "100" is
% the correct usage. Use "hundreds of frames" when space allows
% is even better.}
\end{packeditemize}

Ideally, congestion control and bitrate adaptation (e.g., GCC~\cite{carlucci2016analysis},  Salsify~\cite{salsify}, and NADA~\cite{zhu2016nada}) are designed to handle bandwidth fluctuations, thereby avoiding congestion-induced packet losses.
However, predicting bandwidth fluctuations in advance is challenging, making
loss-resilient methods necessary when decoding frames under packet loss.
% Our evaluation shows \name can be combined with 

% Congestion control and bitrate adaptation logics are often used by video clients employ to cope with bandwidth fluctuation, such as GCC~\cite{carlucci2016analysis},  Salsify~\cite{salsify}, and NADA~\cite{zhu2016nada}.
% Ideally, these adaptation logics should reduce packet loss.
% However, because it is difficult to predict bandwidth drops or packet loss in advance, they cannot completely eliminate packet loss.
% \name is complementary to this line of work, and our results show \name improves loss resilience under different congestion control logics (Figure~\ref{fig:appgcc-delay-simulation}). %\yc{is it okay to put the forward reference to appendix here?}

\passed{\jc{need to smooth the text here}}
%Here, we define {\em packet loss} on a given frame as any packets not received before the receiver decodes the frame. \yc{the prior sentence is replicated from intro}
%, \eg every 50ms at 20 frames per second (fps).
We define {\em packet loss per frame} as any packets not received before the receiver is expected to decode the frame.
In other words, even if a packet is not dropped, it can still be counted as packet loss if it arrives too late. 
%It is important to clarify that our notion of packet loss rate is {\em not}
%network-level loss rate (which rarely exceeds 1\%). 
It is important to note that our notion of packet loss differs from network-level loss.
Even with low network loss (which typically remains below 1\%), 
real-time video may still encounter a high packet loss rate (e.g., over 50\%) in certain frames, as corroborated by recent research in this space~\cite{cheng2023abrf,rudow2023tambur,cheng2020deeprs}.
%\yc{In other words, it's different from network packet loss which rarely exceeds 1\%. For instance, ....}.
%\fyy{Avoid italicizing too many words (unless it's really necessary).}
%It can be consecutive or random and can result from packet drops or latency jitter. 
%This definition is consistent with recent papers~\cite{rudow2023tambur,cheng2023abrf,holmer2013handling}. 
%\fyy{Just use italics (without bold font) to emphasize. Bold font should be conservatively used.}
%Even if a packet is not dropped, it can still be counted as a packet loss, if it arrives too late. 
%Therefore, it is not rare for real-time video clients to have a high (over 50\%) packet loss rate on some frames, even if the network has a very low packet loss rate, which 
% The high packet loss rate per frame 
%also corroborates recent papers in this space~\cite{cheng2023abrf,rudow2023tambur,cheng2020deeprs}.
%\fyy{avoid over-emphasizing in this paragraph; use italics and bold only when necessary}
%\passed{\yc{I changed ``before the receiver decodes the frame'' to ``before the receiver should decode the frame''. Because in traditional codecs, the receiver can only decode the frame after all the packets arrive, so it's a bit confusing to say ``packets not received before receiver decodes the frame''}}
%\fyy{This paragraph is very lengthy in explaining the "packet loss" considered in this paper. Please cut it.}

\tightsubsection{Related work}
\label{subsec:related}
\vspace{-4pt}

% As retransmission delay is difficult to conceal by the buffer, particularly in long-latency networks, 
Various loss-resilient schemes have been studied. 
\passed{\jc{Yihua, please add more citations. }}
\passed{\jc{this section should also be made more technical. Yihua, ping me if you can spend some time on this part}\yc{added a bit words, but maybe not enough}}

\myparashort{Forward error coding (FEC)}adds redundancy at the {\em encoder}
before the data is sent to the network. 
This is also known as error-resilient channel coding. 
Examples include Reed-Solomon codes, LDPC~\cite{mackay1997near}, fountain and rateless codes~\cite{mackay2005fountain,castura2006rateless}, streaming codes~\cite{badr2017fec,rudow2023tambur}, and recent ones based on DNNs~\cite{7926071, 8919799}.
\cmedit{There are also hierarchical and multilevel FEC~\cite{tan1999multicast, tan2001video}, which organizes FEC into multiple layers and protects each layer with different redundancies.}{Addressing comments from reviewer A: Hierarchical/staged/multilevel FEC exists. You should at least mention this.}
FEC is also used to protect frame metadata or the base layer in SVC (also known as UEP~\cite{zhao2010rd, abdallah2019h}).
%Some video codecs also support data partitioning and apply unequal protection using forward error correction (also known as UEP~\cite{zhao2010rd, abdallah2019h}), \ie having higher redundancy for more important information such as codec configuration and frame metadata.
However, in order to pick a suitable rate of redundancy, they need to estimate how many packets will be lost in advance. 
If the loss rate is underestimated, the redundancy will be insufficient to recover missing packets. %and the receiver has to %either wait for more packets or retransmitted packets.
On the other hand, adding excessive redundancy results in a higher bandwidth overhead
and, in turn, a lower video quality.
%, added, the quality will be lowered, even if all packets are received.
%once the received packets are decodable, receiving more packets does not improve the quality, so even if all packets are received, quality will be lower than without redundancy. 

\myparashort{Postprocessing error concealment}reconstructs missing data in lost packets at the {\em decoder}. 
These methods generally consist of two components.
First, the encoded packets should be decodable
when only a subset of the packets is received.
%decoding the partially received frame requires partitioning a video frame into independently decodable packets.
This is accomplished through 
%Modern video encoders use a few techniques to achieve this, 
INTRA-mode macroblock encoding~\cite{660830}, slice interleaving~\cite{ismaeil2000efficient}, or flexible macroblock ordering~\cite{lambert2006flexible}.
However, these approaches often compromise the encoder's ability to exploit redundancies across neighboring MBs, as adjacent MBs are either encoded in INTRA mode or split into different packets (in a checkerboard manner~\cite{kumar2006error, lambert2006flexible} or based on ROI detection~\cite{tan2011new}).
Therefore, these methods impair compression efficiency, causing the encoded frame size to inflate by 10\%--50\%~\cite{kumar2006error, wenger2002scattered, 4494082, dhondt2004flexible}. 

% First, to make the partially received packets decodable, some macroblocks (MBs) need to be set in \texttt{INTRA} mode~\cite{660830, lambert2006flexible}, and adjacent MBs need to be split into different packets in a checkerboard manner~\cite{kumar2006error, lambert2006flexible} or based on ROI detection~\cite{tan2011new}. This reduces the compression efficiency, 
% %limits the encoder's ability to exploits redundancies across neighboring MBs, so 
% and the encoded frame size typically inflates by 10\%-50\%~\cite{kumar2006error, wenger2002scattered, 4494082, dhondt2004flexible}. 
% INTRA mode, FMO, slice interleaving and the size inflation problem

Then, the decoder reconstructs lost data based on the received packets,
%data in the lost packets based on the data in the received packets,
using classic heuristics (\eg motion vectors interpolation~\cite{kolkeri2009error, zhou2010efficient, wang2013novel} and intra-block refreshing~\cite{kumar2006error} in H.264) or neural-network-based inpainting~\cite{xiang2019generative,kang2022error, sankisa2018video, mathieu2015deep,reparo}.
Recent work~\cite{reparo} use vision transformers~\cite{dosovitskiy2020image, arnab2021vivit} to directly predict the missing bits in the lost packets before frame decoding.
%However, since the encoder in these schemes is agnostic to the postprocessing on the decoder side, 
%it will aggressively compress each frame, and as a result, 
%each encoded packet contains little redundancy that can help reconstruct the missing
%motion vectors or residuals.
However, due to the encoder's lack of awareness of the decoder's postprocessing,
each encoded packet contains limited redundancy and information that could aid in reconstructing missing
motion vectors or residuals.
As a result, the reconstruction process is forced to guess the missing data when a packet is lost.
Even recent techniques still see a notable drop in video quality (\eg PSNR drops from 38 dB to 25 dB at a 20\% packet loss~\cite{nam2020novel}).
%, partly due to the
%well-studied notion of ``constrained hallucination''~\cite{liu2020brief, santos2020single}.

\name takes a different approach than FEC and error concealment. 
%\name jointly optimizes (via training) both the (neural-based) encoder and decoder, unlike error concealment that relies only on decoder-side postprocessing. Meanwhile, \name optimizes the encoder and decoder on a range of packet loss rates, unlike FEC that requires a pre-determined redundancy rate. 
Unlike error concealment that relies only on decoder-side postprocessing,
\name jointly optimizes (via training) both the (neural) encoder and decoder.
Unlike FEC that requires a pre-determined redundancy rate, \name's codec is optimized
across a range of packet loss rates. 

\mypara{Other schemes} While there exist other techniques that might help mitigate the impact of packet loss, their primary goals are \textit{not} loss resilience.
%, so they are less resilient to packet loss than the aforementioned schemes. 
Nevertheless, for the sake of completeness, we discuss some notable schemes here and also quantitatively compare \name against several of them in \S\ref{sec:eval}.

Scalable video coding (SVC)~\cite{schwarz2007overview, schierl2007mobile, swift} and fine-granularity scalability (FGS)~\cite{li2001overview,ma2022deepfgs} 
%\amz{same citation twice here}
aim to optimize rate-distortion (RD) tradeoff---video quality achieved by a single encoded bitstream under different received bitrates.
%Recent work also shows that neural encoders can improve SVC compression~\cite{swift}.
SVC encodes a video in multiple quality layers and sends data {\em layer by layer}.
This is feasible for on-demand video~\cite{swift,grad} but in real-time video, all packets of a frame are sent together to reduce frame delay (\S\ref{subsec:background}).
When a packet loss occurs to a base layer, it will block the decoding of any higher layers.
%so a packet loss can occur to any layer which can block the decoding of any higher layer.
% , and a new packet may not always improve quality, if the lower layers are not fully received.
For this reason, SVC is rarely used to improve unicast real-time video (though it is used in multicast video to serve users with heterogeneous network capacities~\cite{schierl2007mobile}).

%so that each newly received packet (or layer) will incrementally improve quality. 
%However, because it imposes a hierarchical order among the packets of a frame, 
%one missing packet can affect all higher quality layers, even if more packets are received in the higher quality layers.
%Though sending each layer reliably one by one is possible in on-demand videos~\cite{swift,grad}, real-time video clients send each frame as one burst to reduce frame delay (as explained in \S\ref{subsec:background}), so packet loss can affect any layer.
%As a result, SVC is rarely used to improve unicast real-time video (though it is used widely by multicast videos to serve users with heterogeneous network capacities~\cite{schierl2007mobile}).

% Reference frame and frame skipping is a form of error concealment though typically has lower quality.
% is another option if other error concealment schemes are not available.
%The encoder and decoder can work jointly to decide when skipping a frame does not hurt quality and the decoding states. 
There are a few alternatives to postprocessing error concealment.
For instance, when loss occurs, Salsify~\cite{salsify} reverts to an older but reliably received frame---instead of the last frame---as the reference frame, so the decoder can safely skip a loss-affected frame without hurting subsequent frames.
However, it needs more bits to encode the same quality than using the last frame 
% (which has a much smaller difference from the current frame) 
as the reference frame, \eg the P-frames between every other frame are 40\% greater in size than between two consecutive frames.
Similar limitations apply to long-term reference frame (LTR)~\cite{zhang2008error}, which makes each P-frame individually decodable if the long-term reference is received,
regardless of packet loss in between or not. 
% However, it shares a similar problem of large size overhead as Salsify.
Voxel~\cite{voxel} skips a loss-affected frame if the encoder indicates that skipping the frame does not affect video quality.
It works well for on-demand video where B-frames can be safely skipped, and the impact of a skipped frame will stop at the next chunk within a few seconds. 
Unfortunately, neither applies to real-time video.

\cmedit{Recently, deep learning has been used in super resolution~\cite{sivaraman2022gemino,kim2020neural,neuroscaler}, SVC~\cite{swift,ma2022deepfgs}, and postprocessing error concealment based on CNNs~\cite{ xiang2019generative, kang2022error, sankisa2018video, mathieu2015deep} or transformers~\cite{reparo, esser2021taming}. 
Super-resolution can reduce packet losses by sending the video in a lower bitrate and enhancing the video quality on the receiver side.
However, it still requires retransmissions to rectify frames impaired by packet loss. 
For SVC and postprocessing error concealment techniques, the aforementioned limitations inherent to these approaches remain, despite the use of deep learning.
}
{
Addressing the comment from reviewer D: My only suggestion is that the paper could provide more discussion about alternative designs.
}

% It has been used to improve compression efficiency (\eg~\cite{dvc,elf,hu2021fvc,swift, NEURIPS2021_96b250a9}), enhance quality (\eg super resolution~\cite{sivaraman2022gemino,kim2020neural,neuroscaler}), scalable video coding (\eg~\cite{swift,ma2022deepfgs}) and postprocessing error concealment (\eg~\cite{reparo, xiang2019generative, kang2022error, sankisa2018video, mathieu2015deep}).
% However, the fundamental limitations of these approaches remain despite the use of deep learning.

Loss resilience has also been studied under specific assumptions, such as availability of multi-path~\cite{meng2022achieving, dhawaskar2023converge}, early retransmission driven by router feedback~\cite{zuo2022deadline}, low-latency networks~\cite{ray2022sqp}, and availability of video gaming states~\cite{he2023neural,he2023real,wu2023zgaming}.
We do not make special assumptions in this work.
%\name does not make any special assumptions. % in this work.

\begin{figure*}[t!]
    % \vspace{-3pt}
    \centering
    \includegraphics[width=0.83\linewidth]{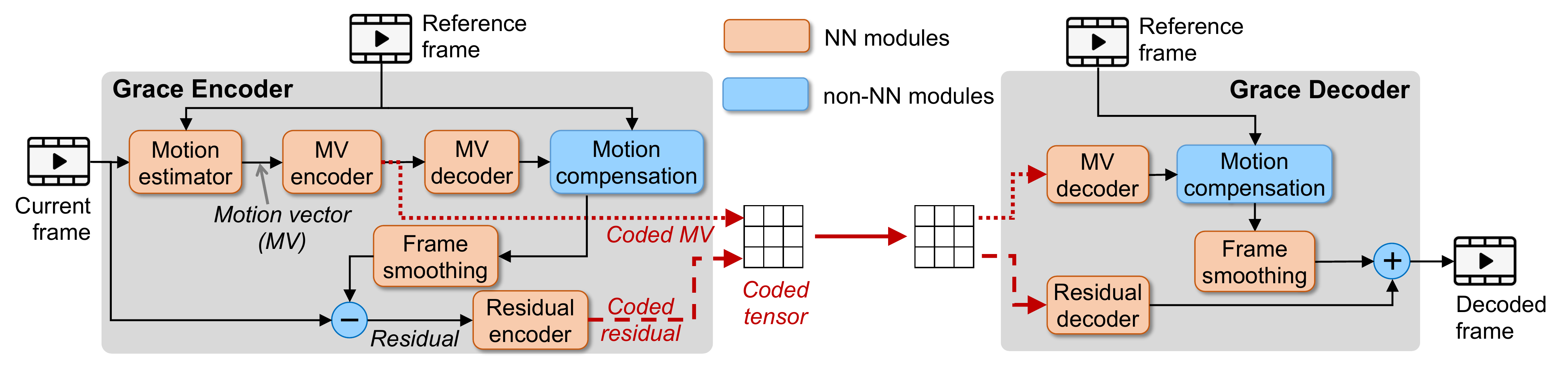}
    \vspace{3pt}
    \tightcaption{Workflow of \name's neural video codec.}
    \label{fig:dvc-arch}
    \vspace{-2pt}
\end{figure*}
% \jc{this section needs more work}

\tightsubsection{Neural video codec background}
\label{subsec:ae-background}
\vspace{-5pt}

%\jc{refer to fig~\ref{fig:dvc-arch}}

Our work is based on neural video codecs (NVCs),  which use learned neural networks ({NN}s), instead of handcrafted logic, to encode and decode video frames~\cite{dvc, swift, hu2021fvc}.
Recent NVCs have demonstrated comparable or even better compression efficiency than traditional video codecs for two reasons: 
\begin{packeditemize}
\item They leverage logical components commonly found in traditional video codecs,
% \footnote{For this reason, most autoencoder-based video coding focuses on inter-frames, the most common frame type in real-time videos. Our work also focuses on inter-frame coding and will include I-frame in \S\ref{subsec:protocol}.}, 
such as motion estimation, warping, and transformative compression (\S\ref{subsec:background}), replacing their handcrafted heuristics with NNs, which can learn more sophisticated algorithms from data.
\item These NVCs exhibit remarkable generalization across a variety of video content because of training on a large corpus of videos (\eg Vimeo-90K~\cite{vimeo-dataset}).
% , so their compression efficiency can reasonably generalize across a variety of video content (which 
This capability to generalize is also observed in our evaluation (\S\ref{sec:eval}).

\end{packeditemize}
% Several factors contribute to this advance.
% First, rather than encoding a frame with a monolithic NN, 

% attracted increasing attention in the multimedia community~\cite{dvc, swift, hu2021fvc}.

% % (We will discuss I-frame in \S\ref{subsec:protocol}, and B-frames are rarely used in real-time videos~\cite{ffmpegstreaming}.)
% It is comprised of steps (\eg motion and residual estimation) typically found in a traditional video codec (\S\ref{subsec:background}) but  uses NNs to replace
% %Most recent video autoencoders~\cite{??,??,??} combine useful concepts from traditional codecs (\eg motion vectors and residuals) and replace 
% handcrafted heuristics for motion estimation, compensation, and \st{transformation} \edit{compression} of residual and motion vectors.
% %, and they achieve compression efficiency comparable to recent video codecs.
% %, and achieve comparable compression efficiency as recent compression algorithms.

%\mypara{Limitations}

%Our work adopts the DVC architecture~\cite{dvc}, a popular NVC. %(depicted in Figure~\ref{fig:dvc-arch}).
%Two properties of NVCs make them particularly suitable for loss resilience.

%Despite their impressive compression efficiency, less attention has been given to making NVCs resilient to packet losses. 
Despite their exceptional compression efficiency, NVCs have received little attention so far in the context of loss resilience.
However, we believe NVCs have the potential to achieve greater loss resilience for the following reasons.
%We have two intuitive reasons to believe that NVCs have the potential to be more loss-resilient than existing approaches.
% \fyy{Reduce unnecessary}
\begin{packeditemize}
%NVCs are particularly suitable for loss resilience because of the following two reasons. 
\item First, unlike traditional codecs that map each pixel (or macroblock) to a distinct motion vector/residual, the highly parameterized NN of NVC's encoder can be trained to map the information of each pixel to multiple elements in output tensor, potentially making lost information recoverable.
%the NVC's encoder, which is a highly parameterized NN, could be trained to produce tensors that are robust to data losses. 
%This is because unlike traditional codecs that map each pixel (or macroblock) to a distinct motion vector and residual, the NVC's NN encoder could be trained to map information of each pixel to multiple elements in its output tensor, potentially making any lost information recoverable.
%bolstering its resistance to packet losses. 
\item Second, the NVC's decoder, comprising convolutional NNs, can be trained to decode not only a direct encoder output but also tensors that resemble those with perturbations such as random noise or zeroing.
In contrast, traditional codecs might fail to decode under similar circumstances.
%the NVC decoder is versatile, capable of decoding not only the direct output from the encoder but also tensors that look similar to that output. 
%Consequently, even if packet loss alters the encoder's output values (such as random zeroing or noise introduction), the NVC decoder can still decode the frame. 
%In contrast, traditional codecs might encounter decoding failures under similar circumstances.
\end{packeditemize}

Nevertheless, NVCs as is still lack tolerance to packet loss.
%still requires traditional schemes (like FEC) to handle packet loss. 
%Their lack of packet loss tolerance is due to two reasons.
Their standard training implicitly assumes that the encoder's output is identical to the decoder's input,
% \footnote{A recent work~\cite{??,??} adds Gaussian random noise\jc{check!} between the encoder and decoder to simulate wireless signal noise, but this is different from packet loss which masks a fraction of the data.}, 
so it does not prepare the NVC to handle data loss between the encoder and decoder. 
Meanwhile, entropy encoding used in conventional NVCs compresses the entire encoder output as a single bitstream, and thus any packet loss will render it undecodeable.

\name is an attempt at transforming NVCs to be resilient to different packet loss rates. % packet loss at different rates.
Our work is related to an emerging line of work on deep joint source-channel coding~\cite{kurka2020deepjscc, pmlr-v97-choi19a, bourtsoulatze2019deep}, which trains an NVC to encode images in a representation robust to signal noises.
\name differs with them on two key fronts. 
First, \name handles video frames, which cannot be treated separately as individual images because any error in one frame can propagate to future frames. 
Second, \name handles packet losses rather than physical-layer signal noises, which can be naturally modeled by differentiable linear transformations~\cite{bourtsoulatze2019deep,gunduz2019machine}.
%they target physical layer protection against signal noises, which can be modeled by linear transformations (as in~\cite{bourtsoulatze2019deep,gunduz2019machine}), but packet losses need to be 
%% is different from (and complementary to) transport-/application-layer protection against packet losses. 
%%While physical-layer noises can be modeled as differentiable linear transformations (as in~\cite{bourtsoulatze2019deep,gunduz2019machine}), 
%In contrast, packet drops are not naturally differentiable and need to be handled differently. % (see \S\ref{subsec:sim-losses}). 
%Second, 

% \tightsubsection{Summary}

In short, traditional error-resilient methods struggle to
maintain video quality across a range of packet losses.
Encoder-based forward error coding (FEC) optimizes quality only for a pre-determined maximum loss rate, whereas decoder-based postprocessing error concealment suffers from suboptimal quality especially at high loss rates.
On the other hand, existing NVCs have the potential to tolerate data perturbations
but are not explicitly designed to handle packet losses.
%\fyy{The last sentence is not clear as the previous subsections have not mentioned "perturbation".}

\begin{comment}

By contrast, a better coding scheme should optimize quality under different packet loss rates. 
We call it {\em data-scalable (DS) coding}.
With the same encoding of each frame, DS coding obtains quality similar to the classic codec (H.265) used in WebRTC without FEC when there is no packet loss; and with higher packet loss rates, the quality gradually degrades but is still better than state-of-the-art loss-resilient schemes.
In the next two sections, we will describe an autoencoder-based solution to approximate the effect of DS coding.

\end{comment}

%the aforementioned solutions either sacrifice the quality in absence of packet loss or fail to get decent quality in presence of high packet loss rates.
%As we will see, \name shows a more desirable relationship between quality and packet loss rate.

%\input{sections/motivate_v11}
%\input{sections/motivate_v10}
%\input{sections/motivate_v9}
%\input{sections/motivate_v8}
%\input{sections/motivate_v7}
%\input{sections/motivate_v5}
%\input{sections/motivate_v4}
%\input{sections/motivate_v3}
%\input{sections/motivate_v2}

%!TEX root = ../main.tex
%!TEX spellcheck = en_US

\vspace{-3pt}
\tightsection{Training \name's neural video codec}
\label{sec:core}

This section outlines the training process of \name's neural video codec (NVC).
%, and the next section will introduce its video delivery framework.
At a high level, \name \textit{jointly trains the neural encoder and decoder under a range of packet losses} to achieve enhanced loss resilience.
%During training, \name improves loss resilience by {\em jointly training the neural encoder and decoder under a range of packet losses}.
%We begin with \name's training process, and then describe how various packet loss rates are simulated during training.
% We begin with how to simulate packet loss, and then
% describe \name training process to optimize quality under various packet loss rates.

\mypara{Basic NVC framework}
Figure~\ref{fig:dvc-arch} depicts the workflow of \name's encoder and decoder (excluding entropy coding and packetization).
The encoder follows a similar logical process as a traditional video encoder (Figure~\ref{fig:codec-workflow}).
It first employs a neural network (NN) to estimate motion vectors (MVs) and encodes them into a quantized tensor using an NN-based MV encoder. Subsequently, the tensor is decoded back into MVs to match those received by the decoder. 
Next, the encoder applies these MVs to the reference frame to generate a motion-compensated frame, 
and uses a frame smoothing NN to increase its similarity with the current frame before calculating the residual differences between them.
%uses a frame smoothing NN to make the motion-compensated frame more similar to the current frame, and calculates the residual differences between the motion-compensated frame and the original frame. 
Finally, an NN-based residual encoder encodes the residuals into another quantized tensor. 
When the encoded MV tensor and the encoded residual tensors are received by the decoder,
they go through the NN-based MV decoder and residual decoder jointly trained with their respective encoders.
%Then, it will do motion compensation and run frame smoothing NN to get the decoded frame.
Appendix~\ref{app:details} provides more details of the tensors and NNs.

\begin{comment}
\mypara{Workflow of \name}
Our work adopts the architecture of a popular neural video codec, DVC~\cite{dvc}.
Figure~\ref{fig:dvc-arch} shows the workflow of \name's encoder (sender) and decoder (receiver).
The encoder first uses a neural network (motion estimator) to estimate the motion and encode the motion into a quantized tensor by MV encoder NN. 
Then it will apply the motion on the reference frame to get the motion-compensated frame.
Before calculating the residual, the encoder uses a frame smoothing NN to make the motion-compensated frame more similar to the current frame.
Finally, it computes and encodes the residual into a quantized tensor by residual encoder NN.
After getting the encoded motion vector and residual tensors, the decoder will first decode their original values. 
Then, it will do motion compensation and run frame smoothing NN to get the decoded frame.
We provide details of the tensors in Appendix~\ref{app:details}.
\end{comment}

Although both the encoder and decoder of \name contain multiple steps, they can be viewed as two differentiable models. 
%\mypara{Terminology}
%The encoder's output is a coded tensor that entails both encoded motion vector (MV) and encoded residuals. 
We denote the encoder by $\Coder$ (with its NN weights $\phi$) and the decoder by $\Decoder$ (with its NN weights $\theta$).
The encoder encodes a frame $\x$ into a {\em \aecode} $\y=\Coder(\x)$, and the decoder decodes $\y$ into a reconstructed frame $\hat{\x}=\Decoder(\y)$.
Traditionally, NVC seeks to minimize the following loss function:% over inputs from training frames \X\X:
\vspace{-0.04cm}
\begin{align}
%\textrm{\bf min~~~~} \mathbb{E}_{\x\sim \X} \Quality(\Decoder(\y), \y, \x)\textrm{, where }\y = \Coder(\x)
%\textrm{\bf min~~} \mathbb{E}_{\x} \Distort(\Decoder(\y), \x)+\alpha\cdot\Size(\y)\textrm{, where }\y = \Coder(\x)\\
%\textrm{\bf min~} 
\mathbb{E}_{\x}[{\color{black}\underbrace{\highlight{white}{$\Distort(\Decoder(\y), \x)$}}_{\text{\sf \footnotesize \textcolor{black!85}{Pixel error}}}}+
\alpha{\cdot}{\color{black}\underbrace{\highlight{white}{$\Size(\Coder(\x))$}}_{\text{\sf \footnotesize \textcolor{black!85}{Encoded size}}}}]\textrm{, where }
{\color{black}\underbrace{\highlight{black}{$\y = \Coder(\x)$}}_{\text{\sf \footnotesize \textcolor{black!85}{No data loss}}}}
%\y = \Coder(\x)
\label{eq:1}
\vspace{-0.12cm}
\end{align}
%\fyy{Can we add a pair of brackets after the expectation symbol? Same for the next equation.}
Here, $\Distort(\hat{\x},\x)$ is the pixel-level reconstruction error of the decoded frame $\hat{\x}$ (by default, L2-norm\footnote{
Note that the L2-norm (or mean squared error) of $\hat{\x}$ and $\x$ is closely related to the PSNR of $\hat{\x}$. To avoid this bias, 
our evaluation in \S\ref{sec:eval} measures the quality improvement in SSIM and subjective user studies.
% To avoid bias towards PSNR, our evaluation in \S\ref{sec:eval} measures the quality improvement in not only PSNR, but also SSIM and user subjective studies.
% Following the common practice of autoencoder training, the training loss function also includes the reconstruction loss of the motion-compensated prediction. 
%Pixel-wise distortion may not be the best way to capture human perception of reconstructed images, but it is often used in autoencoder training for its differentiability. We will evaluate the trained autoencoders in terms of more standard metrics, such as SSIM and PSNR (in RGB and YUV).
} of $\hat{\x}-\x$), and $\Size(\y)$ is the entropy-coded data size of $\y$ in bit-per-pixel (BPP).
The parameter $\alpha$ governs the size-quality tradeoff: a higher $\alpha$ leads to a smaller frame size, $\Size(x)$, but higher distortion (\ie poorer quality) of the reconstructed frame $\hat{\x}$.
% will tend to have better visual quality in $\Decoder(\z)$ but also higher bitrate. 
%In Eq.~\ref{eq:1}, s
As all the functions---$\Coder$, $\Decoder$, $\Distort$, and $\Size$ (approximated by a pre-trained NN~\cite{dvc})---are differentiable, the NN weights $\phi$ and $\theta$ can be trained jointly via gradient descent to minimize Eq.~\ref{eq:1}.

\begin{figure}[t]
    \centering
    \hspace{10pt}\includegraphics[width=0.95\linewidth]{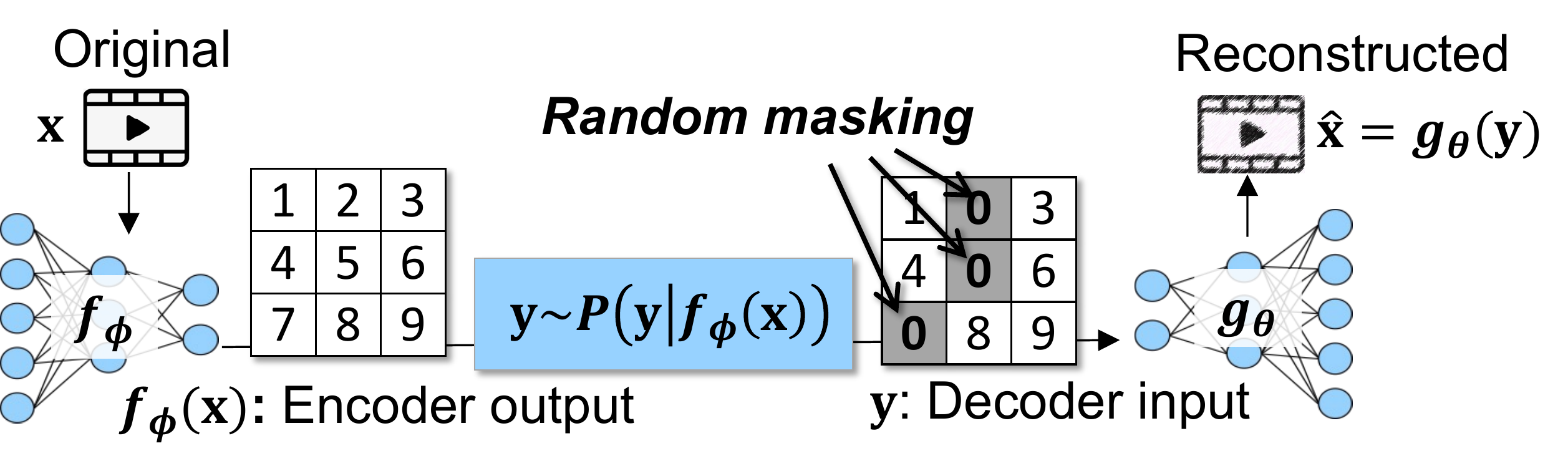}
    \vspace{-10pt}
    \tightcaption{Unlike traditional NVC training that assumes no data loss between the encoder and decoder, \name applies ``random masking''---setting a fraction of randomly selected elements to zeros---to the encoder's output.
    %Such random masking is equivalent to packet loss as shown in \S\ref{subsec:packetization}.
    % \fyy{I prefer not to use tightcaption, but if we were to keep it, let's at least leave ample vertical space around it. It's too crowded currently.}
    }
%    Contrasting the proposed loss-resilient design of autoencoders with traditional autoencoders.}
    \label{fig:grace-training}
    \vspace{-2pt}
\end{figure}

\mypara{Simulating packet loss during training}
We begin by pre-training an NVC using Eq.~\ref{eq:1}, which we refer to as \namepretrain, and then fine-tune it by introducing simulated packet losses in the following manner. 
\name simulates the impact of packet losses by randomly ``masking''---zeroing selected elements---in the encoder's output, $\Coder(\x)$, as shown in Figure~\ref{fig:grace-training}.
The fraction of zeroed elements is dictated by a distribution,
$\loss(\y | \Coder(\x))$, which represents the probability distribution of the resulting tensor $\y$ after random masking $\Coder(\x)$.
For instance, with a 33\% loss rate, $\loss(\y | \Coder(\x))$ is the probability of $\y$ 
arising from the random masking of 33\% of elements in $\Coder(\x)$, as illustrated in Figure~\ref{fig:grace-training}.
%The loss rate is sampled from a packet loss rate distribution (explained shortly).
%During training, \name simply simulates packet loss at a rate randomly selected from a distribution of packet loss rates (explained shortly), rather than running a packet-level simulator. 
%\passed{\amz{why not use a simulator? seems like it would introduce the burst-y distribution you explicitly induce later, but perhaps is not differentiable? Might be helpful to explain explicitly, this is one of the big contributions of this section IMO.}}
%We use $\loss(\y|\Coder(\x))$ to denote the probability distribution of the resulting tensor $\y$ after random zeroing $\Coder(\x)$.
%Given a packet loss rate distribution, 
Formally, \name jointly trains the encoder and decoder NNs to minimize: 
\vspace{-0.04cm}
\begin{align}
%\textrm{\bf min~~~~} \mathbb{E}_{\x\sim \X} \Quality(\Decoder(\y), \x)\textrm{, where }
%\color{blue}\underbrace{\highlight{blue}{\y∼\loss(\y|\Coder(\x))\y\sim \loss(\y | \Coder(\x))}}_{\text{\sf \footnotesize \textcolor{blue!85}{Simulating packet loss}}}
\mathbb{E}_{\x}[\Distort(\Decoder(\y), \x)+\alpha{\cdot}\Size(\Coder(\x))]\textrm{, where }
\color{blue}\underbrace{\highlight{newblue}{$\y\sim \loss(\y | \Coder(\x))$}}_{\text{\sf  \footnotesize \textcolor{blue}{Simulate packet loss}}}
\label{eq:2}
\vspace{-0.12cm}
\end{align}
The key difference from the traditional objective in Eq.~\ref{eq:1} is the distribution function $\loss$ (highlighted in blue), which captures the distribution of decoder input under packet loss.

To train the weights of $\phi$ and $\theta$ under the random perturbations of $\loss$, we employ the REINFORCE trick~\cite{Kingma2014} (commonly used in reinforcement learning~\cite{zhang2021sample, peters2008reinforcement}) to approximate the gradient through Monte Carlo sampling. A more detailed mathematical formulation is included in Appendix~\ref{app:reinforce}.
%\jc{Yihua, i remember Anton wrote a mathematical formulation and proof of this. can you bring it back in appendix? i remember multiple reviewers have asked for more details.}

%It should be noticed that the random probability distribution $\loss$ may not be expressed as a  function of $\phi$ and is thus not differentiable with respect to $\phi$.
%To train the weights in $\phi$, we use the REINFORCE trick~\cite{Kingma2014} to approximate this gradient by Monte Carlo Sampling. This is commonly used in reinforcement learning~\cite{zhang2021sample, peters2008reinforcement}.

%\jc{how about we get directly to "how packet loss affects the training objective of neural codecs" next?
%it's important to explain "how" to simulate packet loss, but it seems to break the flow. 
%also, it seems more natural to explain how packet loss is simulated in training and how it is done in testing together?}

\begin{figure}[t!]
    \centering
    \includegraphics[width=0.77\linewidth]{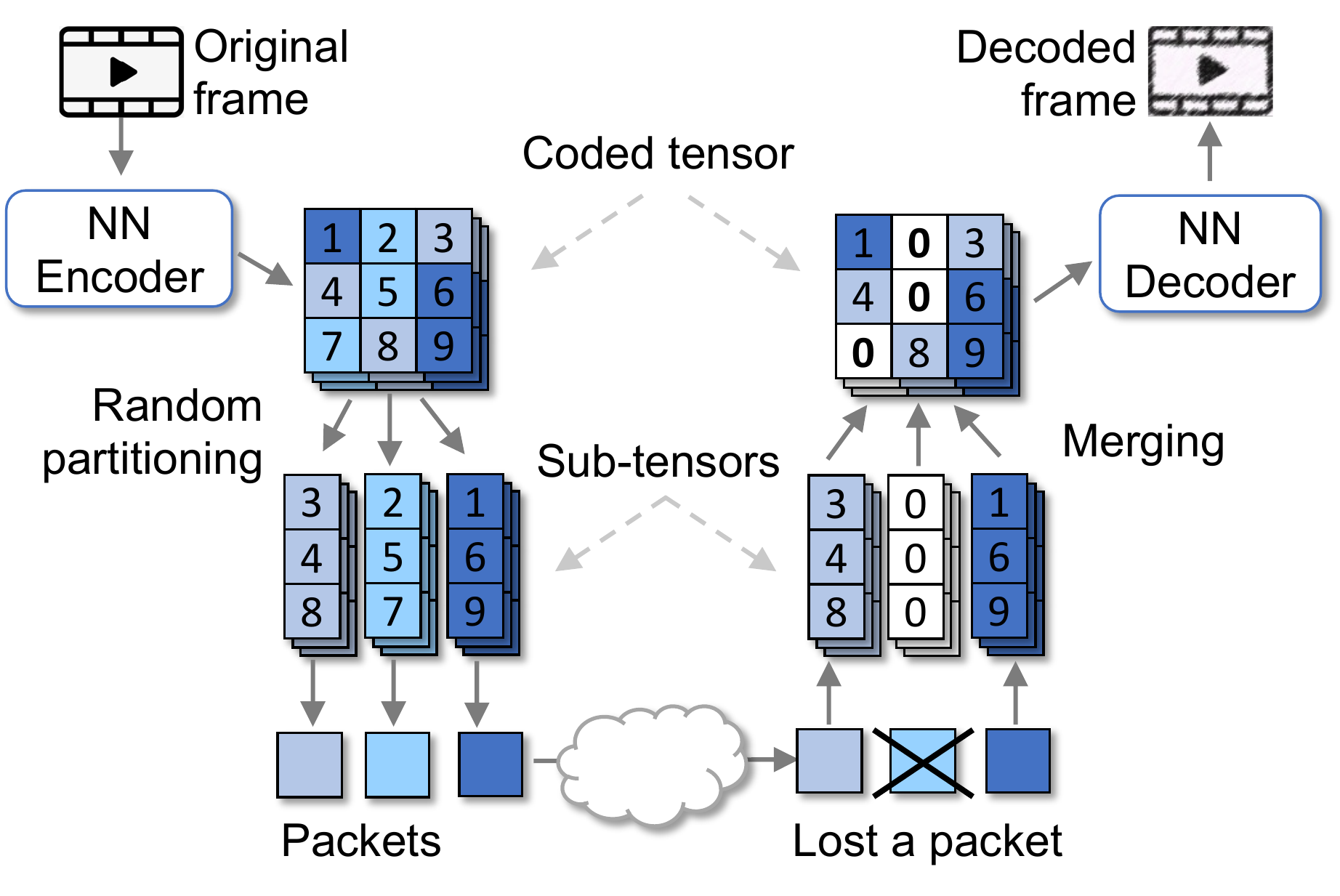}
    \tightcaption{\name's reversible randomized packetization.
    The tensor elements mapped to a lost packet will be set to zeros.}
    \label{fig:per-packet}
    \vspace{-3pt}
\end{figure}

\mypara{Choosing simulated packet loss rates}
% \yc{talking about our consideration, and move the detailed things into implmentation}
% \fyy{This section sounds like having many implementation details. If not critical design, consider moving it to the implementation section?}
%A key design choice in \name is the function \loss that governs what random data erasure rate the autoencoder learns to handle during training. 
To prepare \name's NVC to handle a wide range of loss rates, 
it is essential to simulate such losses in training.
%we should simulate a wide range of loss rates in training.
One approach is to select loss rates uniformly at random from [0, 100\%) and apply them to the encoder output $\Coder(\x)$.
However, the resulting NVC turns out to perform poorly, especially when dealing with low loss rates. 
Notably, even when high loss rates (\eg over 80\%) are introduced only in a small fraction of training samples, we empirically observe a significant drop in video quality under low loss rates while the quality improvement under high loss rates is only marginal.
This phenomenon could be attributed to the encoder's tendency to incorporate more redundant information to prepare for high loss rates, adversely affecting video quality at low loss rates.
%This might be because the encoder will learn to add more redundant information to prepare for high loss rates, which hurts the quality at low loss rates. 
%Since a high packet loss rate (\eg over 80\%) is relatively rare~\cite{rudow2023tambur} d
Therefore, a practically effective distribution should cover both low and high loss rates, with a slight emphasis on low losses. 
Our final choice of loss rate distribution is described in \S\ref{subsec:impl}.
%provides the details about the final loss rate distribution used in our training.
%We leave the exploration of different loss rate distributions to future work.

%\mypara{Apply packet loss during inferencing}
\mypara{Packetization during inferencing}
% \yc{recall in training, we don't do packetization and just apply random masking. So during inference we ....}
% \fyy{The first sentence is not clear, so I didn't fully understand why and when the reversible randomized packetization is needed,
% and how it's related to simulated losses or real losses.}
%Since we use random masking to simulate packet losses during training, we also need to ensure that the impact of actual packet loss is the same as such random masking.
Recall that during training, we simulate packet loss by applying random masking rather than replicating the actual packetization and packet dropping process.
Therefore, it is important to ensure that the impact of actual packet loss during runtime mirrors the effects of random masking.
To achieve this, \name employs reversible randomized packetization as shown in Figure~\ref{fig:per-packet}.
%Figure~\ref{fig:per-packet} shows how \name splits the encoder's output into individual packets.
\name's sender first splits the encoded tensor of a frame (both the encoded MVs and encoded residual) into multiple subtensors using a uniform random mapping. 
We use a reversible pseudo-random function to generate the mapping so that the receiver can correctly recover the original tensor with the same random seed. 
Specifically, we map the $i^{\textrm{th}}$ element to the $j=(i\cdot p\mod n)^{\textrm{th}}$ packet at the $[(i\cdot p-j)/n]^{\textrm{th}}$ position, where $n$ is the number of packets and $p$ is a prime number.
If a packet is lost, the decoder assigns zero to each element whose position is mapped to the lost packet. 
Consequently, an $x$\% packet loss rate has the effect of randomly zeroing $x$\% of the values in the encoder's output tensor.
\footnote{\cmedit{That said, such reversible random packetization requires a frame containing multiple packets. Therefore, \name's encoder controls the packet size such that each frame has at least 2 packets, since real-time video packets don’t need to be as large as 1.5~KB~\cite{sharma2023estimating} in practice.}{Addressing the comment from reviewer B: Reversible randomized packetization requires that you have several packets of data for each frame, or you'd have to buffer them in the decoder.}}
\S\ref{subsec:packetization} explains how each subtensor is losslessly compressed via entropy encoding into the bitstream of a packet, but this lossless compression is bypassed during training for efficiency purposes.

\begin{figure}[t!]
     \centering
     \hspace{0.2cm}
     \includegraphics[width=0.82\linewidth]{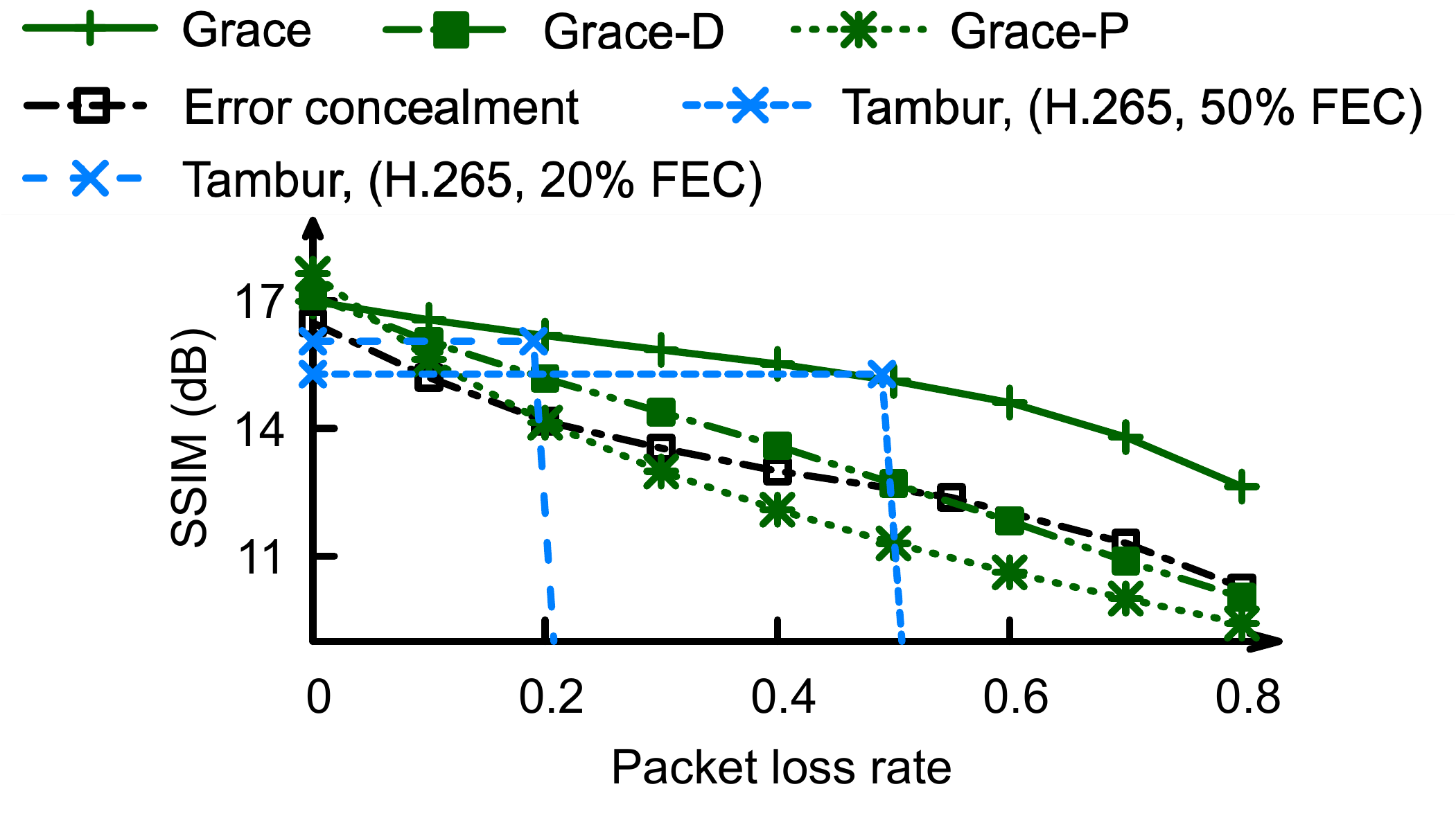}
     \tightcaption{Although \namedecoder and \namepretrain attain slightly better quality than \name in the absence of packet loss, they are much less resilient to loss than \name.}
     \label{fig:ssim-loss-abl}
\end{figure}

\myparaq{Why is \name more loss-resilient}
%Jointly training the encoder and decoder under different packet loss rates enhances \name's loss tolerance for two main reasons.
Unlike decoder-side error concealment, the joint training ensures that the {\em encoder} is also aware of packet losses. 
Empirically, we observe that \name's encoder tends to produce more non-zero values in its output than an NVC pre-trained on the same dataset but without simulated packet loss.
This increase in non-zero values can be viewed as more ``redundancy,'' as the encoder
embeds each pixel's information into multiple elements in its output tensor,
assisting the decoder in discerning loss-affected elements (from intended zeros) and reconstructing video better under packet losses.
%As the neural encoder embeds each pixel's values into multiple elements in its output tensor, this increase in non-zero values within encoder output can be viewed as more redundancy, helping the decoder discern loss-affected elements and facilitate better video reconstruction under packet losses. 
\S\ref{subsec:eval:micro} empirically shows that training only the decoder with simulated loss cannot reach the same level of loss resilience %as jointly training both the encoder and the decoder
(Figures~\ref{fig:ssim-loss-abl} and~\ref{fig:late-visual-example}).

\tightsection{Real-time video framework}
\label{sec:delivery}

%\jc{
%\begin{itemize}
%\item coding of i-frames (compression)
%\item bitrate control (speed)
%\item speedup on cpu (speed)
%\item end-to-end protocol (why no synchronization, interaction with tcp and abr) (resilience)
%\end{itemize}
%}

%\jc{a table mapping the techniques to effects
%\begin{itemize}
%\item compression: i-frames, per-packet entropy
%\item speed: bitrate control, cpu speedup
%\item resilience: retraining, end-to-end protocol
%\end{itemize}
%}

With the training techniques detailed in \S\ref{sec:core}, \name's NVC acquires the ability to withstand simulated packet losses.
This section describes the integration of this NVC into a real-time video delivery framework:
\name entropy-encodes the neural encoder's output into packets (\S\ref{subsec:packetization}), streams frames under packet loss (\S\ref{subsec:protocol}), and accelerates encoding and decoding across various devices (\S\ref{subsec:fast-coding}).
%\item How to entropy encode NVC encoder's output into packets (\S\ref{subsec:packetization}), 
%\item How to stream a sequence of frames (\S\ref{subsec:protocol}), and
%\item How to speed up encoding/decoding on resource-constrained devices without sacrificing loss resilience (\S\ref{subsec:fast-coding}).
%\item How we implement each component in \name (\S\ref{subsec:impl})
%\end{packeditemize}
% \yc{I moved entropy coding to 4.1, since I feel this is more smooth: first talking about tensor -> packets for each frame, then discuss how we stream sequence of frames, finally some optimizations}

\tightsubsection{Entropy encoding the encoder's output}\label{subsec:packetization}

As mentioned in \S\ref{sec:core}, \name splits the encoder's output into subtensors using a reversible-random function, with each subtensor corresponding to an individual packet.
Similar to classic codecs such as H.265 and VP9, each subtensor undergoes lossless compression into a bitstream through arithmetic (entropy) coding.
An arithmetic encoder uses an underlying symbol distribution to compress the values in the tensor.
%at each position of the tensor.
%Unlike classic codecs that use hand-tuned heuristics such as CABAC, 
Instead of relying on hand-tuned heuristics (\eg CABAC~\cite{CABAC} in H.265), 
we adopt the method described in~\cite{dvc}, training a distribution estimator in conjunction with the neural encoder and decoder to better estimate the symbol distribution of each encoder output.
%have a better distributional representation of the values in the encoder output. 
Since \name decodes individual packets independently, the symbol distribution of a packet must be sent as part of the packet to the decoder, which implies that
%Since the decoder uses the same symbol distributions to decode each packet, the symbol distribution must be sent together as part of a packet. 
%However, putting such distribution in each packet incurs an overhead.
the size overhead of symbol distributions increases with more packets. 

%and may greatly inflate the total size if the encoder's output is split into many packets. 
%However, because \name divides the encoder's output tensor into subtensors, each being separately entropy encoded into a packet, thus each packet must include a copy of its own symbol distribution.
%This will greatly increase the size if the data of a frame is split into many packets. 

\name reduces this overhead by employing a simpler symbol distribution that requires fewer bits to store within each packet. 
%, which governs the tradeoff between the size of a symbol distribution and the size of the entropy-encoded data.
%A coarser-grained symbol distribution needs fewer bits to store but may increase the size of the entropy-encoded data. 
%\name's insight is that we can 
Specifically, \name trains the neural encoder to regularize the distribution of values in each encoder's output channel (224 channels in total) to conform to a zero-mean Laplace distribution.
In doing so, the symbol distribution only needs to store the variance for each channel while still effectively compressing the encoder's output tensor. 
As a result, the symbol distribution now requires only $\sim$50 bytes per packet to store, a reduction from 40\% of the packet size to 5\%, without notably affecting the compression efficiency.

\vspace{-2pt}
\tightsubsection{Streaming protocol}
\label{subsec:protocol}
\vspace{-8pt}

%\jc{this subsection is a bit too long. tighten it please.}

% To extend the P-frame DS autoencoder trained in \S\ref{subsec:training} for multiple frames, the first challenge is potential out-of-sync {\em states} between the encoder and decoder when packet loss occurs.
%Two challenges arise when extending the P-frame DS autoencoder trained in \S\ref{subsec:??} to the delivery of multiple frames: (1) potential out-of-sync coding state between the encoder and decoder when packet loss occurs, and (2) how to encode I-frames and how frequently to add I-frames.

\mypara{Basic protocol of \name}
The encoder of \name encodes new frames at a fixed frame rate.
When any packet for the next frame arrives, the decoder immediately attempts to decode the current frame.
Unless all packets of the current frame are lost (which triggers a request for resending the frame), the decoder will decode the current frame using whatever packets have been received.
%the received packets, even if some packets of the frame are lost. 
We refer to a frame decoded using partially received packets as an {\em incomplete frame}.
%The decoder will decode the current frame if all the packets of the frame arrive, or some packet of the current frame arrives and it receives a packet from a new frame.
% when any packet of the next frame is received, the decoder will decode the current frame, %using the neural decoder, 
% if at least one packet of the current frame is received.
%If no packet of the current frame is received, the decoder will request the sender to send the frame again.
However, while \name can decode incomplete frames with decent quality, using these incomplete frames as reference images for decoding future frames causes the encoder's and decoder's states to be ``out of sync,'' \ie the next frame will be decoded based on a different reference image than the one used during encoding.
%However, this basic protocol may degrade the quality of frames not impacted by packet loss.
%This is because although \name's decoder can decode the incomplete frame with decent quality, decoding an incomplete frame makes the encoder/decoder state ``out-of-sync'', \ie the encoding of the next frame will use a different reference image than its decoding.
This inconsistency causes error to propagate~\cite{wang1998error} to future frames
even if all their packets arrive without loss.
%(an issue known as ``error  propagation''.

\begin{comment}
\st{This is because if a reference frame is decoded using a subset of packets, the encoder/decoder state will be ``out-of-sync''---the encoding of the next frame will use a different reference image than its decoding.
For instance, the sender uses the decoded $6^{\textrm{th}}$ frame as the reference to encode the $7^{\textrm{th}}$ frame, expecting the receiver can decode the $7^{\textrm{th}}$ frame using the same reference. }
\st{However, if \name's decoder decodes the $6^{\textrm{th}}$ frame based on a subset of its packets 
%(Figure~\ref{fig:state-sync}(a)),
it will be different from the reference image used by the encoder.
As a result, the $7^{\textrm{th}}$ frame will have slightly lower quality even if there's no packet loss because of such inconsistency in reference frames.}
\st{Similarly, since the $7^{\textrm{th}}$ is also out of sync with the sender's state, the $8^{\textrm{th}}$ frame will suffer too, and so forth.}

\yc{I've removed the long text example and Figure~\ref{fig:state-sync}(a). They used a lot of space to discuss a widely known concept, \ie error propagation. }
\end{comment}

%\begin{figure*}[t!]
%    \centering
%         \centering
%%        \includegraphics[width=0.99\linewidth]{figs/per-packet-encoding-v3.pdf}
%        \includegraphics[width=0.9\linewidth]{figs/state-sync-v1.pdf}
%    \tightcaption{}
%    \label{fig:per-packet}
%\end{figure*}

One strawman solution to resolve error propagation is to synchronize the encoder and decoder on each frame.
However, 
%this will negate the benefit of \name's, as 
the encoding of each frame would be blocked until it knows which packets are used to decode the previous frame. This synchronization delay would render pipeline encoding, transmission, and real-time decoding infeasible.

\mypara{Optimistic encoding with dynamic state resync}
\name employs two strategies to prevent out-of-sync states from blocking the encoder or the decoder.
% \name follows a different method so that the out-of-sync states will block neither the encoder nor the decoder.
% The idea is two-fold. 

First, the encoder {\em optimistically} assumes all packets will be received and encodes frames accordingly, taking advantage of \name decoder's tolerance to packet losses for a small number of frames.
For instance, \S\ref{subsec:eval:resilience} shows that \name is resilient against packet loss across 10 consecutive frames.

\begin{figure}[t!]
    \centering
    \includegraphics[width=\columnwidth]{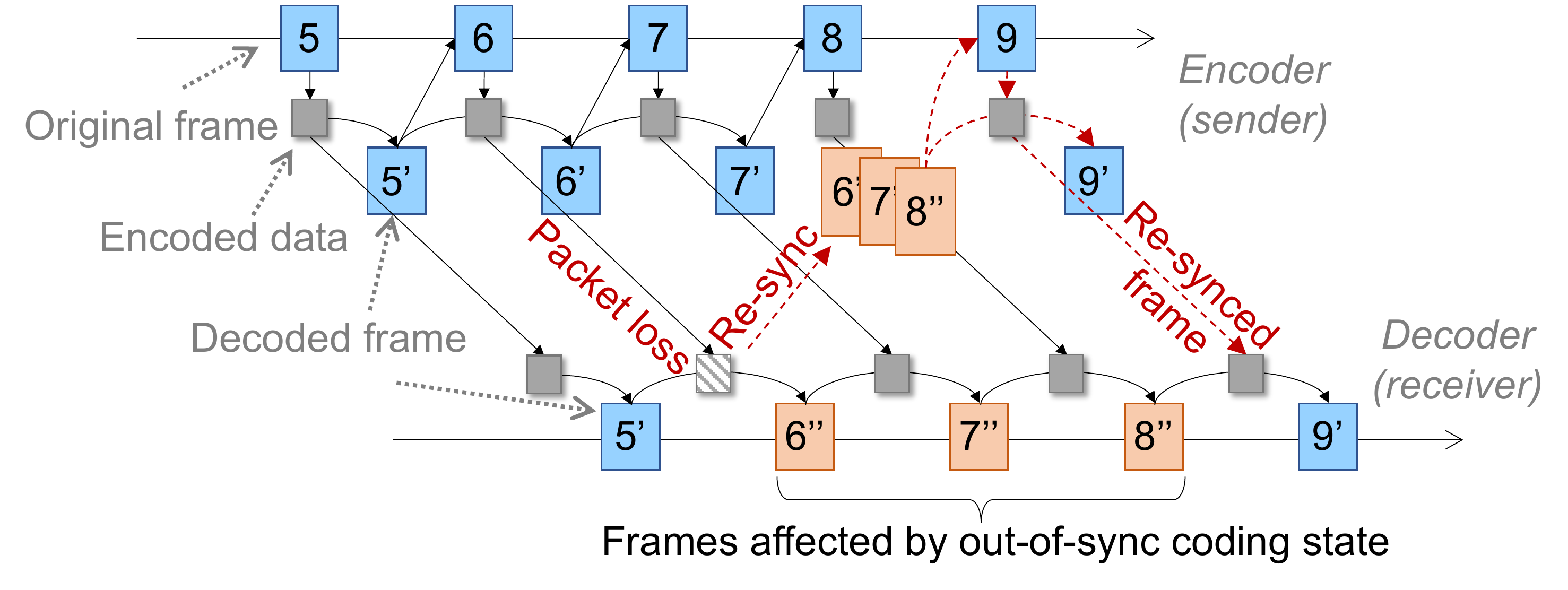}
    \vspace{-10pt}
    \tightcaption{
    %Example of \name's optimistic encoding and state resync. 
    Packet loss introduces discrepancies between the encoder's and the decoder's reference frames. \name's state resync efficiently rectifies these discrepancies without causing interruptions for either the encoder or the decoder.
    %\fyy{This figure is better than before,
    %but still too crowded with too many arrows. Is there a way to further simplify it?}
    %Examples of how to deal with loss-affected frames: 
    %(a) delayed by retransmission, (b) ignoring out-of-sync encoder/decoder states, and (c) 
    %\name's frame optimistic encoding and state resync.\yc{the font of this figure should be larger}
    }
    \label{fig:state-sync}
    \vspace{-2pt}
\end{figure}

Second, when receiving an incomplete frame, the decoder, without stopping decoding new frames, requests the encoder to dynamically resynchronize the state in the following manner.
%In \name, neither encoder nor decoder will be blocked by incomplete frames.
% When an incomplete frame is decoded, the receiver notifies the sender which packets have been used to decode recent frames without stopping decoding new frames. 
Upon receiving a resync request, the encoder re-decodes the recent frames starting from the incomplete frame, using only the subset of packets received by the decoder (as indicated in the resync request), to compute the latest reference frame used by the decoder.
% to reconstruct the same reference frame seen by the decoder. 
%As explained shortly, such re-decoding can be greatly sped-up.
As illustrated in Figure~\ref{fig:state-sync}, if the encoder is about to encode the $9^{\textrm{th}}$ frame and learns that the $6^{\textrm{th}}$ frame has been decoded using partially received packets, it then quickly re-decodes frames from the $6^{\textrm{th}}$ to the $8^{\textrm{th}}$. The $8^{\textrm{th}}$ frame now aligns with the receiver's observation and thus is used as the reference frame for encoding the $9^{\textrm{th}}$ frame.

A potential speed bottleneck is the re-decoding of frames during state resynchronization (\eg the $6^{\textrm{th}}$ to $8^{\textrm{th}}$ frames in Figure~\ref{fig:state-sync}). 
Fortunately, the encoder can re-decode these frames much faster than the regular decoding process by running only the motion decoder and the residual decoder.
%skipping both motion estimation and frame smoothing NNs in Figure~\ref{fig:dvc-arch}.
The insight is two-fold.
First, motion estimation, motion encoding, and residual encoding can be skipped because these frames have already been decoded once at the encoder side, so the re-decoding only needs to estimate the incremental changes caused by the lost packets. 
Second, while skipping the frame smoothing NN may impact the compression efficiency of the last frame (\eg the $9^{\textrm{th}}$ frame in Figure~\ref{fig:state-sync}), it only affects a single frame since the next frame will still be optimistically encoded. 
Appendix~\ref{app:resync} provides more details on the dynamics re-decoding,
and \S\ref{subsec:eval:micro} analyzes its runtime overhead.
%Overall, it saves 93\% of encoding time per frame.\jc{can we have a forward pointer for this? seems strange to show an evaluation result here}

% The insight is two-fold.

%allowing it to re-decode these frames using a different but faster method.
%The insight is that the decoding does not have to have the highest quality, which allows us to skip the motion-smoothing NN in motion compensation (Figure~\ref{??}) which significantly reduces the delay by \fillme\%.\jc{need a few words on why}
%For this optimization to work, the receiver also needs to re-decode the frames from 6-8 without the motion-smoothing NN, but since this is much faster, it will not add much new overhead. 

%, but it leverages the loss resilience of the new autoencoders.
% Like \name, NACK and Salsify allow senders to burst packets without waiting for acks, and when packet loss happens, NACK retransmits lost packets which delays the decoding of future frames and Salsify rewinds the encoder to an old state to generate the next P-frame, allowing the decoder to skip loss-affected frames.
% However, \name's key difference is that 

\name's approach of optimistic encoding and dynamic state resynchronization capitalizes on a key advantage of \name's NVC---it does not need to skip or block the decoding processing for loss-affected frames; instead, it can decode them with decent quality {\em while} the encoder's and decoder's states are out-of-sync for a few frames, thus reducing frame delay.
% This makes \name's optimistic encoding with dynamic state resync 
This approach differs from NACK (negative acknowledgement) in WebRTC~\cite{holmer2013handling}, which requires blocking the decoding of loss-affected frames, and from Salsify's state synchronization~\cite{salsify}, which skips all loss-affected frames.

%Although \name's optimistic encoding with dynamic state resync may sound similar to NACK (negative ack) in WebRTC and the encoder/decoder state synchronization in Salsify, but it has a key difference. 
%The receiver in \name does not need to skip or block the decoding of loss-affected frames; instead can still decode loss-affected frames with decent quality before encoder/decoder states are synchronized, so it reduces frame delay while maintaining high quality. 
\passed{\amz{This is also framed in a roundabout way, but is also a key insight of this work IMO -- using your packetization allows for decoding lossy frames with less delay. Switching these sentences ("Another benefit of \name is that the receiver does not need to skip... This differentiates our dynamic state resync from NACK and Salsify")} \yc{reorganized the contents}}

\vspace{-2pt}
\tightsubsection{Fast coding and bitrate control}
\label{subsec:fast-coding}
\vspace{-8pt}

%\jc{the i-patch and i-frame discussion has been moved to appendix. Yihua, please do a thorough check to (1) make sure all essential discussion on iframes are still covered, to avoid comments like they only explained p-frames and what if i-frame is not loss resilient, etc, 
%and (2) redirect all mentioning of iframe and ipatch to the appendix section}

\begin{figure}[t!]
    % \vspace{-12pt}
    \centering
    \includegraphics[width=0.93\linewidth]{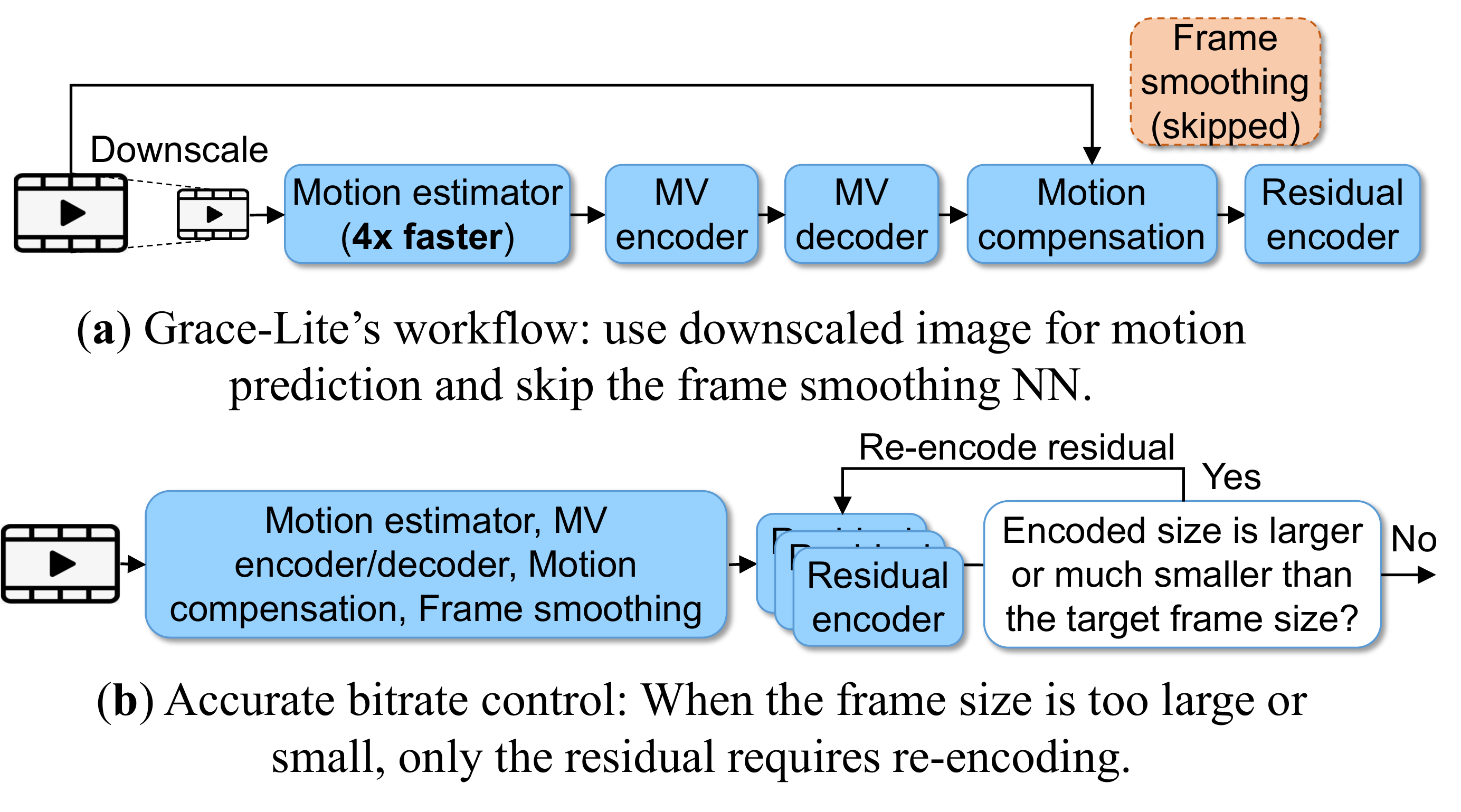}
    \vspace{4pt}
    \tightcaption{\name adapts the NVC for efficient execution on CPUs and accurate bitrate control.}
    \label{fig:speedup}
    \vspace{-2pt}
\end{figure}

\mypara{Fast encoding and decoding}
Using a standard GPU runtime with PyTorch JIT compiling~\cite{torchcompile}, \name meets the latency requirement for real-time video communication on GPUs. 
As shown in \S\ref{subsec:eval:micro}, \name encodes and decodes 720p video at 31.2 and 51.2~fps respectively, on an NVIDIA A40 GPU (5$\times$ cheaper and 3$\times$ slower than A100).
However, \name's NVC remains too heavy to run on laptops with CPUs and mobile phones.
%resource-constrained devices such as laptops and mobile phones.
To address this, we develop \namelite, a lightweight version of \name that incorporates three optimizations (Figure~\ref{fig:speedup}a): 
{\em (i)} motion estimation NN operates on $2\times$ downsampled frames, speeding up the motion estimation by $4\times$; 
{\em (ii)} frame smoothing NN is skipped; % in Figure~\ref{fig:dvc-arch}; 
{\em (iii)} the floating point precision in NNs is reduced from 32 bits to 16 bits, making the inference $2\times$ faster.
% To speed up \name on such devices, we present \namelite, a lightweighted version of \name.
% As shown in Figure~\ref{fig:speedup}a, \namelite implements the following optimizations:
% \begin{packeditemize}
%     \item We run the motion estimation NN on a $2\times$ downsampled frame, speeding up the motion estimation by $4\times$. 
%     \item We skip the frame smoothing NN in Figure~\ref{fig:dvc-arch}. 
%     \item We reduce the NN floating point precision from 32-bit to 16-bit, making the inference $2\times$ faster.
% \end{packeditemize}
These optimizations allow \name to encode and decode frames on an iPhone 14 Pro at 26.3 and 69.4~fps when compiled with the CoreML~\cite{coreml} library, while maintaining similar loss resiliency as \name (\S\ref{subsec:eval:micro}).
% Though \namelite's loss resilience and compression efficiency is slightly worse than \name, it still outperforms other loss-resilient baselines, including neural error concealment that cannot run in real-time on such devices. 

% Our insight is by trading off a bit of compression efficiency, the encoding/decoding can be drastically sped up---the above optimization speeds up the encoding by more than $10\times$ while only having a small impact on compression efficiency. 
% \S\ref{subsec:eval:micro} shows a more detailed study about the effect of the optimizations.

\mypara{Accurate bitrate control}
%Any video encoder must encode a frame at a size barely below the target frame size.
Video encoders are expected to encode frames to match the target frame sizes.
% Typically this can be done either by a heuristic to predict which quality level can yield the right frame size
%  encoding (as in H.264~\cite{OPENH264}) or encoding the frame multiple times at different quality levels (as in Salsify~\cite{salsify}). 
Similar to Salsify~\cite{salsify}, \name encodes a frame multiple times at different quality levels but in a faster way than encoding the frame from scratch each time
% doing the whole encoding process on a frame multiple times 
(illustrated in Figure~\ref{fig:speedup}b). 
To achieve this, \name trains multiple neural encoders, each with a different $\alpha$ (in Eq.~\ref{eq:2}) to enable different quality-size tradeoffs.
During the training phase, adjustments are confined to the residual encoders and decoders, leaving other NN weights fixed. 
Thus, once a frame is encoded, both the motion vector and residual are reusable, with the residual undergoing further encoding through different encoders, each producing a different frame size. 
This procedure, taking under 3~ms, can encode a frame multiple times using residual encoders with distinct $\alpha$ values. In practice, residual sizes can vary from 0.1$\times$ to 10$\times$ the MV size, allowing \name to cover a wide range of bitrates (\S\ref{subsec:eval:resilience}).

\vspace{-2pt}
\tightsubsection{Implementation of \name}\label{subsec:impl}
% \fyy{Put the implementation in a new section?}
\vspace{-5pt}

\name is implemented in $\sim$2000 lines of Python code, including its NVC, packetization, bitrate adaptation, and state synchronization protocol. 
%We will make our codebase and testing script available upon publication.
%Our codebase and testing scripts are available at \url{https://github.com/ApostaC/Grace}.

\mypara{Training}
\name's NVC model architecture is based on a recent work called DVC~\cite{dvc}.
We fine-tune \name from the pre-trained DVC model on the Vimeo-90K~\cite{vimeo-dataset} dataset, under the following distribution of simulated 
\cmedit{per frame packet loss (\S\ref{subsec:background})}{Addressing the comment from reviewer B: Packet loss rate is too high {10\%, 20\%, 30\%, 40\%, 50\%, 60\%}.}:
with an 80\% probability, the loss rate is set to 0\%; with a 20\% probability, the loss rate is randomly selected from \{10\%, 20\%, 30\%, 40\%, 50\%, 60\%\}
\footnote{\cmedit{The packet loss rate should follow a uniform distribution covering a continuous range of losses (e.g., [0, 60\%]). However, we empirically observe that using a discrete loss rate distribution makes the model converge faster without sacrificing the loss resilience.}{Addressing the comment from reviewer B: In your training, you have shown a wide distribution of loss, whereas the loss rates are very discrete. This seems counter-intuitive}}.
\cmedit{By using this loss distribution, \name can be resilient to a wide range of loss rates without assuming the underlying network loss pattern.}{The loss rate pattern is assumed to be uniform, and the Internet is hardly the case. Loss happens mostly in bursts (due to congestion) and the loss rate should change over time. }
To achieve accurate bitrate control (\S\ref{subsec:fast-coding}), we first fine-tune an NVC with a default $\alpha$ ($2^{-7}$) using Eq.~\ref{eq:2}. Subsequently, we perform fine-tuning with 11 $\alpha$ values spanning from $2^{-8}$ to $2^{-15}$,
specifically to refine the residual encoder and decoder for bitrate adaptation.
%The second fine-tuning only changes the residual encoder and decoder for fast bitrate adaptation (\S\ref{subsec:fast-coding}).
% The fine-tuning involved two steps: initial fine-tuning of a base model with different packet losses (\S\ref{sec:core}), followed by employing 11 discrete $\alpha$s (Eq.~\ref{eq:2}), spanning from $1/64$ to $1/32768$, to refine the residual encoder and decoder for bitrate adaptation (\S\ref{subsec:fast-coding}). 
With a learning rate of $10^{-4}$, each fine-tuning step takes about 1--2 hours on an Nvidia A40 GPU.

\mypara{Delivery}
We use \textit{torchac}~\cite{torchac} for entropy encoding and decoding.
\name utilizes PyTorch JIT compilation~\cite{torchcompile} when running on GPUs, while \namelite leverages CoreML~\cite{coreml} for inferencing on mobile devices.
Both \name and \namelite operate using 16-bit floats at runtime.
\name uses BPG~\cite{bpg} to encode and decode I-frames every 1000 frames, and can be integrated with any congestion control (CC) algorithms.
Due to space limitations, we provide more details about I-frames and CCs in Appendix~\ref{app:ipatch} and ~\ref{app:cc}.

\tightsection{Evaluation}
\label{sec:eval}

Our key findings are as follows:
\begin{packeditemize}
\item {\bf Loss resilience:} \name's quality under no packet loss is on par with H.264/H.265 and gracefully declines with higher loss rates. 
Under 20--80\% packet loss, \name improves the SSIM by 0.5--4 dB
%at \fillme~Mbps
compared with other loss-resilient baselines across diverse videos.
%\jc{this should be backed up with some concrete numbers, such as improve SSIM by xxx-xxx dB at \fillme packet loss rate}

\item {\bf Better video smoothness:} 
Under bandwidth fluctuations in real network traces, \name reduces the number of video freezes over 200~ms by up to 90\%, tail frame delay by up to 2--5$\times$, and non-rendered frames by up to 95\%.
%has almost {\em zero} video stalls for over 200ms, which is up-to 90\% lower than loss-resilient baselines.
%Also, \name reduces the tail frame delay by 2-4$\times$ and the ratio of non-rendered frames by 95\%.
Our user study also confirms a 38\% rated score for \name.
%that \name's rated score is 38\% higher than the baselines. %\jc{please update this with new results.}

\item {\bf Speed:} Our implementation of \name encodes/decodes 480p video at 65.8~fps/104.1~fps and 720p video at 33.6~fps/44.1~fps using Nvidia A40 GPU, 1.5--5$\times$ faster than recent neural video codecs~\cite{swift, yang2020learning,shi2022alphavc, dvc}. 
% off-the-shelf autoencoders~\cite{swift, dvc}. 
With the optimization detailed in \S\ref{subsec:fast-coding}, \name can encode/decode 720p video at 26.2~fps/69.4~fps on an iPhone 14 Pro with marginal quality degradation.
%\yc{this number is inaccurate since it does not consider h264 mv}
\end{packeditemize}

\vspace{-2pt}
\tightsubsection{Setup}
\label{subsec:eval:setup}
\vspace{-3pt}

% We build \name's autoencoder based on DVC~\cite{dvc}, a state-of-the-art video autoencoder. 
% We fine-tuned \name on the vimeo-90K~\cite{vimeo-dataset} dataset from the pretrained model of DVC. The fine-tuning takes two steps: we first fine-tuned a base model with packet losses applied during training.  
% Then, we employed 11 distinct $\alpha$s (in Eq.~\ref{eq:2}), ranging from $1/64$ to $1/32768$ to fine-tune the residual encoder and decoder. This enables \name to support various bitrates (\S\ref{subsec:fast-coding}). 
% We set the learning rate at $10^{-4}$, and each step of fine-tuning typically takes 1-2 hours on an Nvidia A40 GPU.
%\fyy{Besides fine-tuning DVC's pretrained model, are there other differences?} 

%Figure~\ref{fig:dvc-arch} shows the architecture of the encoder and decoder. On the sender side, the encoder neural network takes the current frame and its reference frame as the input, and outputs the encoded motion vectors and residuals (\S\ref{subsec:background}). On the receiver side, the decoder neural network decodes the received motion vector and residuals, then reconstructs the new frame based on the reference frame.
%\fyy{Why repeat Grace's design in the experimental setup? It's better to treat this paragraph as "implementation" and explain how DVC is adapted for this purpose.}

\mypara{Testbed implementation}
Our testbed  2 Nvidia A40 GPUs to run the video encoding and decoding with \name's NVC (each using one GPU).
We use a packet-level network simulator to compare \name with baselines under various network conditions. The simulator uses a configurable drop tail queue to mimic congestion-induced packet losses and uses a token bucket scheme to simulate bandwidth variation every 0.1 seconds. Google Congestion Control (GCC)~\cite{carlucci2016analysis}, a standard WebRTC algorithm widely used in real-time video applications, is used to determine the target bitrate of video codecs at each frame.
It is worth noting that GCC is responsive to bandwidth drops and packet losses, as it tends to send data conservatively to avoid video delays and stalls caused by packet losses.
% \jc{Yihua, can we bring back the result with Salsify CC? given there are many recent papers on improving cc for conferencing, people may wonder if GCC is too outdated..}
% \yc{claim the following things to protect ourself: GCC is SOTA and widely used, GCC is very responsive to bandwidth drop and loss, (maybe we can also mention network have random loss)}
The simulator includes 
%emulates a complete real-time video communication application, including 
encoding, packetization, rate adaptation, and decoding.
We set the default frame rate at 25 fps (on par with typical RTC frame rates~\cite{macmillan2021measuring}), though \name can encode at a higher frame rate (\S\ref{subsec:eval:micro}).
\cmedit{Instead of replaying stationary traffic/loss traces, the testbed can simulate dynamic packet loss rates under real-world bandwidth fluctuations.}{Addressing the comment from reviewer A: The approach assumes that background traffic is stationary enough to be learned. It is unclear to what degree this holds true in practice.}
% (matching state-of-the-art real-time video measurements~\cite{macmillan2021measuring}). 
% The packets are transmitted over a network simulated by the simulator. 
It records each decoded frame and its delay, including encoding, transmission, and decoding. We have confirmed our simulator's accuracy regarding frame delay via a real-world validation experiment in the appendix (\S\ref{app:validation}). %(\S\ref{subsec:eval:e2e}).

\begin{table}[]
\footnotesize
\begin{centering}
\begin{tabular}{ccccc}
\toprule
Dataset                                           & \begin{tabular}[c]{@{}c@{}}\# of\\ videos\end{tabular} & Length (s) & Size                                                & Description                                                                             \\ \midrule
{\color[HTML]{000000} Kinetics}                   & 45                                                     & 450 & \begin{tabular}[c]{@{}c@{}}720p\\ 360p\end{tabular} & \begin{tabular}[c]{@{}c@{}}Human actions and\\interaction with objects\end{tabular} \\ \midrule
\multicolumn{1}{l}{{\color[HTML]{000000} Gaming}} & 5                                                      & 100 & 720p                                               & \begin{tabular}[c]{@{}c@{}}PC game  recordings\end{tabular}                            \\ \midrule
{\color[HTML]{000000} UVG}                        & 4                                                      & 80 & 1080p                                               & \begin{tabular}[c]{@{}c@{}}HD videos (human, \\nature, sports, etc.)\end{tabular}   \\ \midrule
{\color[HTML]{000000} FVC}                        & 7                                                      & 140 & 1080p                                               & \begin{tabular}[c]{@{}c@{}}In/outdoor video calls\end{tabular}                  \\ 
\midrule
\textbf{Total}                                             & \textbf{61}                                                     & \textbf{770} &                                                     &   \\                                                                                     
\bottomrule
\end{tabular}
\vspace{6pt}
\tightcaption{Dataset description.}
\label{tab:dataset}                               
\end{centering}
\end{table}

\mypara{Test videos}
Our evaluation uses 61 videos randomly sampled from four public datasets, summarized in Table~\ref{tab:dataset}.
%\footnote{We have shared the test videos in an anonymous link: \url{https://drive.google.com/file/d/1iQhTfb7Kew_z97kDVoj2vOmQqaNjBK9J/view?usp=sharing}}
The total content length is 770 seconds where each video is 10--30 seconds long, matching the setup of similar works~\cite{salsify, rudow2023tambur, conti2021not}. %\jc{Yihua, please fix}
%The video content includes human actions in different environments, video game recordings, nature scenes, sports, as well as indoor and outdoor video calls. 
%The detailed description of the datasets is in Table~\ref{tab:dataset}.
Importantly, these videos are obtained from entirely different sources than the training set, and they span a range of spatial complexity and temporal complexity (detailed in Appendix~\ref{app:siti}), as well as multiple resolutions.
This diversity allows us to assess \name's average performance across different contents and study how content affects its performance.

%allow us to thoroughly assess the generalization capabilities of autoencoders. They encompass various spatial and temporal dynamics (discussed in Appendix~\ref{app:siti}), across multiple resolutions, ensuring a minimum of 5 videos per resolution. The reported performance constitutes an average across all video frames.

\mypara{Network traces}
We test \name and the baselines on 16 real bandwidth traces, eight of which are LTE traces from the Mahimahi network-emulation tool~\cite{netravali2015mahimahi, mahimahi-traces},
and the rest are broadband traces from FCC (July 2021)~\cite{fcc-dataset}.
\cmedit{The traces are in the format of bandwidth timeseries.}{Addressing the comment from reviewer A: Please give more information about the traces you used. // note: we also updated the citation links}
The bandwidth fluctuates between 0.2~Mbps to 8~Mbps in the traces.
By default, we set the one-way propagation delay to 100~ms and the queue size to 25 packets. We also vary these values in \S\ref{subsec:eval:e2e}.

\begin{figure*}[t!]
     \centering
     \includegraphics[width=\linewidth]{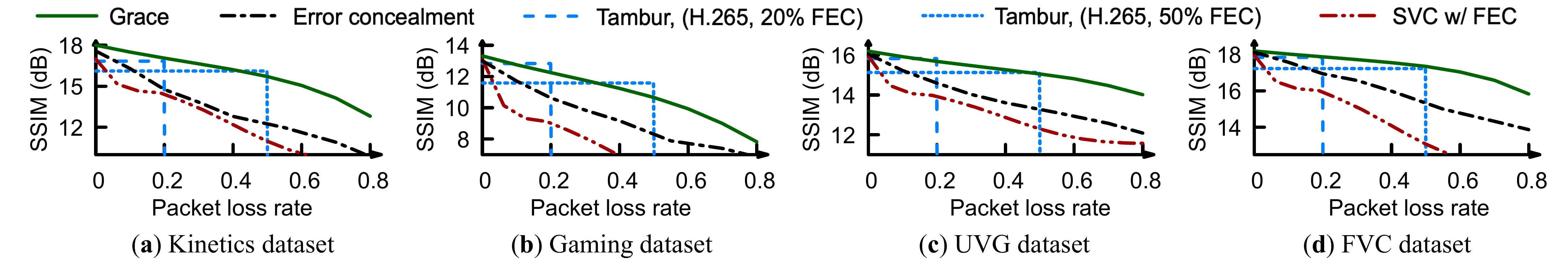}
     \vspace{-12pt}
     \tightcaption{
     Video quality achieved by different schemes under varying packet loss rates at the same encoded bitrate (6~Mbps). %, grouped by testing datasets. 
     %\cmedit{Updated the FEC line. It should be straight up and down.}{Addressing the comment from reviewer A: In Figure 8, why are the dropoffs for the Tambur lines not straight up and down?}
     % When packet loss impacts one frame (bitrate is 6~Mbps), \name has better quality than the baselines across datasets.
     }
     \label{fig:loss_by_dataset}
\end{figure*}

\mypara{Baselines}
We employ H.265 (through FFmpeg v4.2.7) as the underlying video codec
for all baselines (except for NVC-based ones) since
%, based on \texttt{libx265}\footnote{The command line we used to encode a video is \texttt{ffmpeg -y -i Video.y4m -c:v libx265 -preset fast -tune zerolatency -x265-params "crf=Q:keyint=3000" output.mp4} where {\em Q} controls the quality of the frame.}
H.265 is recognized with comparable or better compression efficiency
than VP8/9 and H.264~\cite{x265vp9compare, grois2013performance} (as confirmed in Appendix~\ref{app:h26xvpx}).
%, as the base codec of all baselines (except NVC-based ones). 
We compare \name against a range of loss-resilient baselines that cover various approaches outlined in \S\ref{subsec:related} (more details in Appendix~\ref{app:baseline}).
\begin{packeditemize}  
  \item {\bf Forward error correction}: 
We use Tambur~\cite{rudow2023tambur}, a state-of-the-art FEC scheme based on streaming codes~\cite{badr2017fec}.
%   with H.265.
Its redundancy rate
%defined as $p/n$ (with $p$ as the number of parity packets and $n$ the total packets, including parity packets~\cite{xiao2012dynamic, lee2022r}),
dynamically adapts based on the measured packet loss in the preceding 2 seconds. 
Compared with regular FEC, streaming codes reduce the number of non-decodable frames when transmitting an equivalent amount of parity packets.
We have also validated that Tambur outperforms WebRTC's default FEC scheme.

  \item {\bf Decoder-side neural error concealment}: 
  We use ECFVI~\cite{kang2022error}, an NN-based error concealment method shown to outperform previous techniques relying on motion estimation recovery~\cite{sankisa2018video} or inpainting~\cite{chang2019free}.\footnote{A parallel effort, Reparo~\cite{reparo}, demonstrates effective error concealment for a particular video type (talking head), but it lacks comparisons with any NN-based baselines and does not provide a public codebase for testing.} 
  To ensure each packet is independently decodable, we apply flexible macroblock ordering (FMO)~\cite{dhondt2004flexible} to split the frame into 64×64-pixel\footnote{A smaller block size such as 16×16 can greatly inflate the frame size~\cite{kumar2006error, wenger2002scattered, 4494082}, while a larger block size such as 256×256 hinders information recovery upon packet loss.
  %, making it difficult for the model to conceal errors.
  We empirically choose the 64×64 block size to balance between frame size and quality.}
  blocks and map them randomly to packets.
  This results in a 10\% increase in the encoded frame size, in line with previous findings~\cite{kumar2006error, wenger2002scattered, 4494082}.
  After decoding an incomplete frame, ECFVI uses NNs to estimate missing motion vectors and enhance the reconstructed frame through inpainting. 

  \item {\bf Scalable video coding (SVC)}: %SVC encodes a video at multiple quality layers and sends data layer by layer. 
  We implement an idealized SVC, designed so that when the first $k$ layers arrive, it achieves the same quality as that of H.265 with the same number of received bytes. 
  This idealized implementation surpasses the state-of-the-art NN-based SVC~\cite{swift}. 
  We also add 50\% FEC to protect the base layer for SVC, following a common practice in real-time video applications~\cite{khalek2012cross}.

  \item {\bf Selective frame skipping}: 
  % One potential design choice of selective frame skipping is optimizing the frame delay by only retransmitting the important frames that will cause huge damage to the video quality when missing. 
  % We implement two recent frame skipping baselines. 
  % for this design choice based on Voxel~\cite{voxel} and Salsify~\cite{salsify}.
  Salsify~\cite{salsify} skips frames affected by loss at the decoder side after the encoder receives the packet loss indication and resends a new P-frame using the last fully received frame as a reference.
  Voxel~\cite{voxel} employs selective frame skipping to mitigate video rebuffering and improve the user's QoE.
%  We provide more details about their implementation in Appendix~\ref{app:baseline}.

%\jc{seems too much space spent on a minor baseline. shrink it please?}
  
  %Unlike traditional codecs (\eg, H.265 or VP9), functional codecs are able to skip incomplete frames when packet loss happens and decode the first complete frame without a significant quality drop. Salsify~\cite{salsify} is a recent work implementing the functional codec. 
  %We implement the Salsify baseline based on H.265 with the following features: {\em (i)} the encoded frame size of the codec will never overshoot the target bitrate calculated by the underlying congestion control algorithm, and {\em (ii)} when packet loss happens, the encoder can dynamically choose a reference frame to make later frames decodable without retransmitting any packets.

%  \edit{We show more details about how to integrate \name and SR in Appendix~\ref{app:sr}.}
  %We use a state-of-the-art SR model, SwinIR~\cite{swinir}, as the tool to improve the video quality for both \name and other baselines after the receiver successfully decodes the frame. \yc{consider move it to 5.5}

\end{packeditemize}

% For the baselines of functional codec (Salsify), SVC, and selective frame skipping (Voxel), we make an idealized assumption that on each frame, the output bitrate of the codec perfectly matches the target bitrate chosen by the congestion control algorithm (explained in \S\ref{subsec:eval:e2e}), \ie never overshooting or undershooting, but the actual quality matches that of the highest \texttt{qp} that leads to a frame greater than the target bitrate. 
% We used an idealized version of functional codec (Salsify), SVC, and selective frame skipping (Voxel) baselines. 
We make another idealized assumption in favor of SVC, Salsify, and Voxel. We assume that their codec's output bitrate on every frame perfectly matches the target bitrate determined by the congestion control algorithm,
%(\S\ref{subsec:eval:e2e}),
\ie no overshoots or undershoots.
This idealization makes these baselines perform slightly better than they would under real-world conditions.

\mypara{Variants of \name} 
To highlight the impact of different design choices, we evaluate \textbf{\namepretrain} and \textbf{\namedecoder}. They are trained the same way as \name, except that \namepretrain does not use simulated loss while \namedecoder freezes the encoder NN weights (\ie fine-tuning only the decoder NN with simulated loss).
%They serve as a reference for alternative ways of simulating packet loss in training. 
They represent alternative ways to simulate packet losses during training.
We also test \textbf{\namelite}, which incorporates the optimizations described in \S\ref{subsec:fast-coding}.
% We also implement and test the following variants of \name proposed in \S\ref{sec:core} and \S\ref{subsec:fast-coding}. \textbf{\namepretrain}: \name trained without seeing any packet loss. \textbf{\namedecoder}: \name trained with packet loss solely on the decoder side. \textbf{\namelite}: \name with the speedup optimizations in \S\ref{subsec:fast-coding}.

Furthermore, we present the quality improvement achieved by the state-of-the-art super-resolution (SR) model~\cite{liang2021swinir} when applied to \name and other baselines.
It is important to note that SR can be applied to any decoded frames, making it
{\em orthogonal} to \name's design space. 
Details of this experiment are provided in Appendix~\ref{app:sr}.

\mypara{Metrics}
Following prior work on real-time video communication~\cite{rudow2023tambur, salsify, garcia2019understanding, garcia2020assessment, ammar2016video}, we measure the performance of a video session across three aspects.
\begin{packeditemize}
    \item {\em Visual quality} of a frame is measured by SSIM.
    % , reported as the average across received frames. 
    Following recent work~\cite{salsify,puffer}, we express SSIM in dB, calculated as $-10\log(1 - \text{SSIM})$ across all rendered frames.
    %{\em Quality} of a video is measured by frame-level SSIM, and we report the average value across received frames. Following the recent related work~\cite{salsify}, we report the SSIM values in dB\footnote{We use $-10 * log(1 - SSIM)$ to calculate the corresponding dB values.}. 
    \item {\em Realtimeness} is measured by the 98th percentile (P98) of frame delay (time elapsed between the frame's encoding and decoding), and non-rendered frames (either undecodable due to insufficient FEC protection or exceeding 400~ms after the frame is encoded).
%    , and thus not displayed by the receiver.
    %is measured by the tail of end-to-end frame delay and the ratio of non-rendered frames. End-to-end frame delay means the time lapse between the beginning of encoding and the end of decoding. Non-rendered frames refer to frames that are not played by the receiver due to not being decodable or arriving too late (more than 200ms after the previous frame).
    \item {\em Smoothness} of the video is measured by video stall, defined as an inter-frame gap exceeding 200~ms, following the industry convention~\cite{macmillan2021measuring}.
%    where a video stall is defined as the fraction of inter-frame gaps that exceed a threshold. 
    % Following the previous work~\cite{macmillan2021measuring} and the convention of WebRTC, we define a video stall as an inter-frame gap over 200~ms. 
    We report the average number of video stalls per second and the ratio of video stall time over the entire video length.
    % is defined as the percentage of time that the video is spent in the stalls.
    %We choose this threshold as it matches the definition in WebRTC. 
\end{packeditemize}

\tightsubsection{Compression efficiency and loss resilience}
\label{subsec:eval:resilience}

\mypara{Loss resilience}
\cmedit{In real world, packet loss per frame (defined in \S\ref{subsec:background}) can span a wide range from 0 to over 80\%~\cite{rudow2023tambur}.}{Addressing the comment from reviewer B: the packet loss rate you use in the evaluation are too high.}
Figure~\ref{fig:loss_by_dataset} compares \name's video quality with the baselines under varying packet loss rates across different test video sets.
For a fair comparison, we fix the encoded bitrate of all baselines at 6 Mbps (with actual differences under 5\%) while ensuring that \name's encoded bitrate {\em never} exceeds that of the baselines.
On average, the quality of \name drops by 0.5~dB to 2~dB in SSIM as the packet loss rate rises from 20\% to 50\%, and by up to 3.5~dB when the packet loss rate reaches 80\%.
These quality drops of \name are notably lower than the baselines, including FEC-based and neural error concealment schemes, at the same packet loss rates. 

\begin{figure}[t!]
     \centering
     \includegraphics[width=0.98\linewidth]{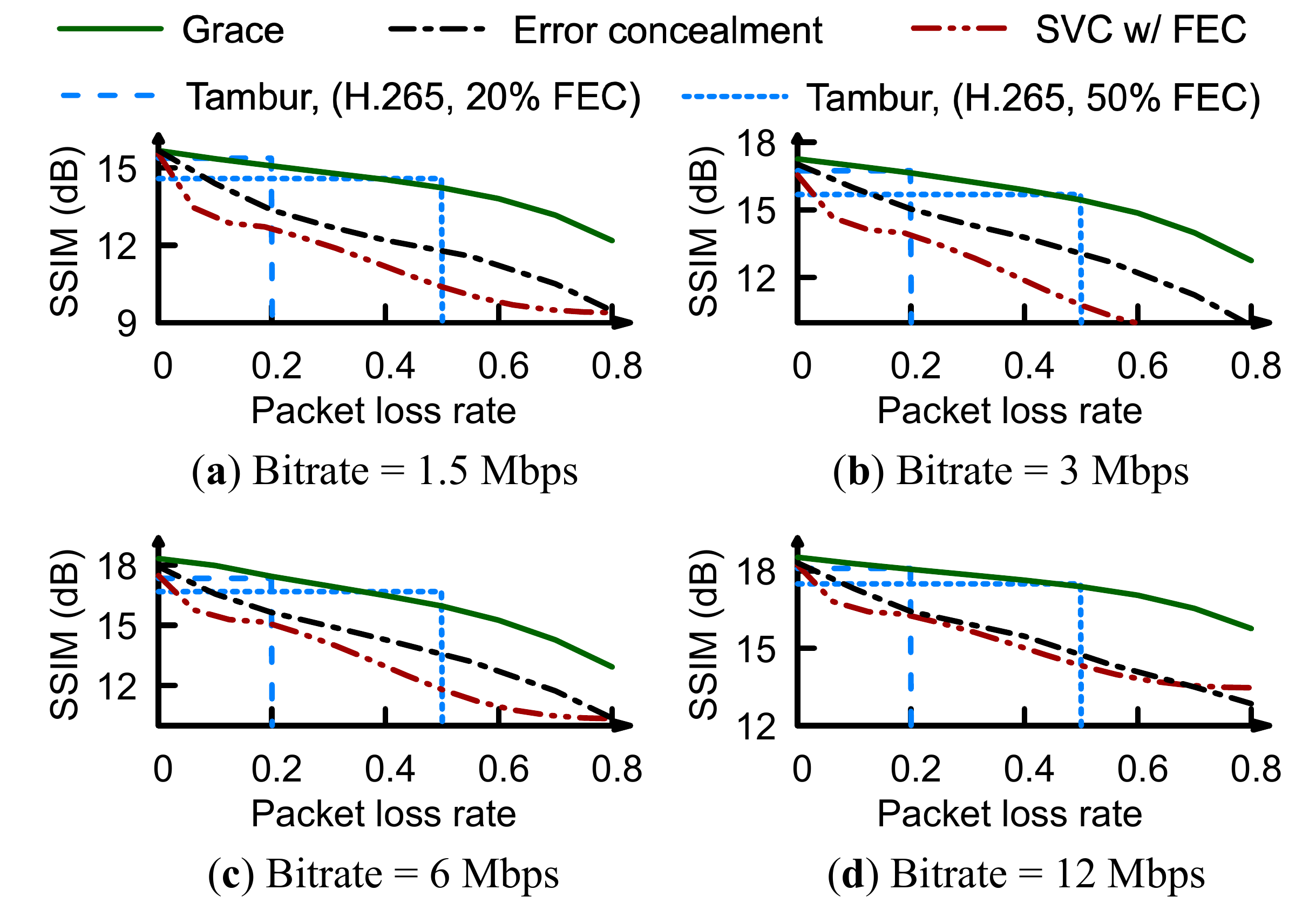}
     % \includegraphics[width=\linewidth]{figs/ssim-loss-bitrate-screenshot.pdf}
     %\vspace{-5pt}
     \tightcaption{
     Video quality of each scheme under different packet loss rates when videos are encoded at different bitrates.
     %\cmedit{Updated the FEC line. It should be straight up and down.}{Addressing the comment from reviewer A: In Figure 8, why are the dropoffs for the Tambur lines not straight up and down?}
     % When packet loss impacts one frame encoded with different bitrates,
    % \name has better loss resilience.
    }
     \vspace{3pt}
     \label{fig:loss_by_bitrate}
\end{figure}

\begin{figure}[t!]
    \centering
    \includegraphics[width=0.98\linewidth]{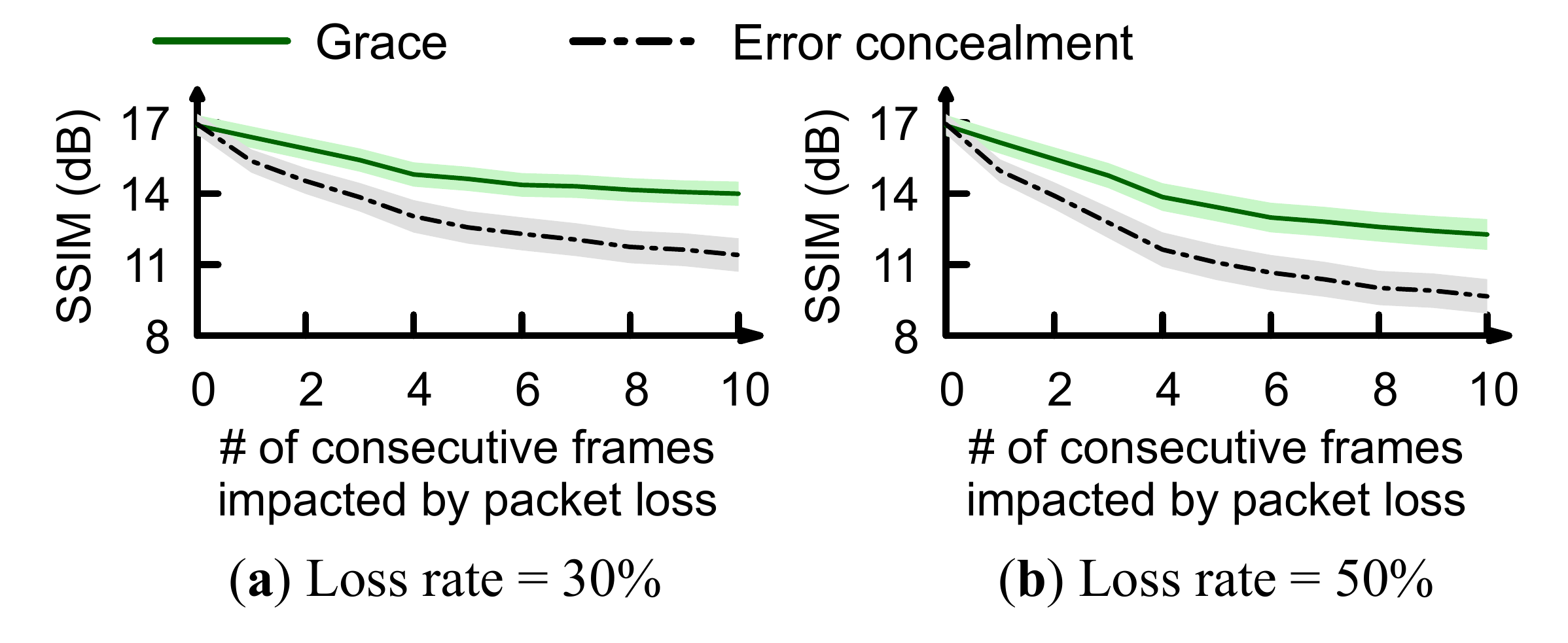}
    \tightcaption{Stress test of applying persistent packet loss on consecutive frames. 
    % degradation when persistent packet losses are retains the frame quality when packet losses happen consecutively on 10 frames without encoder/decoder synchronization.
    }
    %\jc{The axis should be "\# of consecutive frames with xxx\% packet loss. it should start from where they diverge"}}
    \label{fig:stress-test}
\end{figure}

\begin{figure}[t!]
    \centering
    % \hspace{10pt}
    \includegraphics[width=0.88\linewidth]{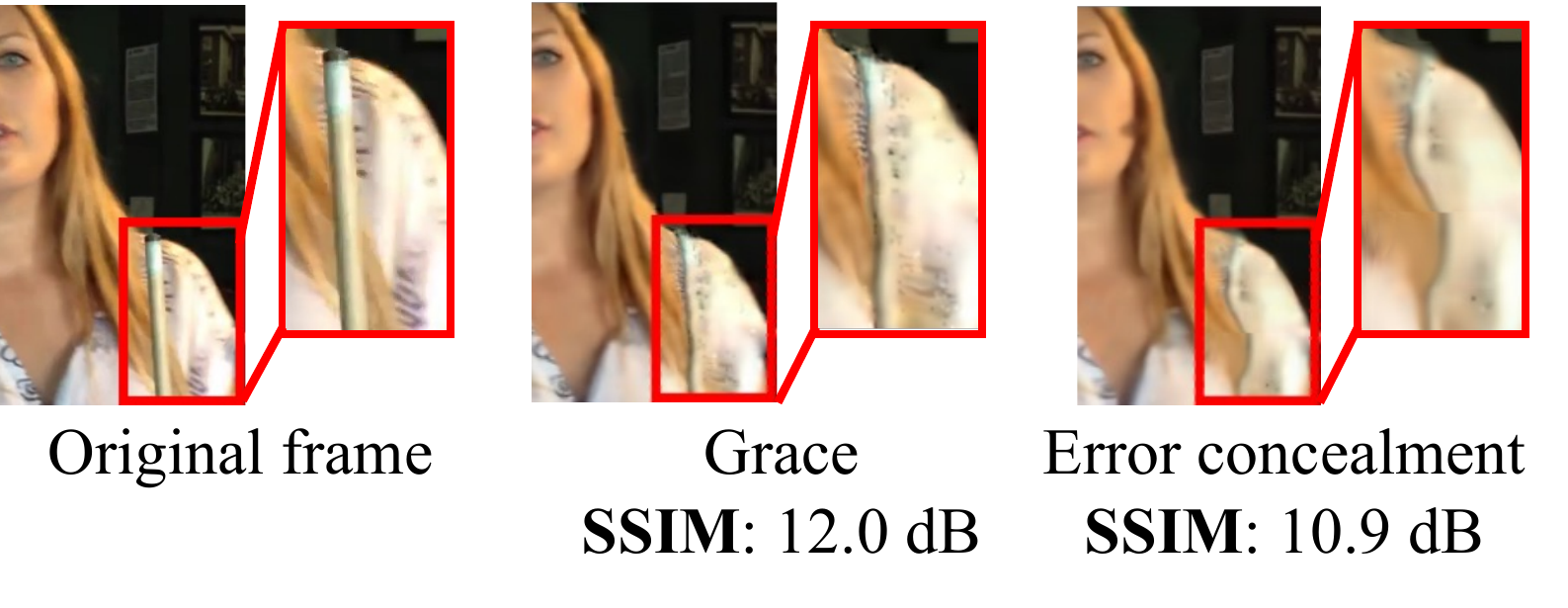}
    \vspace{-2pt}
    \tightcaption{Sample images decoded by \name and error concealment under a 50\% packet loss on three consecutive frames. \name achieves less image distortion.}
    \vspace{3pt}
    \label{fig:early-visual-example}
\end{figure}

Figure~\ref{fig:loss_by_bitrate} shows the average quality across all test videos when the encoded bitrates of all schemes are set to 1.5, 3, 6, and 12~Mbps. 
Compared with the baselines, \name achieves a more graceful and less pronounced quality decline as packet loss increases.
Figure~\ref{fig:stress-test} further stress tests the loss resilience of \name against neural error concealment (the most competitive baseline), when a 30\% or 50\% packet loss is applied to 1 to 10 consecutive frames without the encoder and decoder synchronizing their states.
Although the figure shows that both methods experience quality degradation, \name markedly surpasses the neural error concealment baseline in these extreme conditions. 
Figure~\ref{fig:early-visual-example} visualizes their decoded images after a 50\% packet loss is applied to three consecutive frames, confirming that the image decoded by \name has less visual distortion.
%than the one decoded by neural error concealment.

\begin{figure}[t!]
    \centering
    \includegraphics[width=0.98\linewidth]{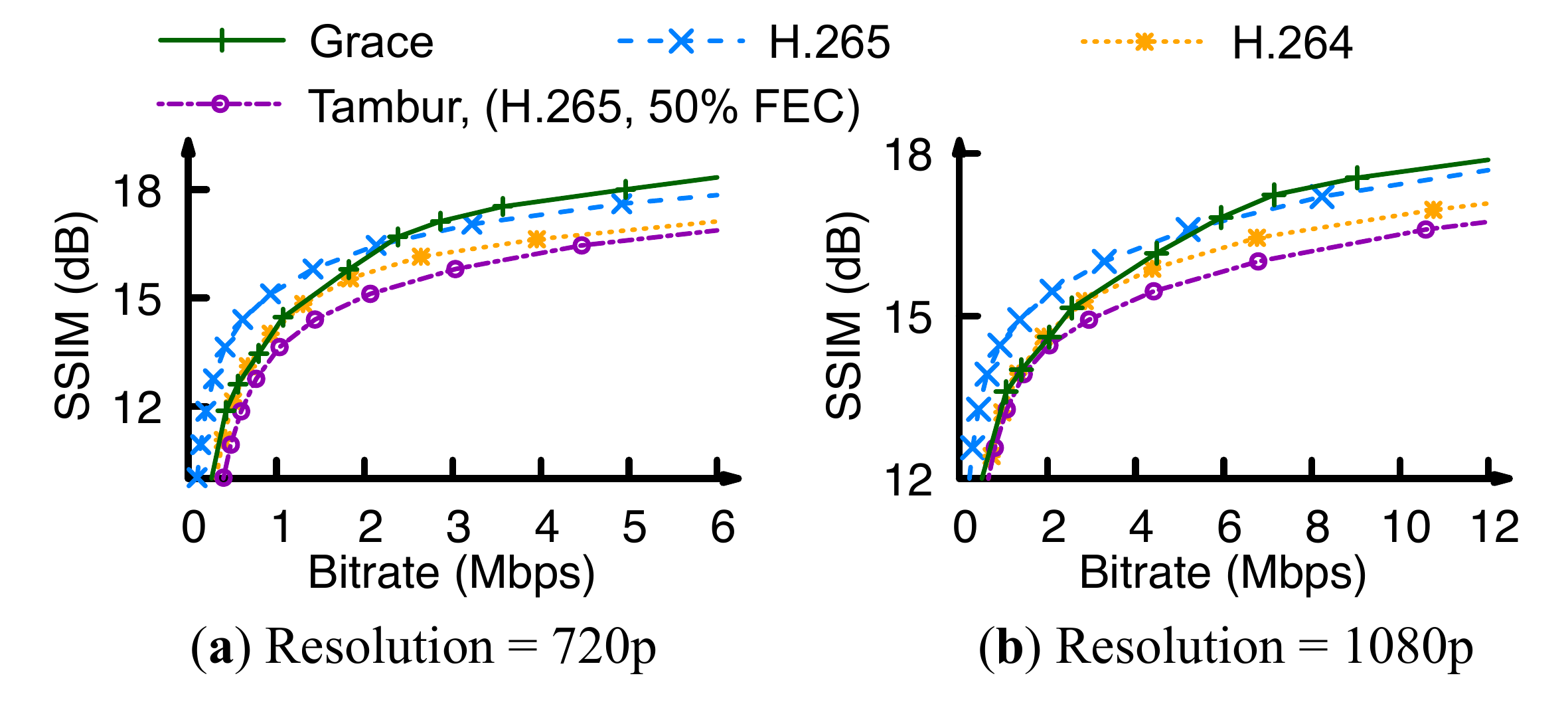}
    %\vspace{-12pt}
    \tightcaption{Quality-size tradeoff of \name on videos with different resolution. Overall, \name is better than H.264 and slightly worse than H.265 in terms of compression efficiency.}
    \vspace{2pt}
    \label{fig:psnr-bpp}
\end{figure}

\mypara{Compression efficiency} 
We verify whether \name's compression efficiency under no packet loss is on par with H.264 and H.265, which are advanced video codecs designed for high compression efficiency rather than loss resilience. 
Figure~\ref{fig:psnr-bpp} groups the test videos by resolution. 
On low bitrates, \name demonstrates similar compression efficiency as H.264 and marginally underperforms H.265 on both 720p and 1080p videos.
On high bitrates (over 3~Mbps for 720p and 6~Mbps for 1080p), \name's compression efficiency matches or even surpasses H.265.
Compared against Tambur with a persistent 50\% FEC redundancy, 
%a state-of-the-art FEC-based scheme, 
\name achieves a better quality-bitrate tradeoff across the entire bitrate range.

\begin{comment}

% For any loss-resilient schemes, 
There is always a tradeoff between compression efficiency and loss resilience, and \name is no exception. 
We assessed \name's compression efficiency (in terms of quality-bitrate trade-off) under zero packet loss, juxtaposed with H.264 and H.265, using the test videos from Table~\ref{tab:dataset}. 
As indicated in Figure~\ref{fig:psnr-bpp}, \name has similar compression efficiency with H.264 and marginally underperforms H.265 on both 720p and 1080p videos. 
In comparison with the baseline using Tambur that applies a constant 50\% redundancy to tolerate up to 50\% packet losses, \name offers a superior quality-bitrate trade-off, implying improved video quality at identical bitrates. 
Furthermore, \name spans a broad bitrate range across all datasets, from 0.25 Mbps to over 10 Mbps, thereby satisfying the bandwidth requirement of different real-time video applications. 

\end{comment}

\begin{figure}[t!]
    \centering
    \includegraphics[width=0.57\linewidth]{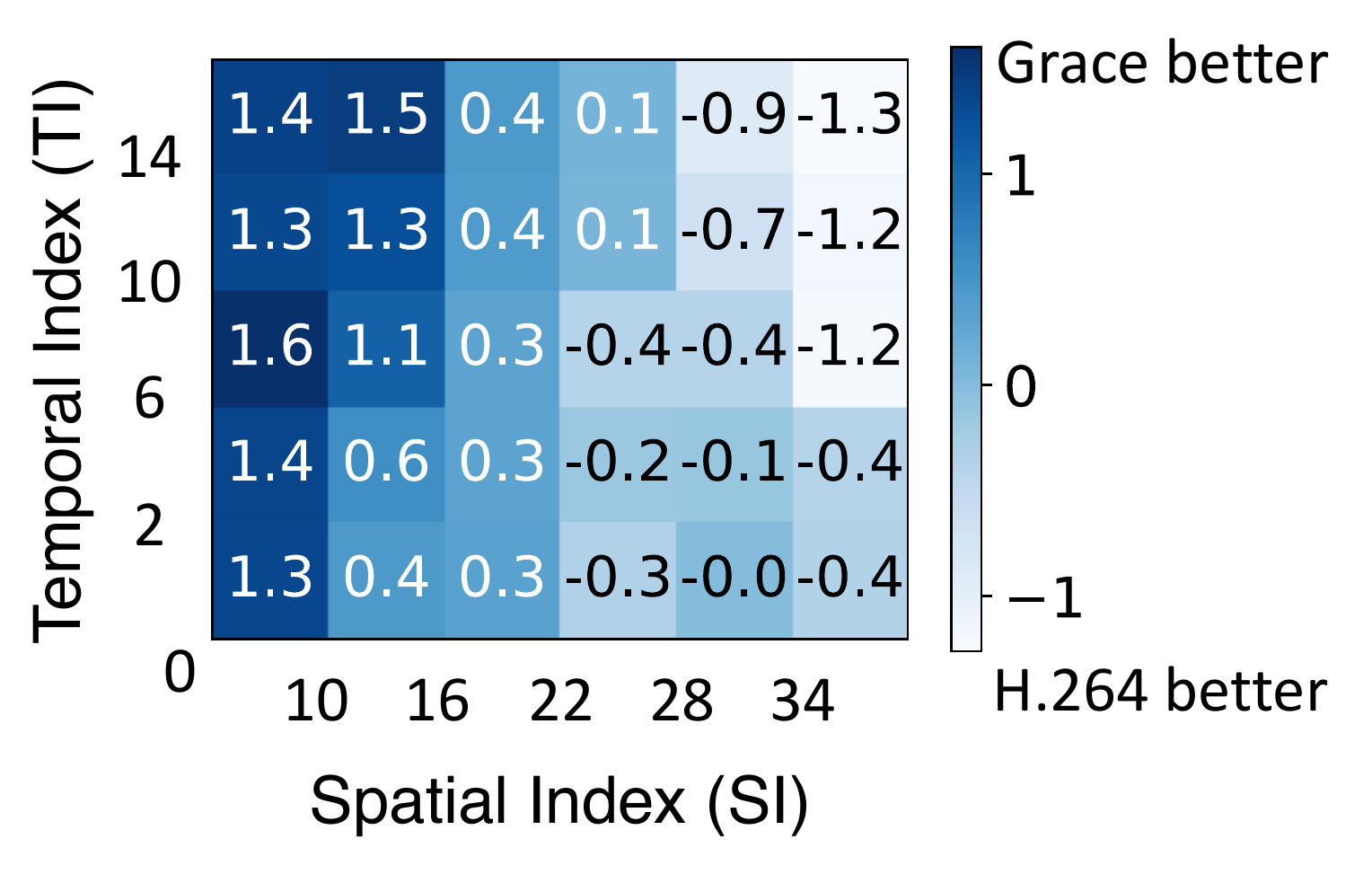}
    \tightcaption{Mean difference in SSIM (dB) between \name and H.264 on videos grouped by SI and TI. At the same bitrate (5~Mbps), \name achieves better video quality than H.264 on low-SI videos but lags behind H.264 on high-SI videos.}
    \label{fig:generalization-study}
\end{figure}

\mypara{Impact of video content on compression efficiency}
To understand the impact of video content on \name's compression efficiency, we group the video content based on spatial index (SI) and temporal index (TI), which are established metrics for assessing the spatiotemporal complexity of videos~\cite{itu1999subjective}.
Figure~\ref{fig:generalization-study} presents the average gain of \name
over H.264 in terms of SSIM for videos in each SI-TI combination, encoded at a bitrate of 5~Mbps.
The results indicate that \name's compression efficiency has a higher advantage over H.264 for videos with low spatial complexity, but this advantage diminishes as the spatial index increases.
\cmedit{For a more thorough understanding of \name's behavior, Appendix~\ref{app:badexample} also shows an example where \name performs poorly.}{Addressing the comment from reviewer A: It would also help if you could give examples where GRACE performs poorly, not just the good examples, to help us understand GRACE's behavior better.}

\tightsubsection{Video quality vs. realtimeness/smoothness}
\label{subsec:eval:e2e}

% To determine the target bitrate of the video codecs used during simulation, we implemented two different congestion control algorithms: Google congestion control~\cite{carlucci2016analysis} (\textit{GCC}, the standard congestion control used in WebRTC) and the congestion control from~\cite{salsify} (\textit{Sal-CC}). 
%We also implemented FEC-protected H.265 and Salsify codec as the baseline codecs (\S\ref{subsec:eval:setup}).

% Figure~\ref{fig:gcc-delay-simulation} and Figure~\ref{fig:gcc-queue-simulation} 
% \name's loss resilience and the ability to decode incomplete frames make it achieve a better quality vs realtimeness or smoothness tradeoff than the baselines, as it does not require any retransmission. 

\begin{figure}[t!]
    \centering
    \includegraphics[width=\linewidth]{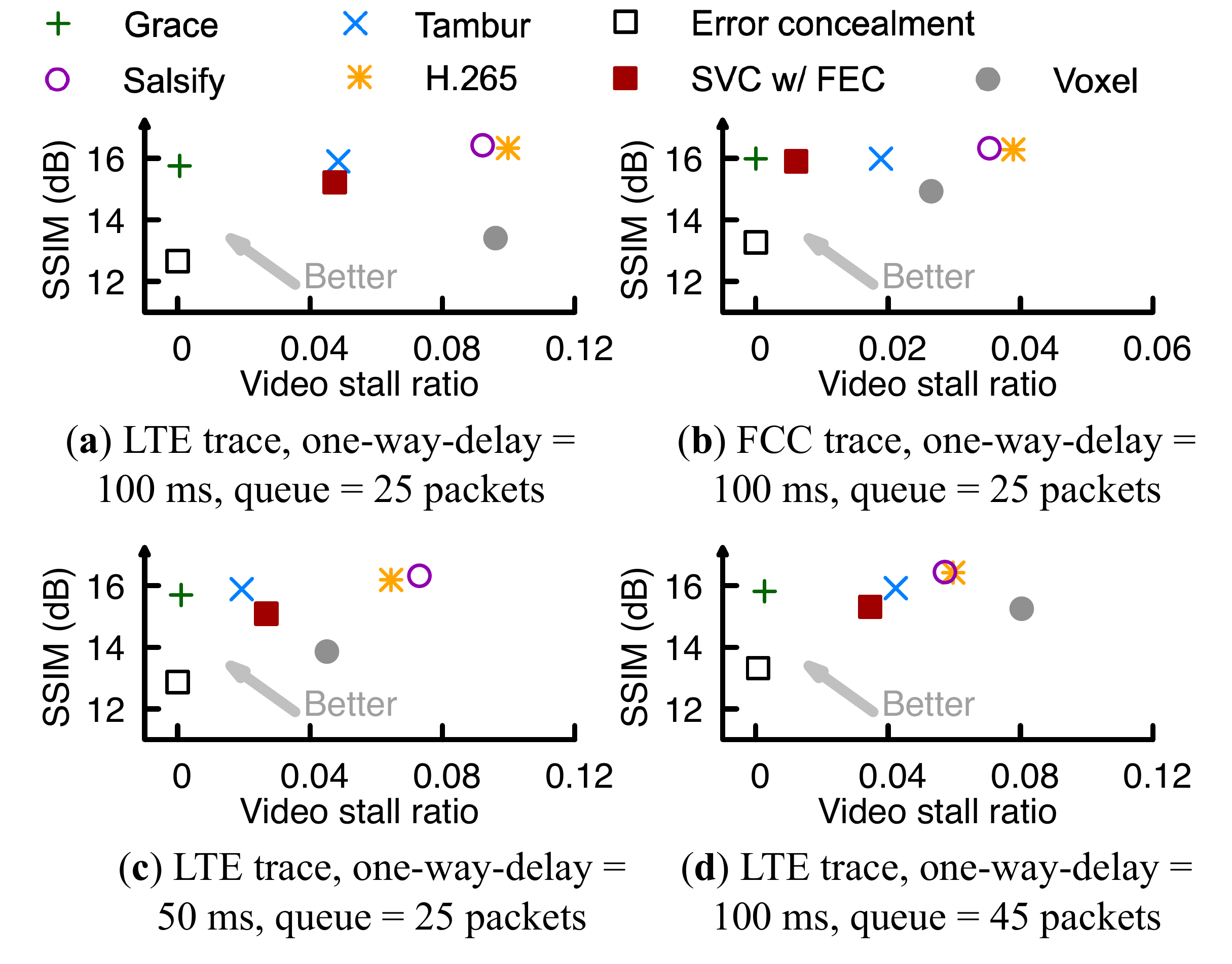}
    \vspace{-10pt}
    \tightcaption{End-to-end simulation results over different network traces, one-way delays and network queue lengths. 
    \vspace{3pt}
    % (a) and (b) compares the impact of the network trace dataset. (a) and (c) compares the impact of one-way-delay. (a) and (d) compares the impact of network queue length. On different setups, \name outperforms baselines in terms of the tradeoff between video stall ratio and video quality.
    }
     \label{fig:gcc-delay-simulation}
\end{figure}

Figures~\ref{fig:gcc-delay-simulation}a evaluates \name against baselines in terms of average quality (SSIM) and video stall ratio (a smoothness metric) using the network traces from the LTE dataset, under a one-way network delay of 100 ms and a drop-tail queue of 25 packets. 
Although the SSIM of \name is slightly lower than that of the baselines with the highest average SSIM, \name significantly reduces the video stall ratio. 

We repeat the test on a different dataset (FCC) under the same network setup 
(Figures~\ref{fig:gcc-delay-simulation}b), with a lower one-way network delay of 50~ms (Figure~\ref{fig:gcc-delay-simulation}c), and with a longer queue length of 45 packets (Figure~\ref{fig:gcc-delay-simulation}d).
% repeats the test on a different dataset (FCC) under the same network setup, \ref{fig:gcc-delay-simulation}c tests it with a lower one-way network delay of 50~ms, and  tests it on a longer queue length. 
In all settings, \name maintains a video stall ratio below 0.5\%, whereas the baselines have 4--32$\times$ more video stalls, except for the error concealment baseline, which yields a 3dB lower SSIM compared with \name. 
This is because when packet loss happens, \name can still decode the frame, while the baselines other than error concealment may experience video stalls due to either skipping frames (\eg Salsify or Voxel) or waiting for retransmission packets (FEC and SVC). %, leading to video stalls.
% (when the packet loss make this frame undecodable), both leads to video stall.

\begin{figure}[t!]
    \centering
    % \begin{subfigure}[t]{0.325\linewidth}
    %      \centering
    %      \includegraphics[width=3\linewidth]{figs/tail-latency-long.pdf}
    %      \tightcaption{Tail frame latency}
    %      \label{fig:tail-latency}
    %  \end{subfigure}
    %  \hfill
    %  \begin{subfigure}[t]{0.325\linewidth}
    %      \centering
    %      % \includegraphics[width=\linewidth]{figs/placeholders/sal-delay-results-queue25delay100.pdf}
    %      \includegraphics[width=\linewidth]{figs/non-rendered-frames.pdf}
    %      \tightcaption{\% of non-rendered frames}
    %      \label{fig:non-rendered-frames}
    %  \end{subfigure}
    %  \hfill
    %   \begin{subfigure}[t]{0.325\linewidth}
    %      \centering
    %      % \includegraphics[width=\linewidth]{figs/placeholders/sal-stall-results-queue25delay100.pdf}
    %      \includegraphics[width=\linewidth]{figs/frequency-of-stalls.pdf}
    %      \tightcaption{Average stalls/sec (\name <0.001)}
    %      \label{fig:frequency-of-stalls}
    %  \end{subfigure}
    %  \hfill
    \includegraphics[width=0.9\columnwidth]{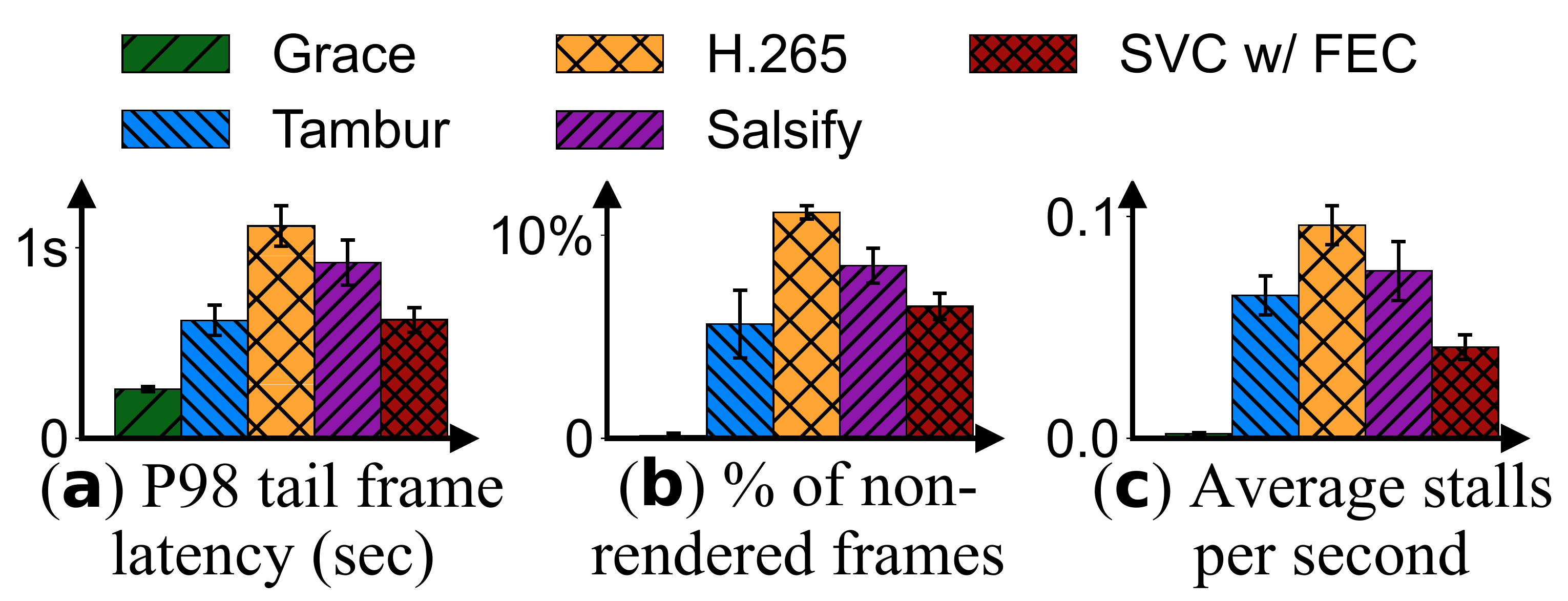}
     \vspace{2pt}
     \tightcaption{\name outperforms other baselines on different metrics of realtimeness and smoothness.
     \cmedit{Updated this figure: added error bar and y-axis}{Addressing the comment from reviewer A: Please label y axes in Figure 15. Also, can you give error/confidence bars?}
     }
     \label{fig:other-metrics}
\end{figure}

Figure~\ref{fig:other-metrics} compares \name with the baselines using other realtimeness and smoothness metrics, with the one-way delay set to 100~ms and the queue length set to 25 packets over the LTE traces.
For clarity, we only include baselines with comparable average SSIMs in this figure (excluding Voxel and error concealment).
While achieving similar video quality, \name reduces the 98th percentile frame delay by a factor of 2--5$\times$ and non-rendered frames by up to 95\%.
In \S\ref{app:salcc}, we also evaluate \name with a different congestion control algorithm---Salsify's CC~\cite{salsify}.

%LTE traces under different network delays (50ms and 100ms) and queue lengths (spanning 15 to 35 packets). 

% Though the SSIM of \name is marginally lower than the baselines with the highest average SSIM, \name significantly reduces the video stall ratio. 

% The primary cause of video stalls is congestion-induced packet loss. When packets are lost, Salsify must skip frames, resulting in increased delay, while H.265 awaits retransmission of lost packets, increasing delay and competing with regular video traffic. 
% Although Tambur has a much smaller video freeze ratio, matching the results in their paper, it does not solve the fundamental limitation of FEC-based solution: when unpredictable congestion losses are higher than the protection rate, it still causes significant video stalls.
% SVC and Voxel also wait for retransmission if the base layer is damaged or a crucial frame is lost. \name, however, can decode a frame as long as any packet from that frame is received.

\begin{figure}[t!]
    \centering
    \includegraphics[width=0.8\linewidth]{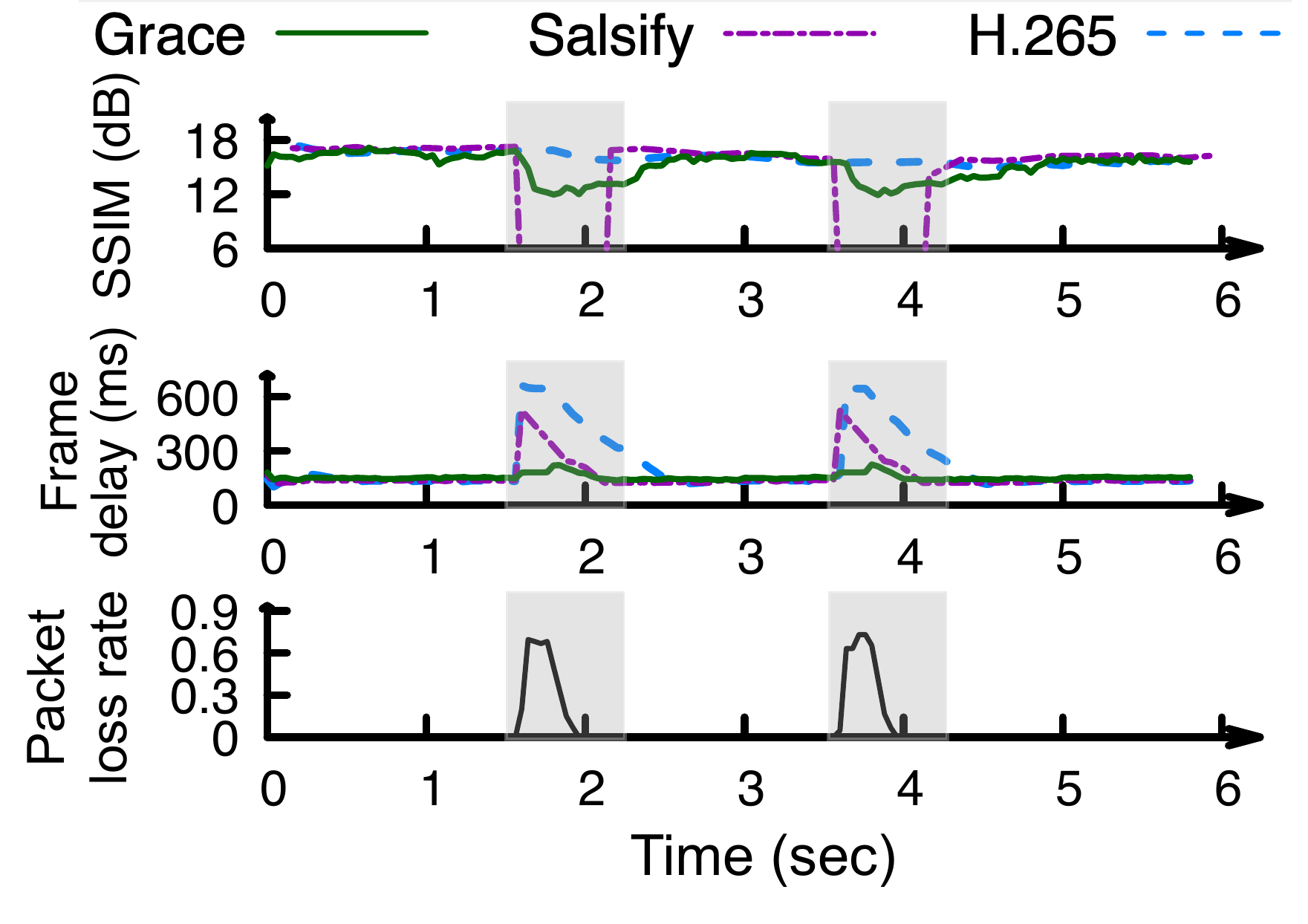}
    \tightcaption{\name achieves lower delay and maintains decent visual quality during sudden bandwidth drops: its delay is lower than both baselines while rendering more frames than Salsify without frame skipping or packet retransmission.}
    \label{fig:timeseries-ours}
\end{figure}

Figure~\ref{fig:timeseries-ours} provides a concrete example of \name's behavior.
% compare \name and the baselines in a concrete bandwidth trace sample to illustrate why it can achieve a better tradeoff (Figure~\ref{fig:timeseries-ours}).
The bandwidth drops from 8~Mbps to 2~Mbps at 1.5 s, lasting for 800 ms, before returning to 8 Mbps (another bandwidth drop occurs at 3.5 s and lasts for the same duration).
During each drop, \name's delay does not experience a sharp increase as the baselines. 
Salsify is the second best owing to its frame skipping while H.265 must wait for retransmissions.
In this experiment, both \name and Salsify use the same CC, leading to
similar qualities on frames not skipped by Salsify.
However, during congestion, \name's quality degrades only marginally without skipping any frames, limiting the drop of SSIM to less than 4 dB even when more than 10 consecutive frames encounter a packet loss of over 50\%. 
With the assistance of state resync (\S\ref{subsec:protocol}), \name's quality resumes quickly (within 1 RTT) after packet losses.

% When we run \name with the models trained without any packet losses, the average SSIM will be decreased from \fillme dB to \fillme dB. This is because the frame quality will be significantly decreased when packet loss happens.

\begin{figure}[t!]
    \centering
    \includegraphics[width=0.95\columnwidth]{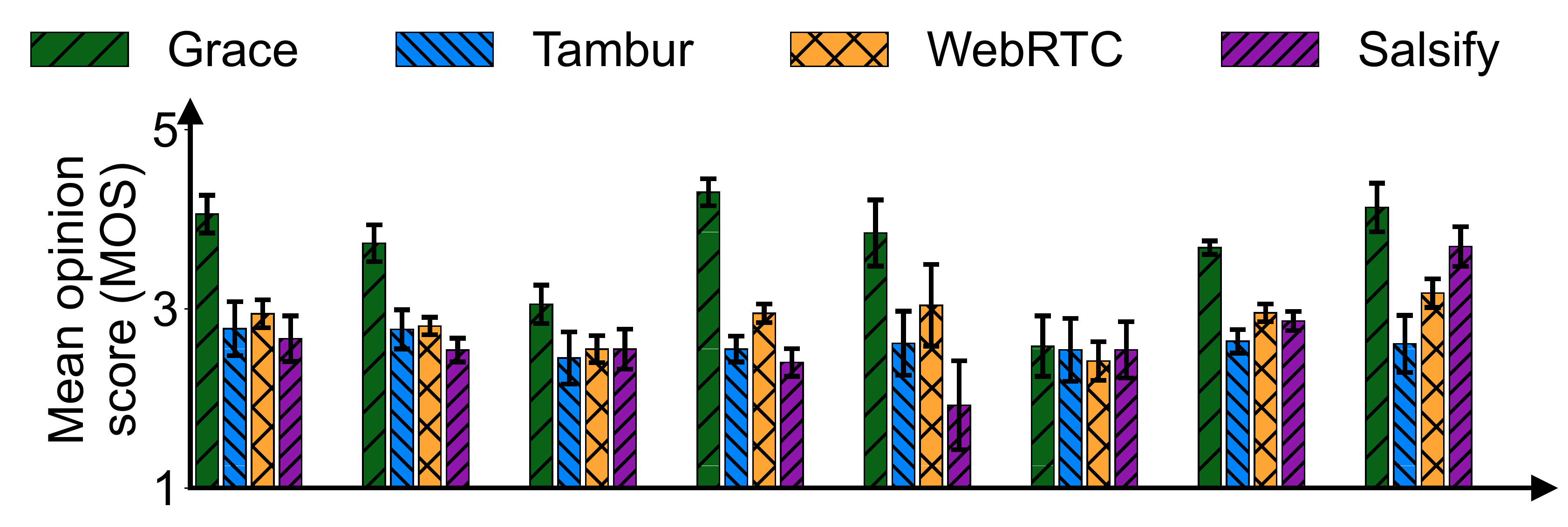}
    %\vspace{-12pt}
    \tightcaption{User study experiment shows that videos streamed by \name are consistently favored by real users.
    \cmedit{The error bar shows the standard deviation of the mean.}{Addressing the comment from reviewer A: What are the ranges plotted in Figure 17?}
    }
    \vspace{1pt}
    \label{fig:user-study}
\end{figure}

\mypara{User study} 
% To show decreasing the ratio of delayed frames and video freezes is worth the minor drop in quality for real users, 
To validate \name's effectiveness, we conducted an IRB-approved user study, collecting 960 user ratings from 240 Amazon MTurk workers~\cite{mturk} using the open-sourced tool VidPlat~\cite{zhang2023vidplat}.
\cmedit{We first choose a few genres based on the real-time video streaming use cases, including cloud gaming, real-time sports events, daily human activities, and video conferencing. Then, we randomly selected 8 video clips from the UGC dataset~\cite{wang2019youtube}.}{Addressing the comment from reviewer A: In the user study, how were the video clips selected?}
\tempstrike{We selected 8 video clips spanning various content types (including cloud gaming, real-time sports events, daily human activities, and video conferencing). }
These video clips were streamed using \name, Salsify codec, WebRTC with default FEC, and H.265 with Tambur.
% with the same setup as the simulation. 
(A screenshot of each video clip is shown in Figure~\ref{fig:user-study-screenshot} in Appendix.)
The sampled videos have a similar distribution of quality, realtimeness, and smoothness as seen in Figure~\ref{fig:gcc-delay-simulation}. 
% , and the user study is properly randomized following the steps described in \cite{sensei}.
Following~\cite{zhang2021sensei}, when an MTurk user signs up for the user study, they are randomly assigned to rate their user experience on a scale of 1--5 for the videos delivered through different methods. 
Figure~\ref{fig:user-study} displays the mean opinion score (MOS) for each video,
confirming that the videos rendered by \name are consistently favored by real users.

\tightsubsection{Microbenchmarking}
\label{subsec:eval:micro}
\vspace{-3pt}

\begin{figure}[t!]
    \centering
    \includegraphics[width=1\linewidth]{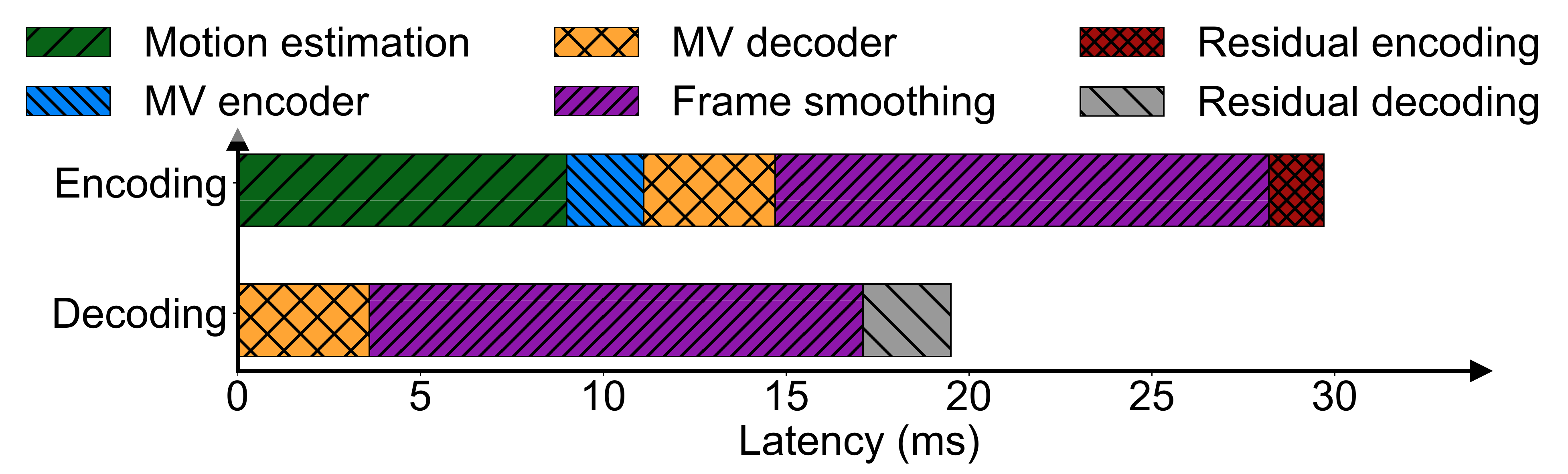}
    \vspace{-8pt}
    \tightcaption{Latency breakdown of GPU-based encoding and decoding of \name on a 720p frame.}
    \vspace{1pt}
    \label{fig:coding-time-gpu}
\end{figure}

\mypara{Encoding/decoding latency breakdown}
Figure~\ref{fig:coding-time-gpu} shows a breakdown of the encoding and decoding delays of \name on an Nvidia A40 GPU (5$\times$ cheaper and 3$\times$ slower than Nvidia A100).
%and breaks down the latency to each component.
\name encodes and decodes a 720p frame within 29.7~ms (33~fps) and 19.5~ms (51.2~fps), respectively.
It can also encode/decode 480p video at 65.8~fps/104.1~fps.
% This equates to frame rates of 33.6~fps and 51.2~fps for encoding and decoding 720p frames, respectively.
% When encoding/decoding a 480p frame, \name achieves 65.8~fps and 104.1~fps, respectively.

% The delay breakdown shows that motion estimation and frame smoothing take most encoding and decoding delays.
This breakdown also carries several implications.
% are the most time-consuming steps, taking 70\% of the encoding time.
First, the fast resync logic (\S\ref{subsec:protocol}) requires the encoder to run the MV decoder and residual decoder, which together only consume 6~ms on a 720p frame,
allowing resync to complete with a minimal increase in encoding delay. 
% Figure~\ref{fig:coding-time-gpu} indicates that the two steps take only 6~ms, 
% which means an encoder running at 25~fps can reencode a frame and finish such resynchronization very quickly.
Moreover, \name may need to encode a frame multiple times as explained in \S\ref{subsec:fast-coding},
%if the first encoded frame is far away from the targeted bitrate (similar to Salsify).
%As explained in \S\ref{subsec:fast-coding},
but the extra overhead only involves residual encoding, which takes only 1.5~ms on a 720p frame.

\begin{comment}

To encode an extra copy under a different bitrate for a 720p video frame, \name only needs 1.5~ms by running residual encoding again, as the most time-intensive portion of encoding is shared across models (\S\ref{subsec:fast-coding}).
% Given that the most time-intensive portion of encoding is shared across models in \name, the additional time required to encode multiple versions is minimal with the bitrate control optimizations in \S\ref{subsec:fast-coding}, averaging merely 1.5ms per extra copy for 720p video (0.5ms for 480p). 
Consequently, \name can encode three distinct copies within 33~ms, enabling the bitrate control as described in \S\ref{subsec:fast-coding}. 
Empirically, we found that over 65\% of frames require a single encoding pass, while 34\% need a second pass to adjust to the target bitrate. 
A marginal subset, less than 1\%, is subjected to a third encoding pass.

\end{comment}

\begin{figure}[t!]
    \centering
    \includegraphics[width=0.87\linewidth]{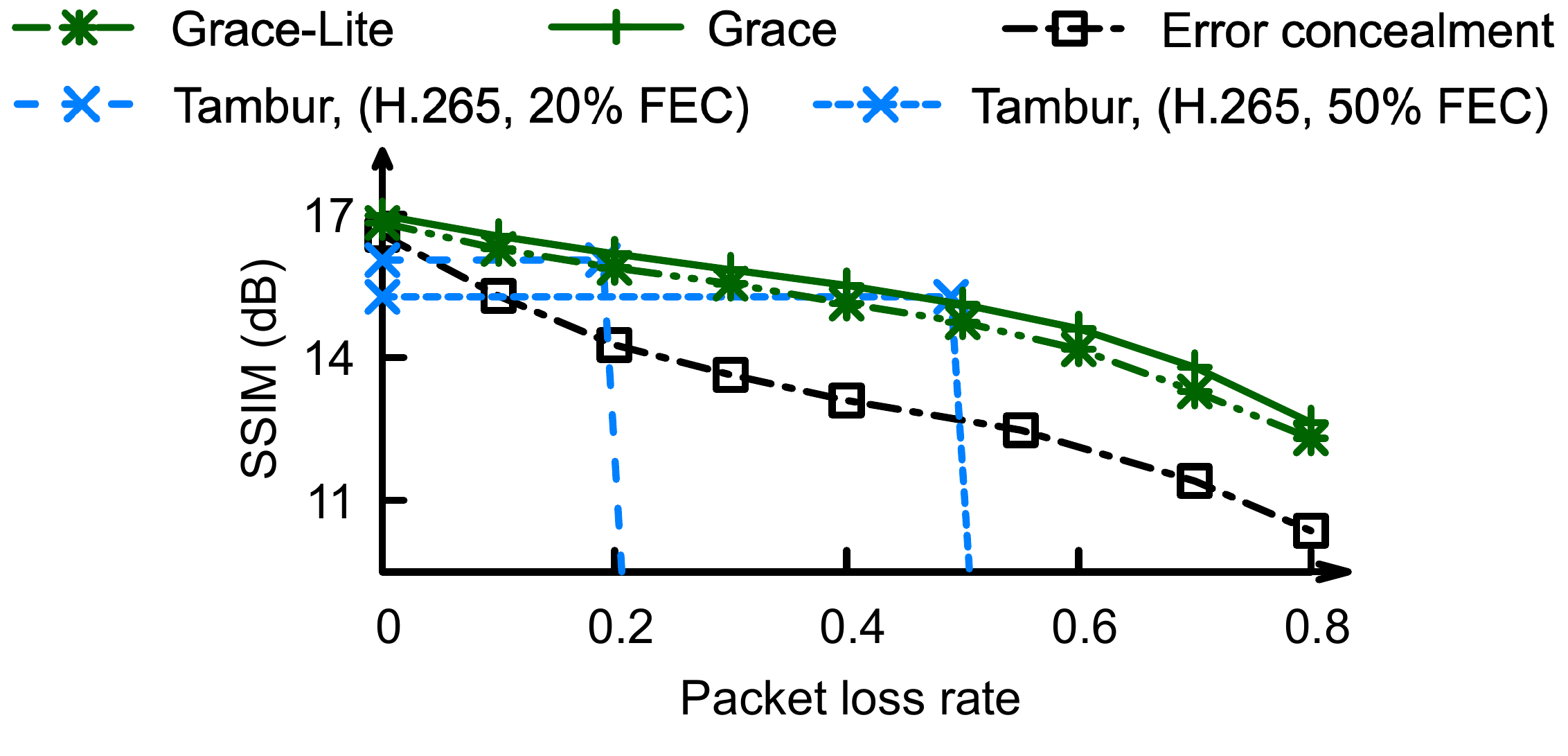}
    \tightcaption{\namelite realizes similar loss resilience to \name and outperforms other baselines.}
    \label{fig:cpu-performance}
    \vspace{-2pt}
\end{figure}

\mypara{Speed optimization in \namelite}
% We compare the encoding/decoding time of \name and \namelite on an iPhone 14 Pro.
% As illustrated in Table~\ref{tab:coding-time-cpu}, the optimizations in \namelite significantly curtail encoding and decoding times. 
With the optimizations described in \S\ref{subsec:fast-coding}, \namelite reduces the encoding delay of a 720 frame on iPhone 14 Pro from 314~ms to 38.1~ms, and the decoding delay from 239~ms to 14.4~ms.
% Without the speed optimizations, \name takes 314.2 and 239.8 ms to encode and decode a 720p frame, respectively; however, these times were dramatically reduced to 38.1 ms and 14.4 ms when using \namelite.
% This result demonstrates that \namelite can stream a 720p real-time video on iPhone 14 Pro. 
We also report \namelite's speed on CPUs with OpenVINO compilation in Appendix~\ref{app:openvino}.
%and for space limitation, we show the results in Appendix~\ref{app:openvino}.
% with OpenVINO on CPUs Appendix~\ref{app:openvino} provides more details about running \namelite on CPUs with the OpenVINO library.
Figure~\ref{fig:cpu-performance} compares the loss resilience of \namelite and \name with the two most competitive baselines in \S\ref{subsec:eval:resilience}---neural error concealment and Tambur.
At the same packet loss, \namelite achieves slightly lower quality than \name, yet it still outperforms other baselines.

\mypara{Impact of joint training}
Figure~\ref{fig:ssim-loss-abl} compares \name with its two variants: \namepretrain and \namedecoder, showing that both variants have lower levels of loss resilience than \name due to not jointly training the encoder and decoder. 
Appendix~\ref{app:abl-simulation} shows an example of frames decoded by the variants.
% When there are multiple consecutive frames impacted by packet loss, \name outperforms both variants by 0.5--2~dB in SSIM as the loss rate changes from 0.1 to 0.8. 
% Figure~\ref{fig:late-visual-example} shows the visual example comparing the decoded image of \name and its two variants when 50\% loss happens on 3 consecutive frames.
% \namepretrain, having never encountered packet loss during training, struggles to faithfully reconstruct the frame when faced with packet loss. 
% \namedecoder does yield a marginal improvement in frame quality, but visible artifacts still exist.
% In contrast, the joint training of both the encoder and decoder under packet loss conditions enables \name to deliver the most superior reconstruction quality, devoid of any prominent artifacts, and also achieves a high SSIM during loss.

% To highlight the benefit of joint training, we train a new \name variant by fine-tuning \namepretrain in the same way as \name except for freezing the encoder NN weights, \ie fine-tuning only the decoder with simulated loss.
% We refer to this variant as \namedecoder.

\tightsection{Limitation}
\label{sec:limit}

% While \name improves video quality across diverse packet loss scenarios, it has the following limitations. 
The current implementation of \name still has several limitations.
%First, \tempstrike{it is} \cmedit{our implementation of \name}{Addressing the comment from reviewer A: "It is not optimized enough..." - is this a limitation of GRACE, or your implementation of GRACE?} not optimized enough to run at 30~fps on very resource-constrained devices that barely sustain a classic video codec.
First, it is not optimized enough to run at 30~fps on very resource-constrained devices that barely sustain a classic video codec.
% for real-time operations on resource-constrained devices, such as budget mobile devices that can barely sustain a classic video codec. 
For instance, achieving real-time encoding and decoding on regular CPUs (\eg Intel Xeon Silver 4216) still requires 32 cores (\S\ref{app:openvino}).
%still requires high-end, 32-core CPUs (Appendix~\ref{app:openvino}).
%cannot run in real-time on resource-constrained devices, \eg budget mobile devices that barely execute a classic video codec.
%Also, to achieve real-time encoding/decoding on CPUs, it currently still requires relatively high-end CPUs with 32 cores (Appendix~\ref{app:openvino}).
Secondly, due to its use of NVC, 
\name may have lower compression efficiency than traditional handcrafted codecs on some video content that deviates a lot from the training data of NVC. 
%does not universally outperform traditional codecs that are heavily optimized for compression efficiency across all video content types. 
For instance, its compression efficiency is worse than H.26x on videos with high spatial complexity (\S\ref{subsec:eval:resilience}). 
In rare instances, \name is observed to fail to accurately reconstruct original frames under high packet losses. 
Third, our focus with \name is on unicast video communication rather than
multiparty conferencing.
%we focus only on unicast video communication, rather than problems in multi-party conferencing (\eg stream forwarding). 
%The goal of this work is to demonstrate the substantial improvement in loss resilience across a wide range of video content and realistic network conditions.
%our approach obtains much better loss resilience across a wide range of video content and realistic network dynamics. 
\tempstrike{
We hope \name can inspire future work to address these limitations. Potential avenues include distilling more lightweight models suitable for less powerful devices, more measurement studies to better understand NVC's generalization 
and extending \name to multiparty conferencing.
We also anticipate that the recent advancements in hardware~\cite{apple-neural-engine, opticflow-acc} will make GPUs more ubiquitous, enabling \name on more devices. % to operate at higher frame rates.
}
We hope \name can inspire future work to address these limitations. Potential avenues include democratizing \name on more devices by embracing the recent advancements in hardware~\cite{apple-neural-engine,opticflow-acc}, 
distilling more lightweight models suitable for less powerful devices.
%and extending \name to multiparty conferencing.
\cmedit{
We acknowledge there is not a good solution to address \name's generalization issue, which is a problem not unique to \name but inherent in general NVCs.
We hope that future measurement studies may shed light on the generalization of NVCs and contribute to their improvement.
}{Addressing the comment from reviewer A: I didn't understand what your proposal was to address the second shortcoming....}

\tightsection{Conclusion}

%This paper presents \name, a new real-time video system that specifically trains autoencoders to tolerate a range of packet loss rates.
%As a result, its quality gracefully degrades in {\em presence} of packet loss, with marginal compromise on quality in {\em absence} of packet loss.

%In some sense, our approach {\em increases} application complexity in order to {\em reduce} friction between networking and networked applications. 
%Traditional optimizations for networking, including recent ML-based ones, have focused on low-level control knobs (like sending rate, bitrate, path selection) which only {\em indirectly} affect what a {\em user} perceives.
%By contrast, our {\em autoencoder-based} approach focuses on maximizing the user-perceived quality of networked applications by training an NN-based data coding layer with packet loss and possibly a broader set of network objectives.

This paper presents \name, a real-time video system designed for loss resilience, 
preserving quality of experience (QoE) for users across diverse packet losses.
\name enhances loss resilience by jointly training a neural encoder and
decoder under a spectrum of packet losses.
It attains video quality on par with conventional codecs in the absence of packet loss,
and exhibits a less pronounced quality degradation as packet loss escalates,
outperforming existing loss-resilient methods.

\tightsection{Acknowledgement}
We thank the anonymous reviewers and our shepherd Dongsu Han. 
This project is supported by NSF CNS 2146496, 2131826, 2313190, 1901466, and UChicago CERES Center.

%-------------------------------------------------------------------------------
% \newpage
\bibliographystyle{plain}
\bibliography{reference}

\begin{thebibliography}{100}

\bibitem{mturk}
{Amazon Mechanical Turk}.
\newblock \url{https://www.mturk.com/}.

\bibitem{ZEROLAT1}
{Aurora5 HEVC Test Results}.
\newblock \url{https://www.visionular.com/en/putting-the-aurora5-hevc-encoder-to-the-test/}.

\bibitem{webrtc-vc-2}
{Bringing Zoom's end-to-end optimizations to WebRTC}.
\newblock \url{https://blog.livekit.io/livekit-one-dot-zero/}.

\bibitem{CABAC}
{Context-adaptive binary arithmetic coding}.
\newblock \url{https://en.wikipedia.org/wiki/Context-adaptive_binary_arithmetic_coding}.

\bibitem{coreml}
{Core ML Documentation}.
\newblock \url{https://developer.apple.com/documentation/coreml}.

\bibitem{webrtc-vr-1}
{Features of WebRTC VR Streaming}.
\newblock \url{https://flashphoner.com/features-of-webrtc-vr-streaming/}.

\bibitem{ffmpegstreaming}
{FFmpeg streaming guide}.
\newblock \url{http://trac.ffmpeg.org/wiki/StreamingGuide}.

\bibitem{lcg}
{Linear Congruential Generator}.
\newblock \url{https://en.wikipedia.org/wiki/Linear_congruential_generator}.

\bibitem{mahimahi-traces}
{Mamahi Cellular traces}.
\newblock \url{https://github.com/ravinet/mahimahi/tree/master/traces}.

\bibitem{fcc-dataset}
{Measuring Broadband Raw Data Releases}.
\newblock \url{https://www.fcc.gov/oet/mba/raw-data-releases}.

\bibitem{webrtc-gaming-1}
{Open Source Cloud Gaming with WebRTC}.
\newblock \url{https://webrtchacks.com/open-source-cloud-gaming-with-webrtc/}.

\bibitem{SITITOOL}
{SI/TI calculation tools}.
\newblock \url{https://github.com/VQEG/siti-tools}.

\bibitem{torchac}
{torchac: Fast Arithmetic Coding for PyTorch}.
\newblock \url{https://github.com/fab-jul/torchac}.

\bibitem{torchcompile}
{Torch.compile tutorial }.
\newblock \url{https://pytorch.org/tutorials/intermediate/torch_compile_tutorial.html}.

\bibitem{x265vp9compare}
{VP9 encoding/decoding performance vs. HEVC/H.264}.
\newblock \url{https://blogs.gnome.org/rbultje/2015/09/28/vp9-encodingdecoding-performance-vs-hevch-264/}.

\bibitem{webrtc-iot-2}
{WebRTC and IoT Applications}.
\newblock \url{https://rtcweb.in/webrtc-and-iot-applications/}.

\bibitem{webrtc-gaming-2}
{WebRTC Cloud Gaming: Unboxing Stadia}.
\newblock \url{https://webrtc.ventures/2021/02/webrtc-cloud-gaming-unboxing-stadia/}.

\bibitem{webrtc-vr-2}
{WebRTC: Enabling Collaboration Augmented Reality App}.
\newblock \url{https://arvrjourney.com/webrtc-enabling-collaboration-cebdd4c9ce06?gi=e19b1c0f65c0}.

\bibitem{webrtc-iot-1}
{WebRTC in IoT: What is the Intersection Point?}
\newblock \url{https://mobidev.biz/blog/webrtc-real-time-communication-for-the-internet-of-things}.

\bibitem{webrtc-vc-1}
{What powers Google Meet and Microsoft Teams? WebRTC Demystified}.
\newblock \url{https://levelup.gitconnected.com/what-powers-google-meet-and-microsoft-teams-webrtc-demystified-step-by-step-tutorial-e0cb422010f7}.

\bibitem{bpg}
{Better Portable Graphics}.
\newblock \url{https://bellard.org/bpg/}, 2014.

\bibitem{zhang2021sensei}
{SENSEI: Aligning Video Streaming Quality with Dynamic User Sensitivity}, author={Zhang, Xu and Ou, Yiyang and Sen, Siddhartha and Jiang, Junchen}.
\newblock In {\em {18th USENIX Symposium on Networked Systems Design and Implementation (NSDI 21)}}, pages 303--320, 2021.

\bibitem{apple-neural-engine}
{Deploying Transformers on the Apple Neural Engine}.
\newblock \url{https://machinelearning.apple.com/research/neural-engine-transformers}, 2022.

\bibitem{opticflow-acc}
{Harnessing the NVIDIA Ada Architecture for Frame-Rate Up-Conversion in the NVIDIA Optical Flow SDK}.
\newblock \url{https://developer.nvidia.com/blog/harnessing-the-nvidia-ada-architecture-for-frame-rate-up-conversion-in-the-nvidia-optical-flow-sdk/}, 2023.

\bibitem{abdallah2019h}
Asma~Ben Abdallah, Amin Zribi, Ali Dziri, Fethi Tlili, and Michel Terr{\'e}.
\newblock {H.264/AVC video transmission over UWB AV PHY IEEE 802.15. 3c using UEP and adaptive modulation techniques}.
\newblock In {\em {2019 International Conference on Advanced Communication Technologies and Networking (CommNet)}}, pages 1--6. IEEE, 2019.

\bibitem{ammar2016video}
Doreid Ammar, Katrien De~Moor, Min Xie, Markus Fiedler, and Poul Heegaard.
\newblock {Video QoE killer and performance statistics in WebRTC-based video communication}.
\newblock In {\em {2016 IEEE Sixth International Conference on Communications and Electronics (ICCE)}}, pages 429--436. IEEE, 2016.

\bibitem{arnab2021vivit}
Anurag Arnab, Mostafa Dehghani, Georg Heigold, Chen Sun, Mario Lu{\v{c}}i{\'c}, and Cordelia Schmid.
\newblock {Vivit: A video vision transformer}.
\newblock In {\em {Proceedings of the IEEE/CVF international conference on computer vision}}, pages 6836--6846, 2021.

\bibitem{badr2017fec}
Ahmed Badr, Ashish Khisti, Wai-tian Tan, Xiaoqing Zhu, and John Apostolopoulos.
\newblock {FEC for VoIP using dual-delay streaming codes}.
\newblock In {\em {IEEE INFOCOM 2017-IEEE Conference on Computer Communications}}, pages 1--9. IEEE, 2017.

\bibitem{blum2021webrtc}
Niklas Blum, Serge Lachapelle, and Harald Alvestrand.
\newblock {WebRTC-Realtime Communication for the Open Web Platform: What was once a way to bring audio and video to the web has expanded into more use cases we could ever imagine.}
\newblock {\em Queue}, 19(1):77--93, 2021.

\bibitem{bourtsoulatze2019deep}
Eirina Bourtsoulatze, David~Burth Kurka, and Deniz G{\"u}nd{\"u}z.
\newblock {Deep joint source-channel coding for wireless image transmission}.
\newblock {\em IEEE Transactions on Cognitive Communications and Networking}, 5(3):567--579, 2019.

\bibitem{carlucci2016analysis}
Gaetano Carlucci, Luca De~Cicco, Stefan Holmer, and Saverio Mascolo.
\newblock {Analysis and design of the google congestion control for web real-time communication (WebRTC)}.
\newblock In {\em {Proceedings of the 7th International Conference on Multimedia Systems}}, pages 1--12, 2016.

\bibitem{8919799}
Fabrizio Carpi, Christian Häger, Marco Martalò, Riccardo Raheli, and Henry~D. Pfister.
\newblock {Reinforcement Learning for Channel Coding: Learned Bit-Flipping Decoding}.
\newblock In {\em {2019 57th Annual Allerton Conference on Communication, Control, and Computing (Allerton)}}, pages 922--929, 2019.

\bibitem{castura2006rateless}
Jeff Castura and Yongyi Mao.
\newblock {Rateless coding over fading channels}.
\newblock {\em IEEE communications letters}, 10(1):46--48, 2006.

\bibitem{chang2019free}
Ya-Liang Chang, Zhe~Yu Liu, Kuan-Ying Lee, and Winston Hsu.
\newblock {Free-form video inpainting with 3d gated convolution and temporal patchgan}.
\newblock In {\em {Proceedings of the IEEE/CVF International Conference on Computer Vision}}, pages 9066--9075, 2019.

\bibitem{cheng2023abrf}
Sheng Cheng, Han Hu, and Xinggong Zhang.
\newblock {ABRF: Adaptive BitRate-FEC Joint Control for Real-Time Video Streaming}.
\newblock {\em IEEE Transactions on Circuits and Systems for Video Technology}, 2023.

\bibitem{cheng2020deeprs}
Sheng Cheng, Han Hu, Xinggong Zhang, and Zongming Guo.
\newblock {DeepRS: Deep-learning based network-adaptive FEC for real-time video communications}.
\newblock In {\em {2020 IEEE International Symposium on Circuits and Systems (ISCAS)}}, pages 1--5. IEEE, 2020.

\bibitem{pmlr-v97-choi19a}
Kristy Choi, Kedar Tatwawadi, Aditya Grover, Tsachy Weissman, and Stefano Ermon.
\newblock Neural joint source-channel coding.
\newblock In Kamalika Chaudhuri and Ruslan Salakhutdinov, editors, {\em Proceedings of the 36th International Conference on Machine Learning}, volume~97 of {\em Proceedings of Machine Learning Research}, pages 1182--1192. PMLR, 09--15 Jun 2019.

\bibitem{660830}
Wen-Jeng Chu and Jin-Jang Leou.
\newblock {Detection and concealment of transmission errors in H.261 images}.
\newblock {\em IEEE Transactions on Circuits and Systems for Video Technology}, 8(1):74--84, 1998.

\bibitem{conti2021not}
Mauro Conti, Simone Milani, Ehsan Nowroozi, and Gabriele Orazi.
\newblock {Do Not Deceive Your Employer with a Virtual Background: A Video Conferencing Manipulation-Detection System}.
\newblock {\em arXiv preprint arXiv:2106.15130}, 2021.

\bibitem{swift}
Mallesham Dasari, Kumara Kahatapitiya, Samir~R. Das, Aruna Balasubramanian, and Dimitris Samaras.
\newblock Swift: Adaptive video streaming with layered neural codecs.
\newblock In {\em 19th USENIX Symposium on Networked Systems Design and Implementation (NSDI 22)}, pages 103--118, Renton, WA, April 2022. USENIX Association.

\bibitem{dhawaskar2023converge}
Sandesh Dhawaskar~Sathyanarayana, Kyunghan Lee, Dirk Grunwald, and Sangtae Ha.
\newblock {Converge: QoE-driven Multipath Video Conferencing over WebRTC}.
\newblock In {\em {Proceedings of the ACM SIGCOMM 2023 Conference}}, pages 637--653, 2023.

\bibitem{dhondt2004flexible}
Yves Dhondt and Peter Lambert.
\newblock {Flexible Macroblock Ordering: an error resilience tool in H. 264/AVC}.
\newblock In {\em {5th FTW PhD Symposium}}. Ghent University. Faculty of Engineering, 2004.

\bibitem{dosovitskiy2020image}
Alexey Dosovitskiy, Lucas Beyer, Alexander Kolesnikov, Dirk Weissenborn, Xiaohua Zhai, Thomas Unterthiner, Mostafa Dehghani, Matthias Minderer, Georg Heigold, Sylvain Gelly, et~al.
\newblock {An image is worth 16x16 words: Transformers for image recognition at scale}.
\newblock {\em arXiv preprint arXiv:2010.11929}, 2020.

\bibitem{esser2021taming}
Patrick Esser, Robin Rombach, and Bjorn Ommer.
\newblock Taming transformers for high-resolution image synthesis.
\newblock In {\em Proceedings of the IEEE/CVF conference on computer vision and pattern recognition}, pages 12873--12883, 2021.

\bibitem{salsify}
Sadjad Fouladi, John Emmons, Emre Orbay, Catherine Wu, Riad~S. Wahby, and Keith Winstein.
\newblock Salsify: {Low-Latency} network video through tighter integration between a video codec and a transport protocol.
\newblock In {\em 15th USENIX Symposium on Networked Systems Design and Implementation (NSDI 18)}, pages 267--282, Renton, WA, April 2018. USENIX Association.

\bibitem{gao2020flow}
Chen Gao, Ayush Saraf, Jia-Bin Huang, and Johannes Kopf.
\newblock {Flow-edge guided video completion}.
\newblock In {\em {Computer Vision--ECCV 2020: 16th European Conference, Glasgow, UK, August 23--28, 2020, Proceedings, Part XII 16}}, pages 713--729. Springer, 2020.

\bibitem{garcia2019understanding}
Boni Garc{\'\i}a, Micael Gallego, Francisco Gort{\'a}zar, and Antonia Bertolino.
\newblock {Understanding and estimating quality of experience in WebRTC applications}.
\newblock {\em Computing}, 101:1585--1607, 2019.

\bibitem{garcia2020assessment}
Boni Garc{\'\i}a, Francisco Gort{\'a}zar, Micael Gallego, and Andrew Hines.
\newblock {Assessment of qoe for video and audio in webrtc applications using full-reference models}.
\newblock {\em Electronics}, 9(3):462, 2020.

\bibitem{grois2013performance}
Dan Grois, Detlev Marpe, Amit Mulayoff, Benaya Itzhaky, and Ofer Hadar.
\newblock {Performance comparison of h. 265/mpeg-hevc, vp9, and h. 264/mpeg-avc encoders}.
\newblock In {\em {2013 Picture Coding Symposium (PCS)}}, pages 394--397. IEEE, 2013.

\bibitem{7926071}
Tobias Gruber, Sebastian Cammerer, Jakob Hoydis, and Stephan~ten Brink.
\newblock {On deep learning-based channel decoding}.
\newblock In {\em {2017 51st Annual Conference on Information Sciences and Systems (CISS)}}, pages 1--6, 2017.

\bibitem{gunduz2019machine}
Deniz G{\"u}nd{\"u}z, Paul de~Kerret, Nicholas~D Sidiropoulos, David Gesbert, Chandra~R Murthy, and Mihaela van~der Schaar.
\newblock {Machine learning in the air}.
\newblock {\em IEEE Journal on Selected Areas in Communications}, 37(10):2184--2199, 2019.

\bibitem{he2023neural}
Zhaoyuan He, Yifan Yang, Shuozhe Li, Diyuan Dai, and Lili Qiu.
\newblock {Neural Video Recovery for Cloud Gaming}.
\newblock {\em arXiv preprint arXiv:2307.07847}, 2023.

\bibitem{he2023real}
Zhaoyuan He, Yifan Yang, Lili Qiu, and Kyoungjun Park.
\newblock {Real-Time Neural Video Recovery and Enhancement on Mobile Devices}.
\newblock {\em arXiv preprint arXiv:2307.12152}, 2023.

\bibitem{holmer2013handling}
Stefan Holmer, Mikhal Shemer, and Marco Paniconi.
\newblock {Handling packet loss in WebRTC}.
\newblock In {\em {2013 IEEE International Conference on Image Processing}}, pages 1860--1864. IEEE, 2013.

\bibitem{hu2021fvc}
Zhihao Hu, Guo Lu, and Dong Xu.
\newblock {FVC: A new framework towards deep video compression in feature space}.
\newblock In {\em {Proceedings of the IEEE/CVF Conference on Computer Vision and Pattern Recognition}}, pages 1502--1511, 2021.

\bibitem{ismaeil2000efficient}
Ismaeil Ismaeil, Shahram Shirani, Faouzi Kossentini, and Rabab Ward.
\newblock {An efficient, similarity-based error concealment method for block-based coded images}.
\newblock In {\em {Proceedings 2000 International Conference on Image Processing (Cat. No. 00CH37101)}}, volume~3, pages 388--391. IEEE, 2000.

\bibitem{itu-g114}
ITU-T.
\newblock {Recommendation {G.114}}, one-way transmission time.
\newblock {\em Series G: Transmission Systems and Media, Digital Systems and Networks, Telecommunication Standardization Sector of ITU}, 2003.

\bibitem{itu1999subjective}
P~ITU-T~RECOMMENDATION.
\newblock {Subjective video quality assessment methods for multimedia applications}.
\newblock 1999.

\bibitem{kang2022error}
Jaeyeon Kang, Seoung~Wug Oh, and Seon~Joo Kim.
\newblock {Error compensation framework for flow-guided video inpainting}.
\newblock In {\em {Computer Vision--ECCV 2022: 17th European Conference, Tel Aviv, Israel, October 23--27, 2022, Proceedings, Part XV}}, pages 375--390. Springer, 2022.

\bibitem{khalek2012cross}
Amin~Abdel Khalek, Constantine Caramanis, and Robert~W Heath.
\newblock {A cross-layer design for perceptual optimization of H. 264/SVC with unequal error protection}.
\newblock {\em IEEE Journal on selected areas in Communications}, 30(7):1157--1171, 2012.

\bibitem{kim2020neural}
Jaehong Kim, Youngmok Jung, Hyunho Yeo, Juncheol Ye, and Dongsu Han.
\newblock {Neural-enhanced live streaming: Improving live video ingest via online learning}.
\newblock In {\em {Proceedings of the Annual conference of the ACM Special Interest Group on Data Communication on the applications, technologies, architectures, and protocols for computer communication}}, pages 107--125, 2020.

\bibitem{Kingma2014}
Diederik~P. Kingma and Max Welling.
\newblock {Auto-Encoding Variational Bayes}.
\newblock In {\em {2nd International Conference on Learning Representations, {ICLR}} 2014, Banff, AB, Canada, April 14-16, 2014, Conference Track Proceedings}, 2014.

\bibitem{kolkeri2009error}
Vineeth~Shetty Kolkeri.
\newblock {\em {Error concealment techniques in H. 264/AVC, for video transmission over wireless networks}}.
\newblock PhD thesis, The University of Texas at Arlington, 2009.

\bibitem{kumar2006error}
Sunil Kumar, Liyang Xu, Mrinal~K Mandal, and Sethuraman Panchanathan.
\newblock {Error resiliency schemes in H. 264/AVC standard}.
\newblock {\em Journal of Visual Communication and Image Representation}, 17(2):425--450, 2006.

\bibitem{kurka2020deepjscc}
David~Burth Kurka and Deniz G{\"u}nd{\"u}z.
\newblock {Deepjscc-f: Deep joint source-channel coding of images with feedback}.
\newblock {\em IEEE Journal on Selected Areas in Information Theory}, 1(1):178--193, 2020.

\bibitem{lambert2006flexible}
Peter Lambert, Wesley De~Neve, Yves Dhondt, and Rik Van~de Walle.
\newblock {Flexible macroblock ordering in H. 264/AVC}.
\newblock {\em Journal of Visual Communication and Image Representation}, 17(2):358--375, 2006.

\bibitem{reparo}
Tianhong Li, Vibhaalakshmi Sivaraman, Lijie Fan, Mohammad Alizadeh, and Dina Katabi.
\newblock {Reparo: Loss-Resilient Generative Codec for Video Conferencing}.
\newblock {\em arXiv preprint arXiv:2305.14135}, 2023.

\bibitem{li2001overview}
Weiping Li.
\newblock {Overview of fine granularity scalability in MPEG-4 video standard}.
\newblock {\em IEEE Transactions on circuits and systems for video technology}, 11(3):301--317, 2001.

\bibitem{li2022towards}
Zhen Li, Cheng-Ze Lu, Jianhua Qin, Chun-Le Guo, and Ming-Ming Cheng.
\newblock {Towards an end-to-end framework for flow-guided video inpainting}.
\newblock In {\em {Proceedings of the IEEE/CVF conference on computer vision and pattern recognition}}, pages 17562--17571, 2022.

\bibitem{liang2021swinir}
Jingyun Liang, Jiezhang Cao, Guolei Sun, Kai Zhang, Luc Van~Gool, and Radu Timofte.
\newblock {Swinir: Image restoration using swin transformer}.
\newblock In {\em {Proceedings of the IEEE/CVF international conference on computer vision}}, pages 1833--1844, 2021.

\bibitem{Liu21_FuseFormer}
Rui Liu, Hanming Deng, Yangyi Huang, Xiaoyu Shi, Lewei Lu, Wenxiu Sun, Xiaogang Wang, Jifeng Dai, and Hongsheng Li.
\newblock {FuseFormer: Fusing Fine-Grained Information in Transformers for Video Inpainting}.
\newblock In {\em ICCV}, 2021.

\bibitem{grad}
Yunzhuo Liu, Bo~Jiang, Tian Guo, Ramesh~K. Sitaraman, Don Towsley, and Xinbing Wang.
\newblock Grad: Learning for overhead-aware adaptive video streaming with scalable video coding.
\newblock In {\em Proceedings of the 28th ACM International Conference on Multimedia}, MM '20, page 349–357, New York, NY, USA, 2020. Association for Computing Machinery.

\bibitem{dvc}
Guo Lu, Wanli Ouyang, Dong Xu, Xiaoyun Zhang, Chunlei Cai, and Zhiyong Gao.
\newblock {DVC}: An end-to-end deep video compression framework.
\newblock In {\em {2019 IEEE/CVF Conference on Computer Vision and Pattern Recognition (CVPR)}}, pages 10998--11007, 2019.

\bibitem{4494082}
Rong Luo and Bin Chen.
\newblock {A Hierarchical Scheme of Flexible Macroblock Ordering for ROI based H.264/AVC Video Coding}.
\newblock In {\em {2008 10th International Conference on Advanced Communication Technology}}, volume~3, pages 1579--1582, 2008.

\bibitem{ma2022deepfgs}
Yi~Ma, Yongqi Zhai, and Ronggang Wang.
\newblock {DeepFGS: Fine-Grained Scalable Coding for Learned Image Compression}.
\newblock {\em arXiv preprint arXiv:2201.01173}, 2022.

\bibitem{mackay2005fountain}
David~JC MacKay.
\newblock {Fountain codes}.
\newblock {\em IEE Proceedings-Communications}, 152(6):1062--1068, 2005.

\bibitem{mackay1997near}
David~JC MacKay and Radford~M Neal.
\newblock {Near Shannon limit performance of low density parity check codes}.
\newblock {\em Electronics letters}, 33(6):457--458, 1997.

\bibitem{macmillan2021measuring}
Kyle MacMillan, Tarun Mangla, James Saxon, and Nick Feamster.
\newblock {Measuring the performance and network utilization of popular video conferencing applications}.
\newblock In {\em {Proceedings of the 21st ACM Internet Measurement Conference}}, pages 229--244, 2021.

\bibitem{mathieu2015deep}
Michael Mathieu, Camille Couprie, and Yann LeCun.
\newblock {Deep multi-scale video prediction beyond mean square error}.
\newblock {\em arXiv preprint arXiv:1511.05440}, 2015.

\bibitem{meng2022achieving}
Zili Meng, Yaning Guo, Chen Sun, Bo~Wang, Justine Sherry, Hongqiang~Harry Liu, and Mingwei Xu.
\newblock {Achieving consistent low latency for wireless real-time communications with the shortest control loop}.
\newblock In {\em {Proceedings of the ACM SIGCOMM 2022 Conference}}, pages 193--206, 2022.

\bibitem{nam2020novel}
Cholman Nam, Changgon Chu, Taeguk Kim, and Sokmin Han.
\newblock {A novel motion recovery using temporal and spatial correlation for a fast temporal error concealment over H. 264 video sequences}.
\newblock {\em Multimedia Tools and Applications}, 79:1221--1240, 2020.

\bibitem{netravali2015mahimahi}
Ravi Netravali, Anirudh Sivaraman, Somak Das, Ameesh Goyal, Keith Winstein, James Mickens, and Hari Balakrishnan.
\newblock {Mahimahi: Accurate Record-and-Replay for HTTP}.
\newblock In {\em {2015 USENIX Annual Technical Conference (USENIX ATC 15)}}, pages 417--429, 2015.

\bibitem{voxel}
Mirko Palmer, Malte Appel, Kevin Spiteri, Balakrishnan Chandrasekaran, Anja Feldmann, and Ramesh~K. Sitaraman.
\newblock Voxel: Cross-layer optimization for video streaming with imperfect transmission.
\newblock In {\em Proceedings of the 17th International Conference on Emerging Networking EXperiments and Technologies}, CoNEXT '21, page 359–374, New York, NY, USA, 2021. Association for Computing Machinery.

\bibitem{peters2008reinforcement}
Jan Peters and Stefan Schaal.
\newblock {Reinforcement learning of motor skills with policy gradients}.
\newblock {\em Neural networks}, 21(4):682--697, 2008.

\bibitem{ray2022sqp}
Devdeep Ray, Connor Smith, Teng Wei, David Chu, and Srinivasan Seshan.
\newblock {SQP: Congestion Control for Low-Latency Interactive Video Streaming}.
\newblock {\em arXiv preprint arXiv:2207.11857}, 2022.

\bibitem{rudow2023tambur}
Michael Rudow, Francis~Y. Yan, Abhishek Kumar, Ganesh Ananthanarayanan, Martin Ellis, and KV~Rashmi.
\newblock {Tambur: Efficient loss recovery for videoconferencing via streaming codes}.
\newblock In {\em {20th USENIX Symposium on Networked Systems Design and Implementation (NSDI 23)}}, pages 953--971, 2023.

\bibitem{sankisa2018video}
Arun Sankisa, Arjun Punjabi, and Aggelos~K Katsaggelos.
\newblock {Video error concealment using deep neural networks}.
\newblock In {\em {2018 25th IEEE International Conference on Image Processing (ICIP)}}, pages 380--384. IEEE, 2018.

\bibitem{schierl2007mobile}
Thomas Schierl, Thomas Stockhammer, and Thomas Wiegand.
\newblock {Mobile video transmission using scalable video coding}.
\newblock {\em IEEE transactions on circuits and systems for video technology}, 17(9):1204--1217, 2007.

\bibitem{schwarz2007overview}
Heiko Schwarz, Detlev Marpe, and Thomas Wiegand.
\newblock {Overview of the scalable video coding extension of the H. 264/AVC standard}.
\newblock {\em IEEE Transactions on circuits and systems for video technology}, 17(9):1103--1120, 2007.

\bibitem{sharma2023estimating}
Taveesh Sharma, Tarun Mangla, Arpit Gupta, Junchen Jiang, and Nick Feamster.
\newblock {Estimating WebRTC Video QoE Metrics Without Using Application Headers}.
\newblock {\em arXiv preprint arXiv:2306.01194}, 2023.

\bibitem{shi2022alphavc}
Yibo Shi, Yunying Ge, Jing Wang, and Jue Mao.
\newblock {AlphaVC: High-Performance and Efficient Learned Video Compression}.
\newblock In {\em {Computer Vision--ECCV 2022: 17th European Conference, Tel Aviv, Israel, October 23--27, 2022, Proceedings, Part XIX}}, pages 616--631. Springer, 2022.

\bibitem{sivaraman2022gemino}
Vibhaalakshmi Sivaraman, Pantea Karimi, Vedantha Venkatapathy, Mehrdad Khani, Sadjad Fouladi, Mohammad Alizadeh, Fr{\'e}do Durand, and Vivienne Sze.
\newblock {Gemino: Practical and Robust Neural Compression for Video Conferencing}.
\newblock {\em arXiv preprint arXiv:2209.10507}, 2022.

\bibitem{tan2011new}
Keyu Tan and Alan Pearmain.
\newblock {A new error resilience scheme based on FMO and error concealment in H. 264/AVC}.
\newblock In {\em {2011 IEEE International Conference on Acoustics, Speech and Signal Processing (ICASSP)}}, pages 1057--1060. IEEE, 2011.

\bibitem{tan1999multicast}
Wai-tian Tan and Avideh Zakhor.
\newblock {Multicast transmission of scalable video using receiver-driven hierarchical FEC}.
\newblock In {\em Packet Video Workshop}, volume~99, 1999.

\bibitem{tan2001video}
Wai-Tian Tan and Avideh Zakhor.
\newblock Video multicast using layered fec and scalable compression.
\newblock {\em IEEE Transactions on circuits and systems for video technology}, 11(3):373--386, 2001.

\bibitem{wang1998error}
Yao Wang and Qin-Fan Zhu.
\newblock {Error control and concealment for video communication: A review}.
\newblock {\em Proceedings of the IEEE}, 86(5):974--997, 1998.

\bibitem{wang2013novel}
Yi~Wang, Xiaoqiang Guo, Feng Ye, Aidong Men, and Bo~Yang.
\newblock {A novel temporal error concealment framework in H. 264/AVC}.
\newblock In {\em {2013 IEEE International Conference on Multimedia and Expo (ICME)}}, pages 1--6. IEEE, 2013.

\bibitem{wang2019youtube}
Yilin Wang, Sasi Inguva, and Balu Adsumilli.
\newblock {YouTube UGC dataset for video compression research}.
\newblock In {\em 2019 IEEE 21st International Workshop on Multimedia Signal Processing (MMSP)}, pages 1--5. IEEE, 2019.

\bibitem{wenger2002scattered}
Stephan Wenger and Michael Horowitz.
\newblock {Scattered slices: a new error resilience tool for H. 26L}.
\newblock {\em JVT-B027}, 2, 2002.

\bibitem{wicker1999reed}
Stephen~B. Wicker and Vijay~K. Bhargava.
\newblock {\em {Reed-Solomon} codes and their applications}.
\newblock John Wiley \& Sons, 1999.

\bibitem{wu2023zgaming}
Jiangkai Wu, Yu~Guan, Qi~Mao, Yong Cui, Zongming Guo, and Xinggong Zhang.
\newblock {ZGaming: Zero-Latency 3D Cloud Gaming by Image Prediction}.
\newblock In {\em Proceedings of the ACM SIGCOMM 2023 Conference}, pages 710--723, 2023.

\bibitem{xiang2019generative}
Chongyang Xiang, Jiajun Xu, Chuan Yan, Qiang Peng, and Xiao Wu.
\newblock {Generative adversarial networks based error concealment for low resolution video}.
\newblock In {\em {ICASSP 2019-2019 IEEE International Conference on Acoustics, Speech and Signal Processing (ICASSP)}}, pages 1827--1831. IEEE, 2019.

\bibitem{vimeo-dataset}
Tianfan Xue, Baian Chen, Jiajun Wu, Donglai Wei, and William~T Freeman.
\newblock {Video enhancement with task-oriented flow}.
\newblock {\em International Journal of Computer Vision}, 127(8):1106--1125, 2019.

\bibitem{puffer}
Francis~Y. Yan, Hudson Ayers, Chenzhi Zhu, Sadjad Fouladi, James Hong, Keyi Zhang, Philip Levis, and Keith Winstein.
\newblock Learning \emph{in situ}: a randomized experiment in video streaming.
\newblock In {\em 17th {USENIX} Symposium on Networked Systems Design and Implementation ({NSDI} 20)}, pages 495--511, Santa Clara, CA, February 2020. {USENIX} Association.

\bibitem{yang2020learning}
Ren Yang, Fabian Mentzer, Luc Van~Gool, and Radu Timofte.
\newblock {Learning for video compression with recurrent auto-encoder and recurrent probability model}.
\newblock {\em IEEE Journal of Selected Topics in Signal Processing}, 15(2):388--401, 2020.

\bibitem{neuroscaler}
Hyunho Yeo, Hwijoon Lim, Jaehong Kim, Youngmok Jung, Juncheol Ye, and Dongsu Han.
\newblock {NeuroScaler: neural video enhancement at scale}.
\newblock In {\em {Proceedings of the ACM SIGCOMM 2022 Conference}}, pages 795--811, 2022.

\bibitem{locki}
Huanhuan Zhang, Anfu Zhou, Yuhan Hu, Chaoyue Li, Guangping Wang, Xinyu Zhang, Huadong Ma, Leilei Wu, Aiyun Chen, and Changhui Wu.
\newblock {Loki: improving long tail performance of learning-based real-time video adaptation by fusing rule-based models}.
\newblock In {\em {Proceedings of the 27th Annual International Conference on Mobile Computing and Networking}}, MobiCom '21, page 775–788, New York, NY, USA, 2021. Association for Computing Machinery.

\bibitem{onrl}
Huanhuan Zhang, Anfu Zhou, Jiamin Lu, Ruoxuan Ma, Yuhan Hu, Cong Li, Xinyu Zhang, Huadong Ma, and Xiaojiang Chen.
\newblock {OnRL: improving mobile video telephony via online reinforcement learning}.
\newblock In {\em {Proceedings of the 26th Annual International Conference on Mobile Computing and Networking}}, MobiCom '20, New York, NY, USA, 2020. Association for Computing Machinery.

\bibitem{zhang2021sample}
Junzi Zhang, Jongho Kim, Brendan O'Donoghue, and Stephen Boyd.
\newblock {Sample efficient reinforcement learning with REINFORCE}.
\newblock In {\em {Proceedings of the AAAI Conference on Artificial Intelligence}}, volume~35, pages 10887--10895, 2021.

\bibitem{zhang2022flow}
Kaidong Zhang, Jingjing Fu, and Dong Liu.
\newblock {Flow-guided transformer for video inpainting}.
\newblock In {\em {European Conference on Computer Vision}}, pages 74--90. Springer, 2022.

\bibitem{zhang2022inertia}
Kaidong Zhang, Jingjing Fu, and Dong Liu.
\newblock {Inertia-guided flow completion and style fusion for video inpainting}.
\newblock In {\em {Proceedings of the IEEE/CVF conference on computer vision and pattern recognition}}, pages 5982--5991, 2022.

\bibitem{zhang2008error}
Qing Zhang and Guizhong Liu.
\newblock {Error resilient coding of H. 264 using intact long-term reference frames}.
\newblock 2008.

\bibitem{zhang2023vidplat}
Xu~Zhang, Hanchen Li, Paul Schmitt, Marshini Chetty, Nick Feamster, and Junchen Jiang.
\newblock {VidPlat: A Tool for Fast Crowdsourcing of Quality-of-Experience Measurements}, 2023.

\bibitem{zhao2010rd}
Zenghua Zhao and Shubing Long.
\newblock {RD-Based Adaptive UEP for H. 264 Video Transmission in Wireless Networks}.
\newblock In {\em {2010 International Conference on Multimedia Information Networking and Security}}, pages 72--76. IEEE, 2010.

\bibitem{concerto}
Anfu Zhou, Huanhuan Zhang, Guangyuan Su, Leilei Wu, Ruoxuan Ma, Zhen Meng, Xinyu Zhang, Xiufeng Xie, Huadong Ma, and Xiaojiang Chen.
\newblock {Learning to coordinate video codec with transport protocol for mobile video telephony}.
\newblock In {\em {The 25th Annual International Conference on Mobile Computing and Networking}}, pages 1--16, 2019.

\bibitem{zhou2010efficient}
Jie Zhou, Bo~Yan, and Hamid Gharavi.
\newblock {Efficient motion vector interpolation for error concealment of H. 264/AVC}.
\newblock {\em IEEE Transactions on Broadcasting}, 57(1):75--80, 2010.

\bibitem{zhu2016nada}
X~Zhu, P~Pan, M~Ramalho, S~Mena, P~Jones, J~Fu, S~D’Aronco, and C~Ganzhorn.
\newblock {Nada: A unified congestion control scheme for real-time media, draft-ietf-rmcat-nada-02}.
\newblock {\em Internet Engineering Task Force, IETF}, 2016.

\bibitem{zuo2022deadline}
Xutong Zuo, Yong Cui, Xin Wang, and Jiayu Yang.
\newblock {Deadline-aware Multipath Transmission for Streaming Blocks}.
\newblock In {\em {IEEE INFOCOM 2022-IEEE Conference on Computer Communications}}, pages 2178--2187. IEEE, 2022.

\end{thebibliography}

\appendix

%!TEX root = ../main.tex
%!TEX spellcheck = en_US

\newpage

\section{Details of NVC architecture and training}

\subsection{More details on \name's NVC model}\label{app:details}

% DVC model to compress P-frames mimics the mpeg encoder and decoder structure, with residual, motion vector encoder and decoder, and motion compensation substituted with convolutional networks.

% residual network

% Motion estimation is realized in three steps. First, we use optical flow estimation using a library to estimate motion vectors from optical between two frames. Then, optical flow is encoded using a motion vector encoder. The motion vector encoder consists of four convolutional layers, separated by three generalized divisive normalization layers. After the encoder, we apply quantization and loss to the motion vector data, which then is fed into the decoder. The motion vector decoder consists of four deconvolution layers, with three inverse generalized divisive normalization layers.

% We take the output of the motion vector decoder and output of residual decoder and feed it into the motion compensation network. Motion compensation network first takes in the reconstructed motion vector and a reference frame, where reference frame is warped using the reconstructed motion vector. Then, the resulting warped frame together with the reference frame and reconstructed motion vector are fed into a convolutional network, which consists of six convolutional layers with average pooling steps in between them. The output of this convolutional network is the reconstructed frame.

Grace uses the exact same model architecture as the original DVC model \cite{dvc}. With an RGB input image of size $C \times H \times W$, where $H, W$ are the height and width of the image, and $C = 3$ is number of channels in RGB images, the encoder neural network will encode the image into a compressed motion vector of size $128 \times (H/16) \times (W/16)$ and a compressed residual of size $96 \times (H/16) \times (W/16)$. Then those two compressed features will be quantized and converted into bytesteam using entropy encoding.

When we finetune the DVC model to get our \name's loss resilient model, we train on the 90k Vimeo Dataset, with batch size of 4, learning rate of $10^{-4}$ and learning rate decay of 0.1, and an Adam optimizer. 

\subsection{Making \name trainable}
\label{app:reinforce}

Since $\loss$ is a non-differentiable random function, the gradient of the expectation of $\Distort$ in Eq.~\ref{eq:2} cannot be directly calculated. 
To address this issue, we use the REINFORCE trick~\cite{Kingma2014} for reparameterization. 
%We express the gradient of Eq.~\ref{eq:2} as
% According to differentiation property of logarithms, we can 
First, given the packet loss distribution $\loss(\y)$, we can apply the differentiation property of logarithms to get

$$ \nabla_\phi \loss(\y) =  \loss(\y) \nabla_\phi \log{\loss(\y)}$$

Therefore, our gradient of the expectation of $\Distort(\Decoder(\y), \x)$ becomes
\begin{multline}
\nabla_\phi \mathbb{E}_{\y\sim \loss(\y)}([\Distort(\Decoder(\y), \x)]) \\ 
= \mathbb{E}_{\y\sim \loss(\y)}([\Distort(\Decoder(\y), \x)\nabla_\phi \log{\loss(\y)}])
\end{multline}
which can be estimated using Monte-Carlo sampling
$\approx \frac{1}{N} \sum_{i=1}^{N} \Distort(\Decoder(\y_i), \x) \nabla_\phi \log{\loss(\y_i)} $.
Since in our application, the loss is an independent and identically distributed random variable, the gradient evaluates to either $0$ or $1$, hence we propagate the gradients for the encoder only for $\Distort(\Decoder(\y_i), \x)$ where $\loss(\y_i) = 1$.

\section{Realtime video framework for \name}
\subsection{Fast re-encoding and re-decoding under loss}\label{app:resync}

In \name's NVC, the most time-consuming components are motion estimation NN and frame smoothing NN, taking 28\% and 42\% of the total encoding time respectively. Fortunately, we do not need to use them during resync (\S\ref{subsec:protocol}).
When the packet loss feedback arrives at the encoder, it takes the following steps to generate a new reference frame to re-sync with the decoder (Assuming the loss feedback is for 6$^{th}$ frame and the encoder is about to encode 10$^{th}$ frame)
\begin{packeditemize}
    \item First, \name re-decodes the motion vector and residuals based on the packet loss feedback for 6$^{th}$ frame. This step needs to run the motion decoder NN and residual decoder NN, which only takes around 18\% of the encoding time.
    \item Second, \name apply the cached motion vector and residuals of 7$^{th}$ frame on the ``reconstructed'' 6$^{th}$ frame to generate the ``reconstructed'' 7$^{th}$ frame. It applies the same logic on 8$^{th}$ and 9$^{th}$ frame and finally gets the ``reconstructed'' 9$^{th}$ frame. We do not run frame smoothing NN since the quality of the reference frame does not have a significant impact on compression efficiency. Therefore, this step does not involve any NN inference. It only needs to apply the motion and add the residuals, which takes 1\% of the encoding time.
    \item Finally, \name uses the ``reconstructed'' 9$^{th}$ frame as the reference frame to encode the 10$^{th}$ frame. It is the same as encoding a frame when there are no packet losses. It will add an extra tag to the frame so that the receiver knows which reference frame to use.
\end{packeditemize}
To summarize, the encoder side's computational overhead is usually less than 10\%. The logic requires the encoder to cache the motion vectors and residuals, but the cached value of frame $x$ can be dropped after receiving the packet loss feedback of that frame.

At the receiver side, when receiving the frame with the extra tag, it will follow the same process as the second step above to generate the same ``reconstructed'' reference frame as the encoder. Again, the overhead is negligible since it does not require NN inference.

\begin{figure}[t!]
    \centering
    \includegraphics[width=0.75\linewidth]{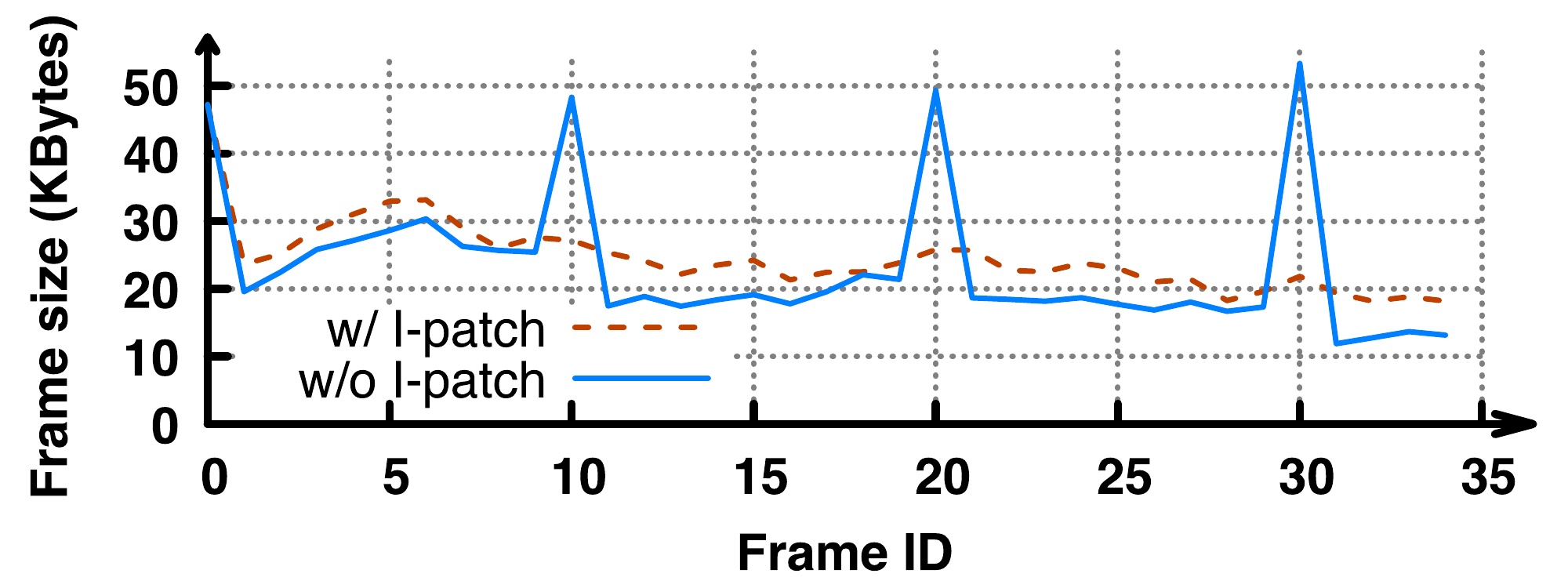}
    \tightcaption{Encoding each P-frame with a small I-patch leads to smoother frame sizes than naively inserting I-frames.}
    \label{fig:keyframe_smooth}
\end{figure}

\subsection{How \name handles I-frames} \label{app:ipatch}

\cmedit{}{Addressing the comment from reviewer B: How does it select an I-frame? Now that you can recover from lost packets, it might be that you can get away with using frequent I-frames.}
\name uses BPG~\cite{bpg} (also used in H.265) to encode and decode I-frames every 1000 frames. 
%\yc{consider move the I-patch part to appendix and just mention I-frame is not critical in some other places}
That said, in many NVCs (including DVC), 
%Placement of I-frames poses a particular challenge for \name's NVCs (both pre-trained and re-trained), because while 
the quality of P-frames will gradually degrade after an I-frame.
%, quality gradually degrades after about 10 frames. 
%for two reasons: (1) I-frames are generally bigger in size than P-frames, and 
%(2)  while the first few P-frames after an I-frame enjoy high quality, quality gradually degrades after about 10 frames. 
By simply adding frequent I-frames (\eg every 10 frames), we can achieve similar average compression efficiency with  H.264 and H.265 when they use an optimal I-frame interval.
% \footnote{Many computer-vision papers on NVCs have conveniently used a small GoP when compared against H.265, but the implication of large I-frame size for congestion control is rarely discussed.}, 
% which strikes a better tradeoff between quality and {\em average} bitrate (comparable to H.264 and H.265 with large I-frame intervals).
However, since I-frames are larger than P-frames, adding too many I-frames causes frequent spikes in frame size. %, so it is difficult for congestion control to utilize network capacity. %at a fixed frame interval. 
%make it difficult for congestion control to send the frames at a fixed interval (inversely proportional to the frame rate).
Instead, \name uses an {\em extra} small square-sized patch as a tiny I-frame, called {\em I-patch}, on {\em every P frame}.
We split each frame into $k$ patches, and for a window of $k$ frames, each frame is sent with an extra I-patch at a different location, so I-patch ``scan through'' the whole frame every $k$ frame. 
By default, $k=30$ though we empirically found any value between 10 and 30 works well.
% the location of the I-patch on a frame changes over time, so that the I-patch will scan through the whole frame size every $k$ frames. 
With I-patch, \name does not need to send any I-frames (except the first frame). 
We use BPG~\cite{bpg} to encode/decode the I-patch. 
% , and found that as long as $k$ is between 10 and 30, the compression efficiency is roughly the same. 
Figure~\ref{fig:keyframe_smooth} shows that when $k=10$, I-patch mitigates the sudden size increase caused by I-frames. 

%\tightsubsection{Encoding a sequence of frames}
%\label{subsec:delivery:iframe}

%To stream a sequence of video frames, a key question is how often I-frames should be inserted in P-frames. 
%(As explained in \S???????????????????????????????????????????????????????????????????????????????????????????????????\ref{subsec:ae-background}, \name uses only I-/P-frames, as in most real-time video systems.)
%For the P-frame NVC we have tried, we observe that the decoded quality can be very high right after each I-frame but it gradually degrades after about 10 frames. 

%To add I-frames frequently without the intermittent bitrate spikes caused by the I-frames, 
% (128$\times$128 to 512$\times$512) 

It is worth noting that though I-patch encoding can also use a loss-resilient NVC, we do not protect their packet loss to simplify the system design. 
This is because if each patch will see an I-patch every $k$ frames, so even if one patch is lost, its impact is confined to the next $k$ frames, and empirically, even this impact is marginal since P-frames are still delivered.

\subsection{Working with congestion control}\label{app:cc}

\cmedit{
\name can be integrated with any existing congestion control (CC) algorithms.
When combined with \name, CC does {\em not} need to retransmit packets, unless no packets of a frame are received.
CC determines the sending rate of packets and the target size of the next frame, while \name decides the content in each packet. 
Therefore, \name would not change the properties of the CC, such as fast convergence, oscillation avoidance, and TCP friendliness.
In real-time video communication, traditional CC algorithms like GCC~\cite{carlucci2016analysis} typically mitigate packet losses by reducing bandwidth use, due to the non-loss-tolerant nature of conventional video codecs.
These codecs necessitate retransmissions when packet loss happens, causing frame delays and video stalls.
Conversely, \name is designed to handle packet losses by decoding the partially received frames with graceful quality.
This capability allows \name to employ a more aggressive congestion control strategy, which, while resulting in occasional packet losses, enhances bandwidth utilization.
An illustration of this approach can be found in Appendix~\ref{app:salcc}, where \name works with Salsify's congestion control (Sal-CC)~\cite{salsify} that yields a higher average sending rate albeit with increased packet loss.
}{Addressing congestion-control-related comments from reviewer A.}

\subsection{Integration in WebRTC}
\label{app:webrtc}
\name is implemented with 3K lines of code, in both Python (mostly for NVC NNs) and C++ (for frame delivery and WebRTC integration). The code and trained model of \name will be made public upon the publication of this paper.
The integration with WebRTC is logically straightforward since \name (including I-frame and P-frame encodings) exposes similar interface as the default codec in WebRTC. 
% \jc{Ziyi, please fill in the details}

We substitute the libvpx VP8 Encoder/Decoder in WebRTC with our \name implementation. When the sender encodes a frame, it parses the image data from the \texttt{VideoFrame} data structure (YUV format) into \texttt{torch.Tensor} (RGB format) and feed it into our \name encoder, which will return the encoded result as a byte array. Then the encoded bytes are stored into an \texttt{EncodedImage} (class in WebRTC) and sent through the network to the receiver as RTP packets. 
% \zy{I could also add several sentences about packetization logic (commented out below)}
% The encoded bytes are then packetized as RTP packets and sent through the network to the receiver. 
We modify the built-in \texttt{RtpVideoStreamReceiver} (class in WebRTC) so that the receiver could flexibly decode the received packets even when not all the packets are received. When the receiver decides to decode the frame, it depacketizes the received packets into encoded data. Then it will use the \name decoder to decode the image into RGB format and then convert it back to YUV for displaying on the receiver side.

\section{Supportive details for \name's evaluation experiments}

\begin{figure}[t]
    \centering
    \includegraphics[width=0.79\linewidth]{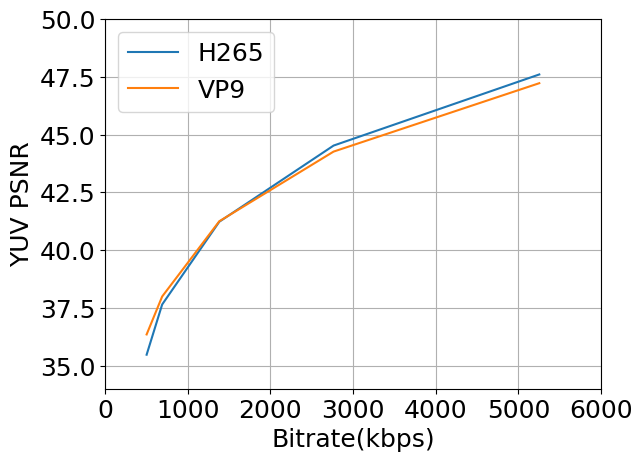}
    \tightcaption{H265 vs VP9 Encoding Efficiency on Kinetics}
    \label{fig:h265vsvp9}
\end{figure}

\tightsubsection{VP9 and H265 Comparison}\label{app:h26xvpx}
In our paper we mainly compared with codecs in the H26x family. Since many prior work used VPx codec, we ran a simple experiment to show they have similar efficiency. We randomly chose 12 videos with resolution 1280x720 from the Kinetics dataset we used and compared encoding efficiency between VP9 and H265. We configured VP9 to use speed/quality tradeoff level 8 and set H265 to very-fast, zero-latency, and no B-frame. We confirm that they have similar performance as shown in Fig \ref{fig:h265vsvp9}.

\tightsubsection{Baseline and testbed implementation details}\label{app:baseline}
We provide the extra implementation details of our baselines here:
\begin{packeditemize}
    \item {\bf Tambur}: To match the implementation in Tambur's paper~\cite{rudow2023tambur}, we force the codec to not encode any I-frames. Following recent work in real-time video coding~\cite{ZEROLAT1,ffmpegstreaming}, we use the \texttt{zerolatency} option (no B-frames) and the \texttt{fast} preset of H.265. The command line we used to encode a video is \texttt{ffmpeg -y -i Video.y4m -c:v libx265 -preset fast -tune zerolatency -x265-params "crf=Q:keyint=3000" output.mp4} where {\em Q} controls the quality of the frame.
    
    \item {\bf Error concealment}: We employ ECFVI~\cite{kang2022error}, an NN-based error concealment pipeline, to mitigate errors from packet losses with H.265 encoding/decoding. 
    When an incomplete frame is received, it starts a 3-step process to compensate for the errors. First, it uses a neural network to estimate the motion vector of the missing part from the previous $N$ frames. Next, the missing pixel values are propagated from the reference frame using the estimated motion vector. Finally, an inpainting neural network is applied to enhance frame quality and minimize error propagation. 
    We set $N = 5$ during our evaluation.

    ECFVI operates under the assumption that packet loss only corrupts portions of a frame, leaving the rest part (corresponding to the arrived packets) decodable. 
    However, as discussed in \S\ref{subsec:packetization}, a single packet loss typically renders an entire frame undecodable in H.264/H.265. 
    To reconcile this, we use flexible macroblock ordering (FMO) technique within the underlying H.265 video codec. 
    This allows different parts of a frame to be encoded and packetized independently into distinct packets. 
    In our baseline implementation, the frame is partitioned into 64$\times$64-pixel blocks and randomly mapped to various packets during packetization. 
    This method introduces a size overhead, as the codec cannot eliminate redundancy among packets. Based on prior works~\cite{kumar2006error, wenger2002scattered, 4494082}, we account for an additional 10\% size overhead to ensure that each packet is individually decodable.
    
    ECFVI is chosen as the baseline for error concealment for two main reasons: {\em (i)} Its 3-step method is recognized as state-of-the-art within the computer vision research area. It surpasses the prior works that only do motion estimation~\cite{sankisa2018video} or inpainting~\cite{chang2019free}. 
    {\em (ii)} Similar methods have been adopted by various recent works such as \cite{gao2020flow}, \cite{li2022towards}, and \cite{zhang2022inertia}, while ECFVI ranking as the most proficient among them.
    {\em (iii)} ECFVI's performance is also on par with or better than other recent error concealment techniques, including those utilizing transformers~\cite{Liu21_FuseFormer, zhang2022flow}.
    
    \item {\bf Voxel} (selective frame skipping): We sort the video frames by the SSIM drop caused by skipping the frame (in real-time video communication, we usually cannot get the quality drop caused by skipping frames in advance. Thus, we are making an idealized assumption that improves the baseline). 
    For 25\% frames with the lowest SSIM drop, we use the default error concealment method in H.264/AVC~\cite{zhou2010efficient} without any packet retransmission, and for the remaining frames (which cause more SSIM drops when skipped), we retransmit all the lost packets. 
    We use a GoP (chunk length) of 4 seconds, which is also used by Voxel. 
    
    \item {\bf Salsify} (functional codec): We implement the Salsify codec based H.265 with the following two key features: firstly, the encoded frame size never surpasses the target bitrate determined by the underlying congestion control algorithm; secondly, upon packet loss, the encoder can dynamically select a reference frame, enabling subsequent frames to be decoded without resending any packets.
\end{packeditemize}

\begin{figure}[t]
    \centering
        \includegraphics[width=0.7\linewidth]{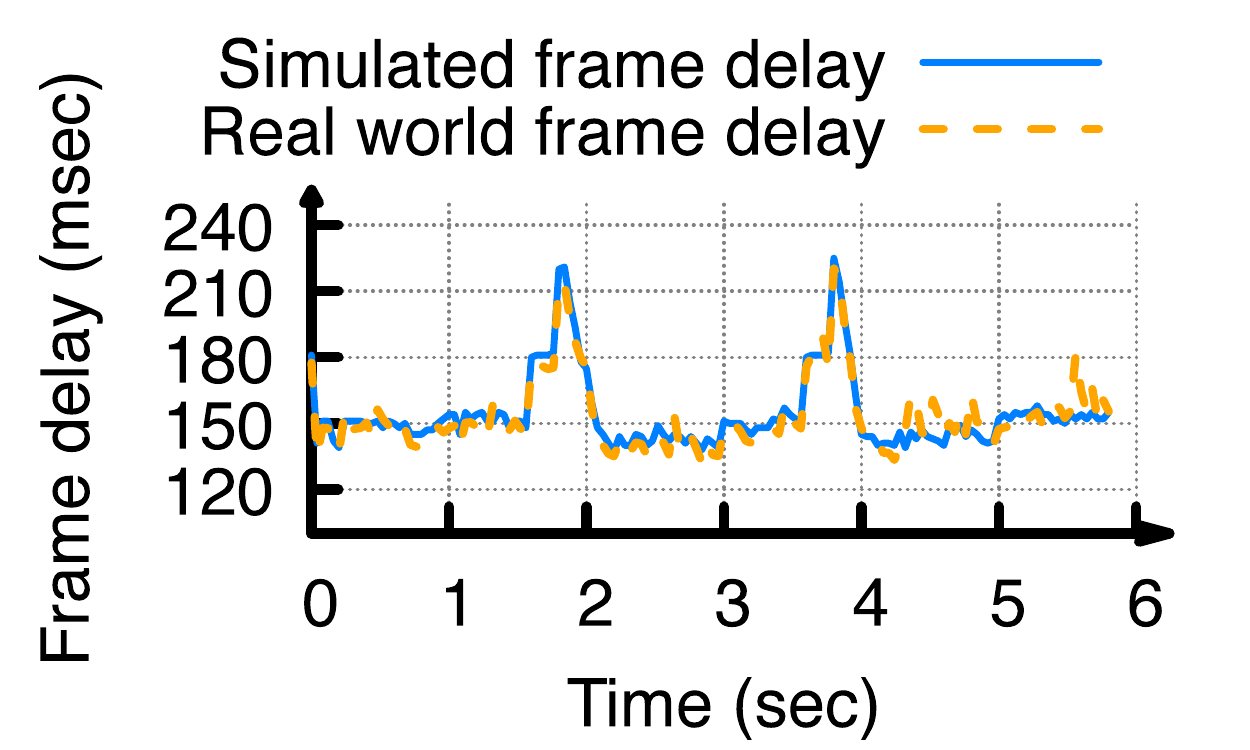}
        \tightcaption{The simulated frame delay of \name is close to the real world measured frame delay}
        \label{fig:validation}
\end{figure}

\tightsubsection{Simulator validation}\label{app:validation}
Our simulator runs on an Ubuntu 18.04 server with 2 Intel Xeon 4210R CPU, and 256GB memory, with 2 Nvidia A40 GPUS. 
To validate that the frame delay measured in simulation matches the real-world numbers, we run a real-world emulation using \name. Being the same as simulation, we use 2 Nvidia A40 GPUs, one for encoding and one for decoding. 
The encoder process encodes the video using \name's encoder and send the encoded packets through an emulated network.
The decoder process decodes the frame using the same logic as mentioned in \S\ref{sec:delivery}.
We compute the real-world frame delay by calculating the difference between the encoding time and the decoding time of a frame.
Figure~\ref{fig:validation} compares the simulated frame delay and real-world measured frame. We use the bandwidth trace same as Figure~\ref{fig:timeseries-ours}. The result validates that our simulated frame delay is accurate.
It is worth noting that we are running real encoding and decoding process in the simulation, hence the calculated frame quality should also be the same as using \name in the real world.

\subsection{Distribution of video content complexity}\label{app:siti}

\begin{figure}[!htb]
    \centering
         \centering
        \includegraphics[width=0.75\linewidth]{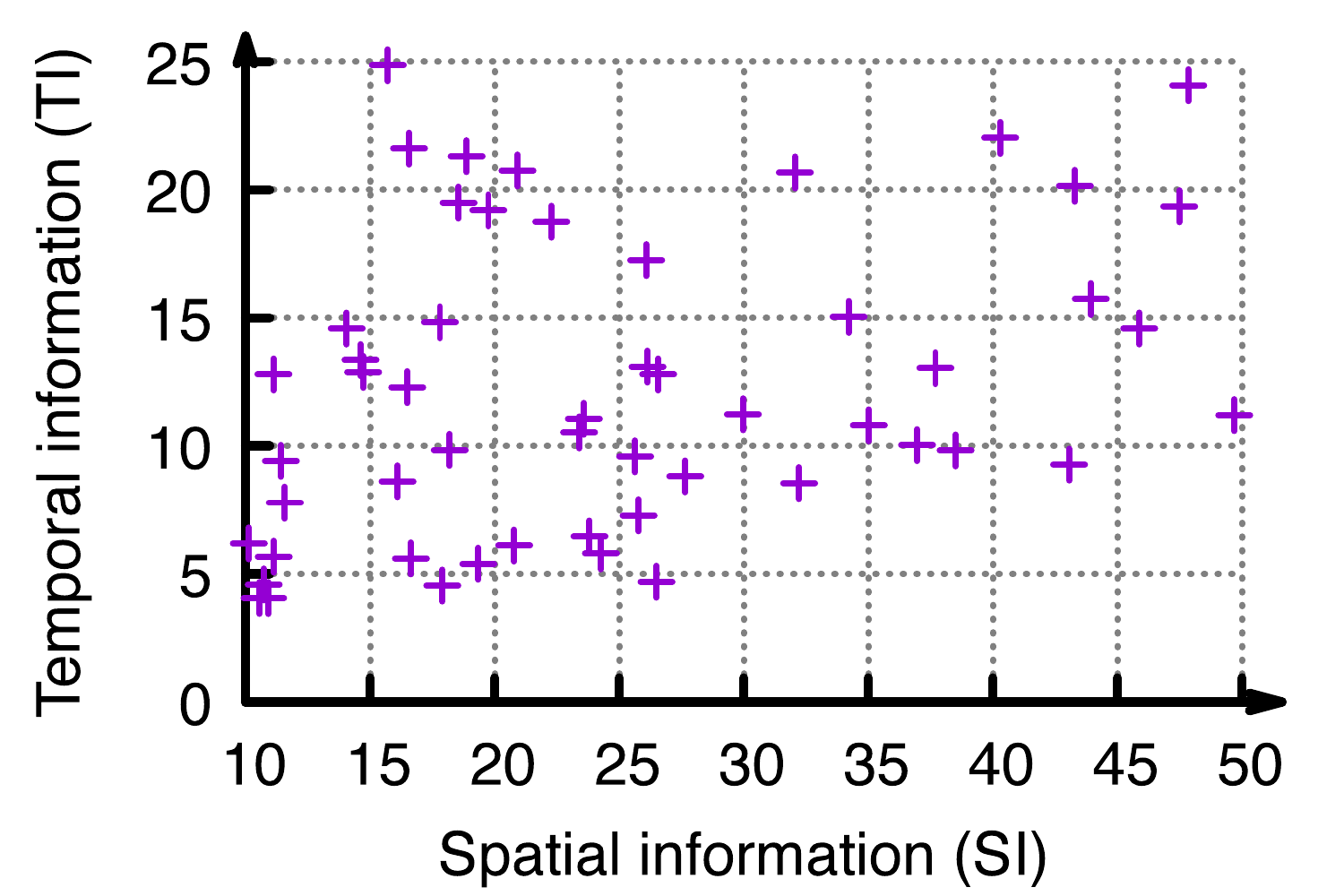}
    \caption{Spatial information (SI) and temporal information (TI) of test videos}
    \label{fig:SITI}
\end{figure}

To validate the test videos that we use cover different content complexities and movements, we calculate the spatiotemporal complexity of the video. We use Spatial Information (SI) and Temporal Information (TI)~\cite{itu1999subjective}, which are frequently-used metrics to measure the spatiotemporal complexity and a larger SI/TI means that the video has a higher spatial/temporal complexity. The metrics are calculated by the tool~\cite{SITITOOL} provided by Video Quality Experts Group (VQEG) and the result is shown in Figure~\ref{fig:SITI}. 

The result validates that {\em (i)} the spatiotemporal complexity of the videos we used covers a wide range: SI is ranging from 15 to 85 and TI is ranging from 3 to 25. %, which corresponds to a recent report.
{\em (ii)} Our test videos covers all the following types: high spatial complexity and high temporal complexity, high spatial complexity but low temporal complexity, low spatial complexity but high temporal complexity, and low spatial complexity and low temporal complexity.

\subsection{Illustration example where \name performs poorly}
\label{app:badexample}
\begin{figure*}[t]
    \centering
    \includegraphics[width=0.99\linewidth]{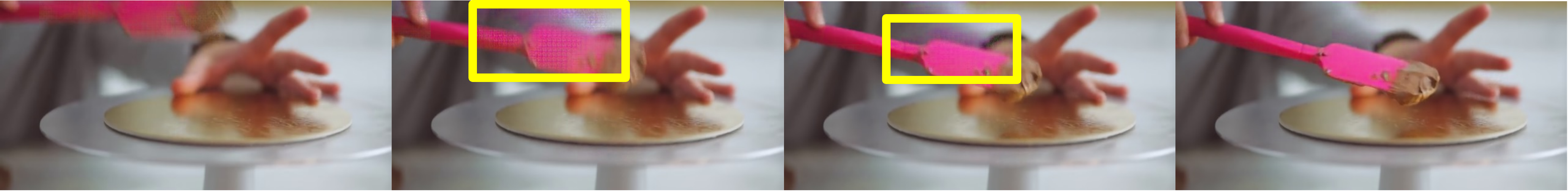}
    \caption{An example where \name performs poorly. It shows four consecutive decoded frames where the pink brush moves down quickly. Some artifacts in The yellow box degrade the frame quality and impact the SSIM.}
    \label{fig:badexample}
\end{figure*}

In some rare cases, \name may suffer from poor quality.
Figure~\ref{fig:badexample} visualizes an example of four consecutive frames when \name performs poorly. 
As shown in the yellow box, the frame decoded by \name has some notable artifacts around the moving object, which degrades the SSIM.

\subsection{Screenshot of videos we used for user study}
\begin{figure*}
\centering
\begin{subfigure}{.24\textwidth}
    \centering
    \includegraphics[width=.95\linewidth]{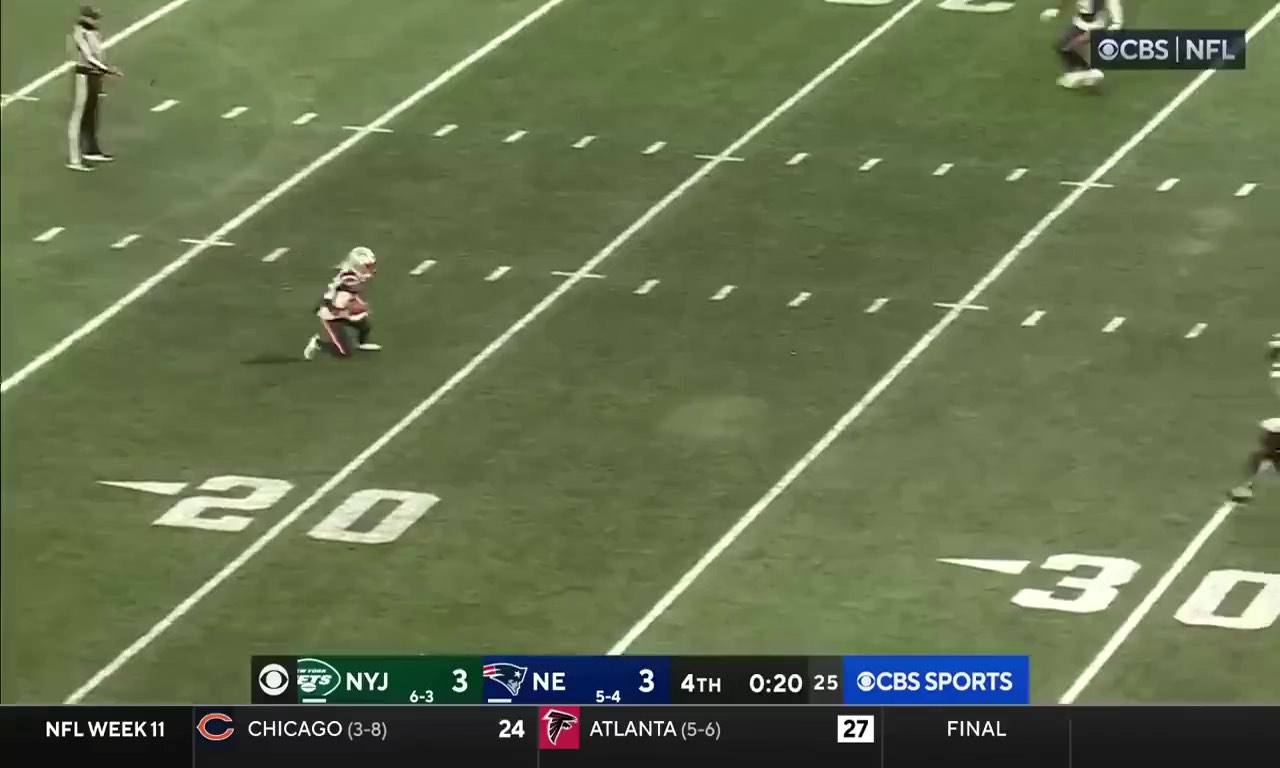}  
    \caption{Sports: Football}
    \label{SUBFIGURE LABEL 1}
\end{subfigure}
\begin{subfigure}{.24\textwidth}
    \centering
    \includegraphics[width=.95\linewidth]{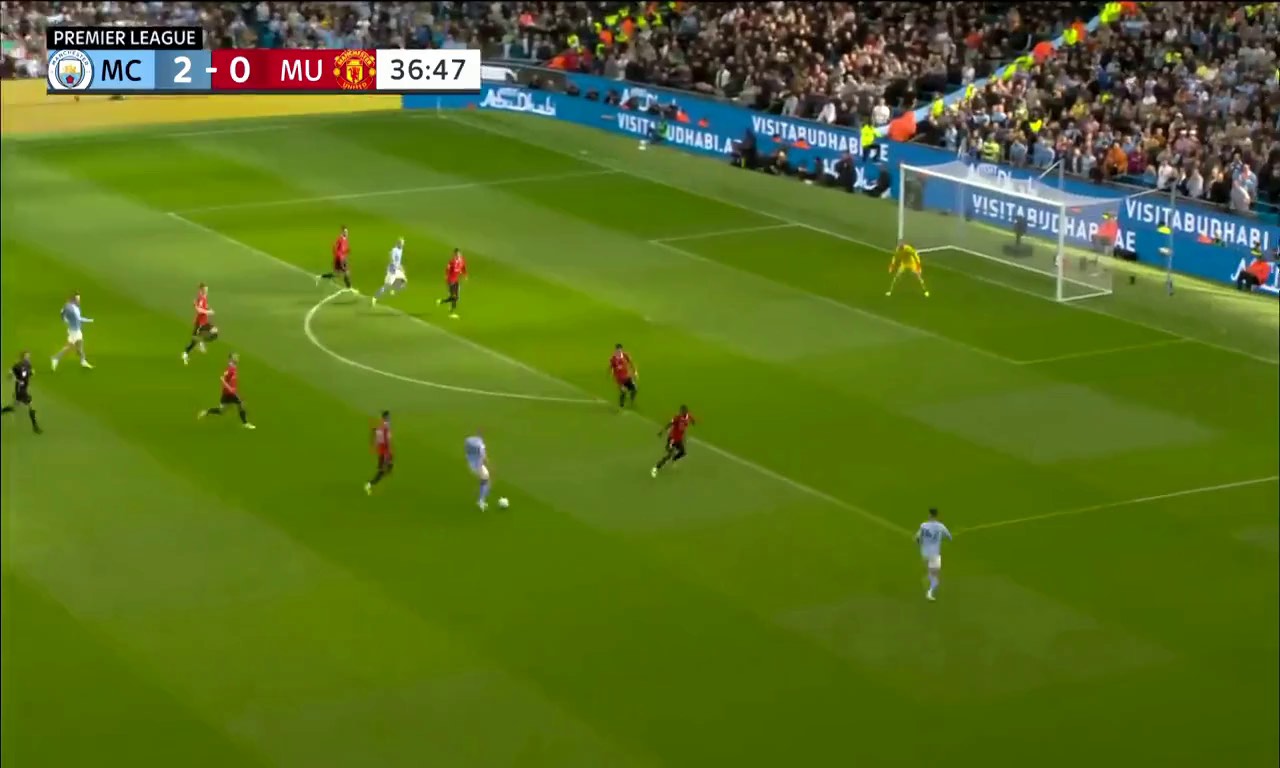}  
    \caption{Sports: Soccer}
    \label{SUBFIGURE LABEL 2}
\end{subfigure}
\begin{subfigure}{.24\textwidth}
    \centering
    \includegraphics[width=.95\linewidth]{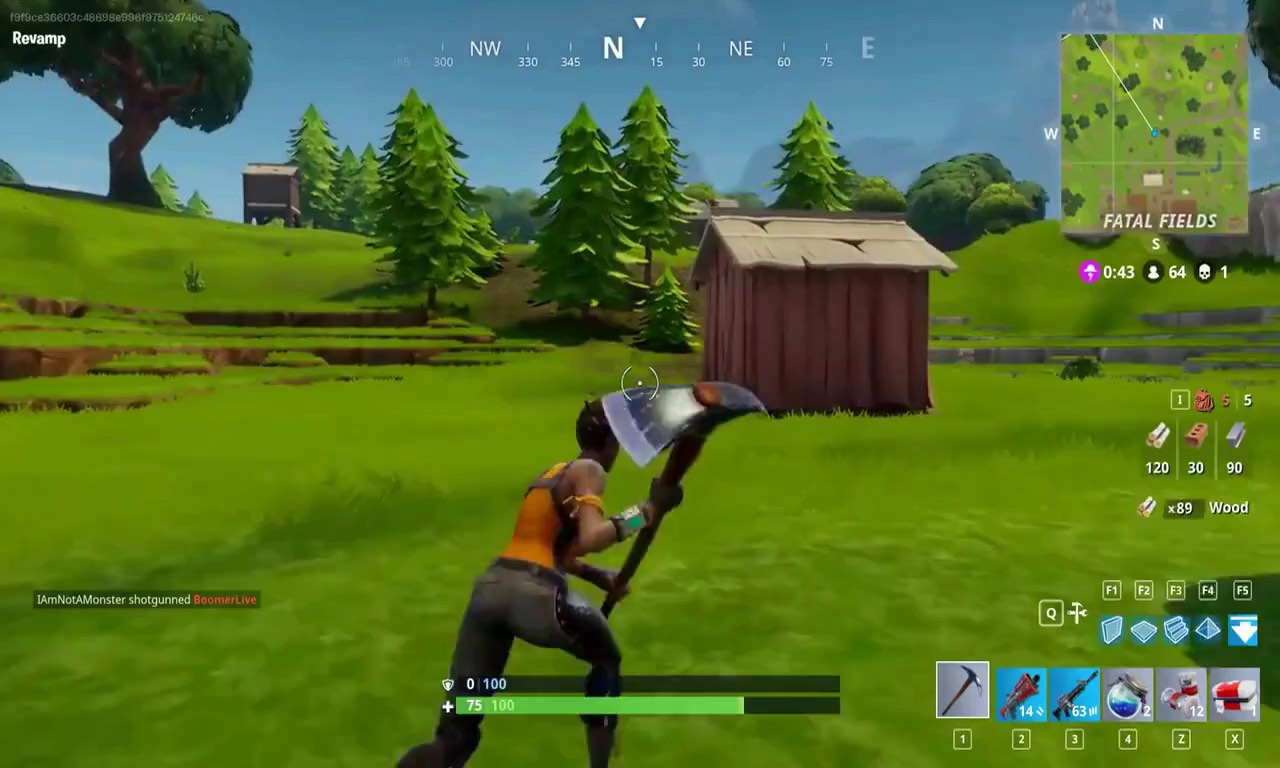}  
    \caption{Gaming: Fortnite}
    \label{SUBFIGURE LABEL 3}
\end{subfigure}
\begin{subfigure}{.24\textwidth}
    \centering
    \includegraphics[width=.95\linewidth]{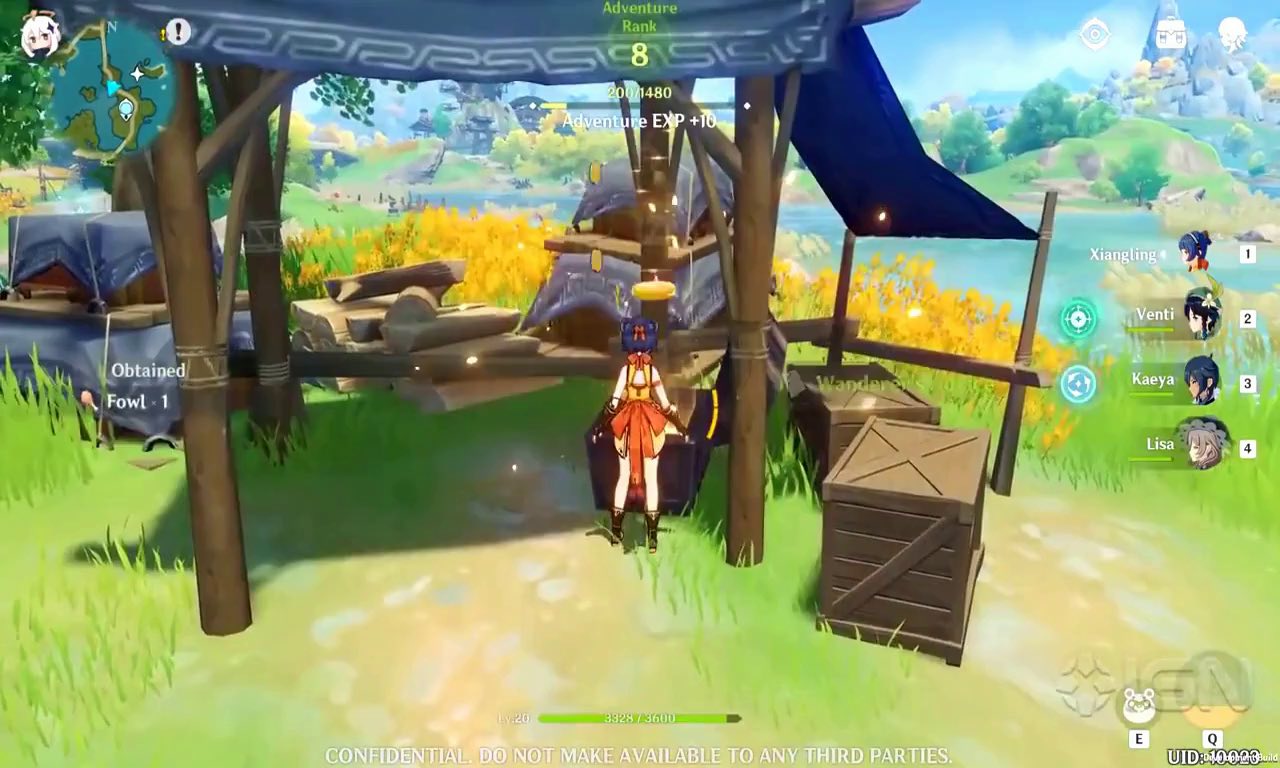}  
    \caption{Gaming: Genshin Impact}
    \label{SUBFIGURE LABEL 4}
\end{subfigure}
\begin{subfigure}{.24\textwidth}
    \centering
    \includegraphics[width=.95\linewidth]{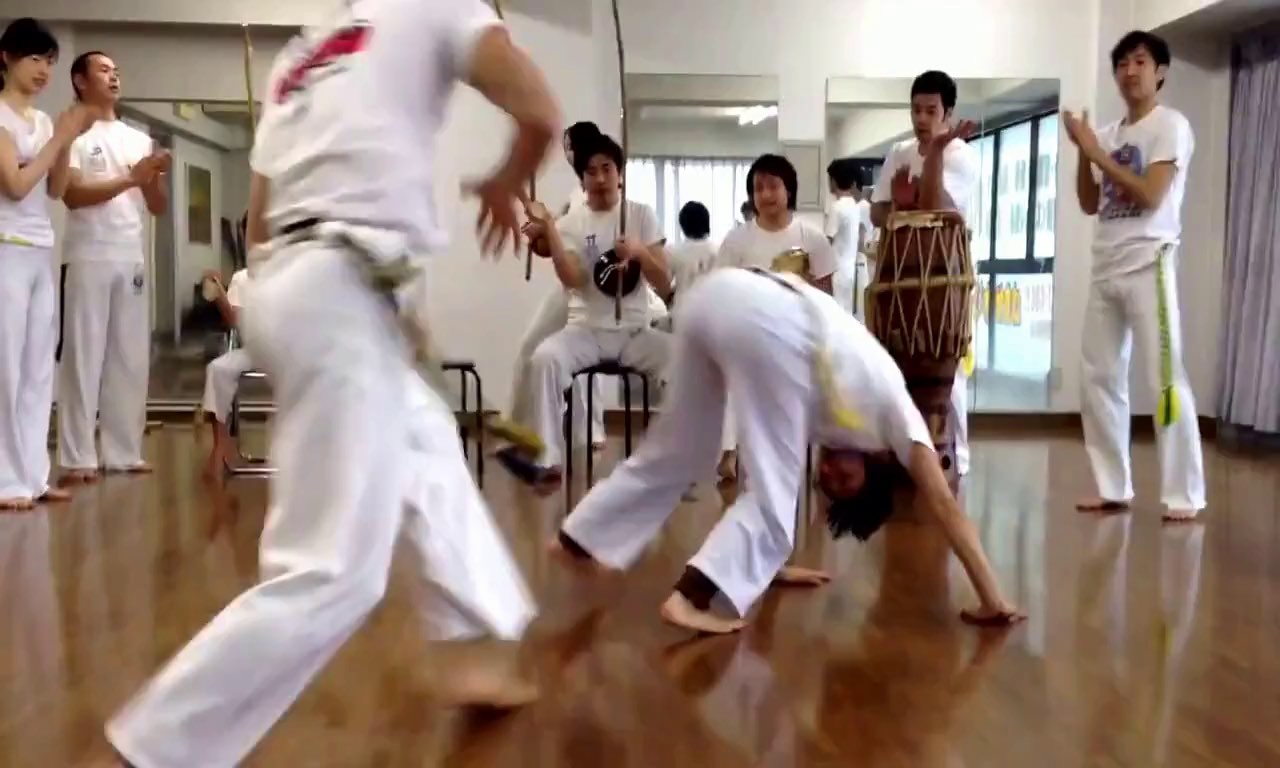}  
    \caption{Daily movement: Taekwondo}
    \label{SUBFIGURE LABEL 5}
\end{subfigure}
\begin{subfigure}{.24\textwidth}
    \centering
    \includegraphics[width=.95\linewidth]{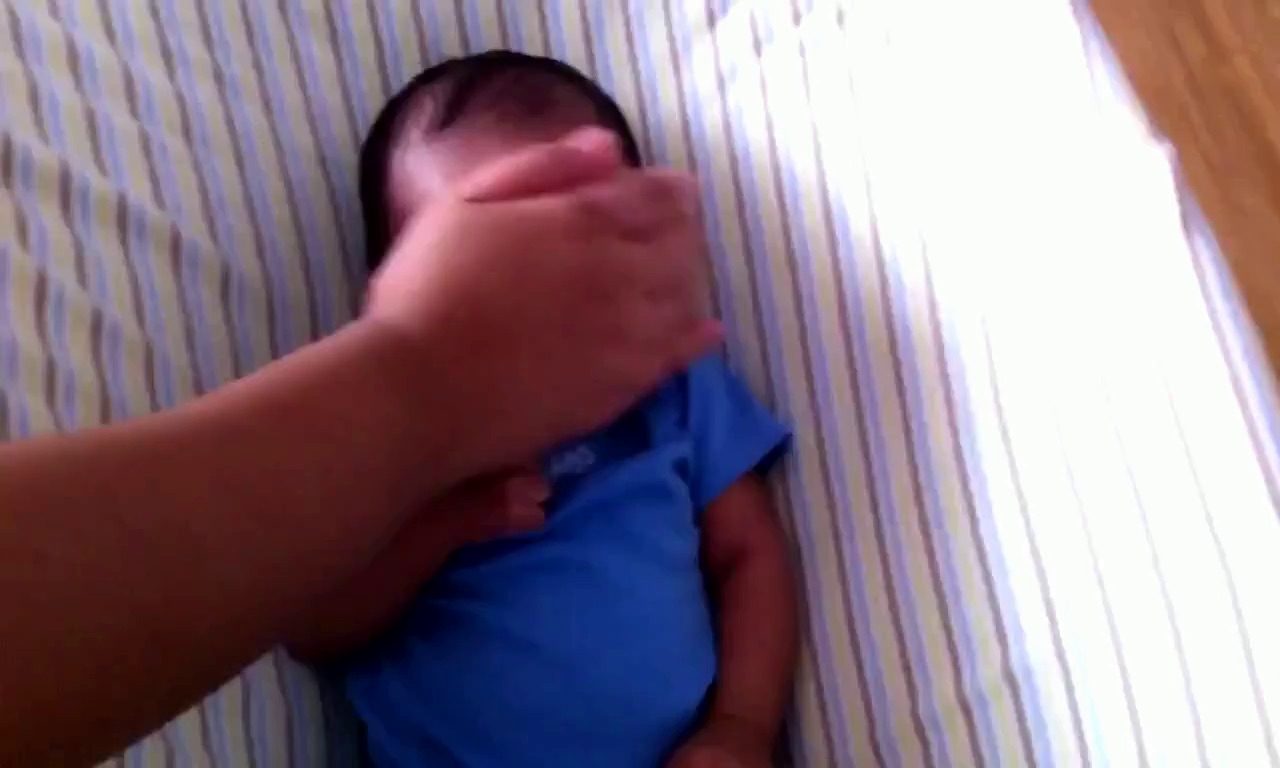}  
    \caption{Daily movement: Baby in crib}
    \label{SUBFIGURE LABEL 6}
\end{subfigure}
\begin{subfigure}{.24\textwidth}
    \centering
    \includegraphics[width=.95\linewidth]{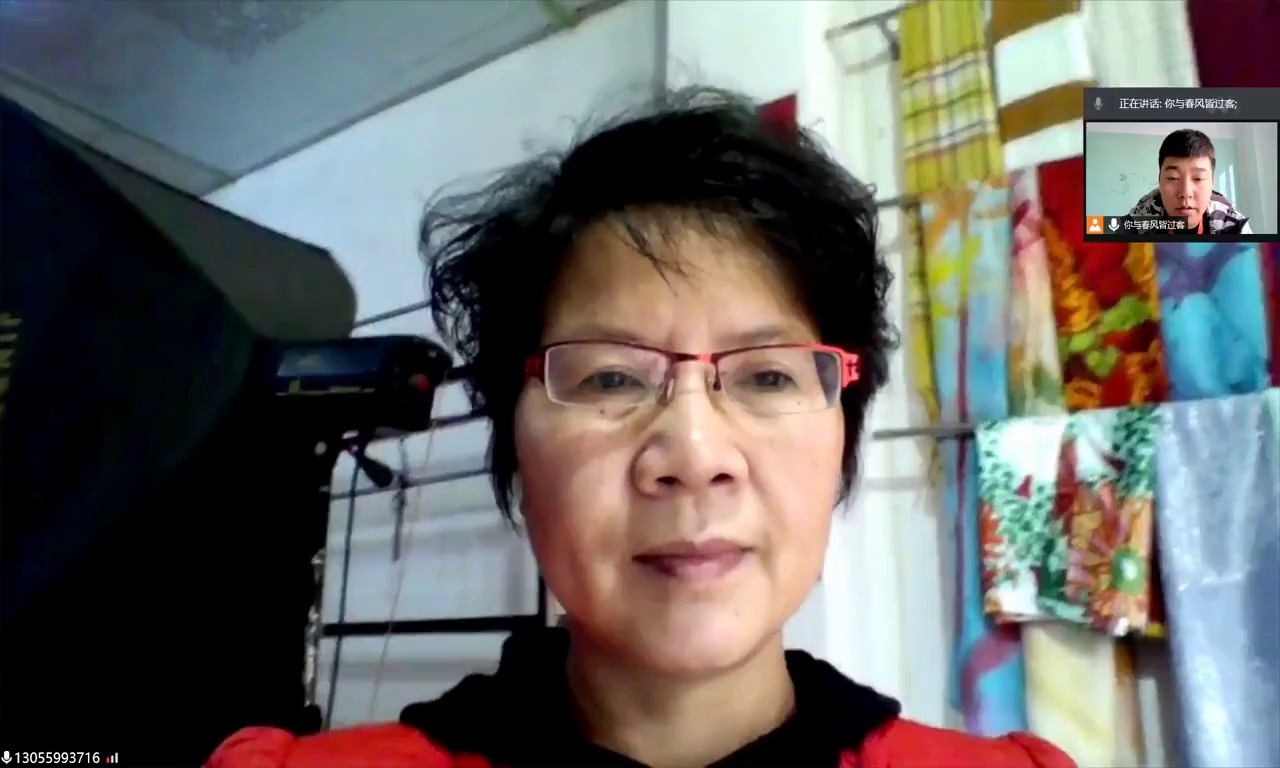}  
    \caption{Talking heads: indoor }
    \label{SUBFIGURE LABEL 4}
\end{subfigure}
\begin{subfigure}{.24\textwidth}
    \centering
    \includegraphics[width=.95\linewidth]{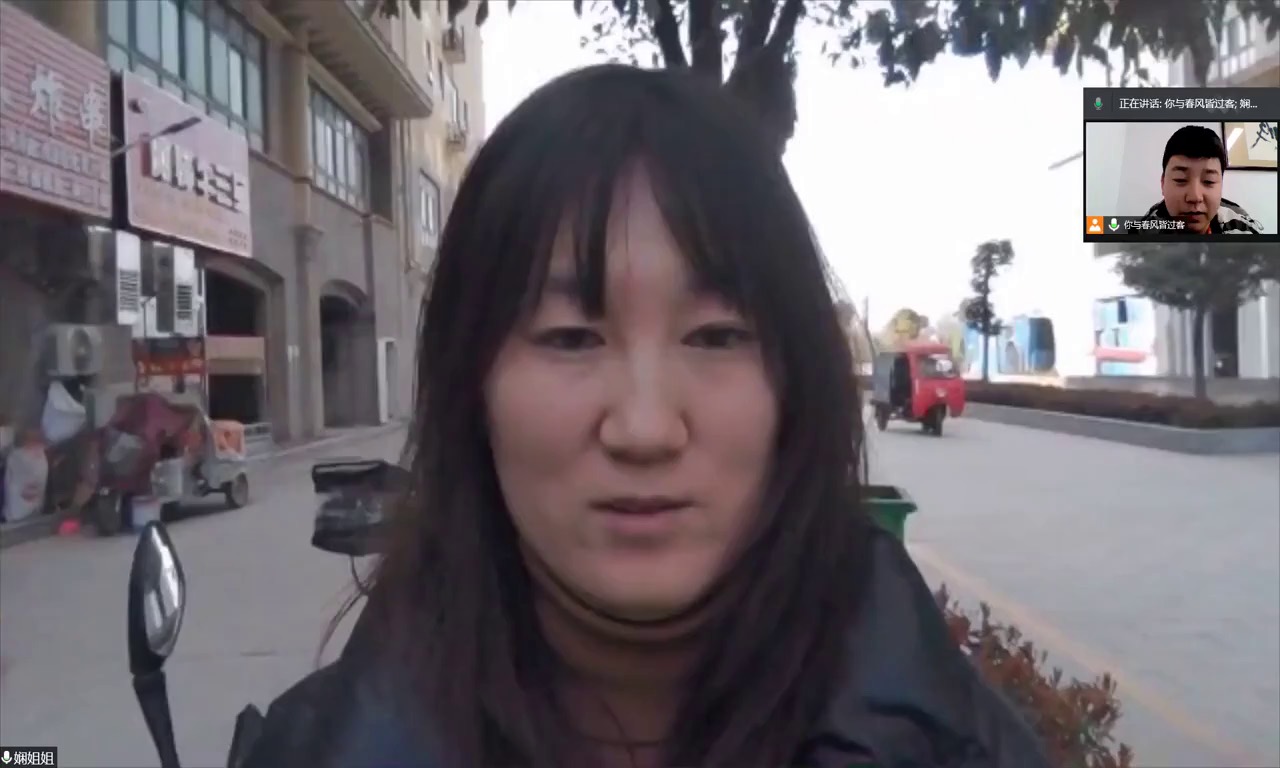}  
    \caption{Talking heads: outdoor}
    \label{SUBFIGURE LABEL 4}
\end{subfigure}
\caption{Summary of videos used in our user study.  They span four categories: sports (a, b), gaming (c, d), daily movement (e, f), and talking heads (g, h). }
\label{fig:user-study-screenshot}
\end{figure*}

Figure~\ref{fig:user-study-screenshot} shows the screenshot of the videos we used for the user study (in \S\ref{subsec:eval:e2e})

% \begin{figure*}[t!]
%     \centering
%     % \includegraphics[width=0.8\linewidth]{figs/timeseries-placeholder.png}
%     % \includegraphics[width=0.75\linewidth]{figs/decoded examples.pdf}
%     \includegraphics[width=0.9\linewidth]{figs/decoded-example-v3.pdf}
%     \tightcaption{Example of decoded images under loss  }
%     \label{fig:decoded-example}
% \end{figure*}

\tightsubsection{Working with other congestion control}\label{app:salcc}

\begin{figure*}[t!]
    \centering
    \begin{subfigure}[t]{0.245\linewidth}
         \centering
         \includegraphics[width=4\linewidth]{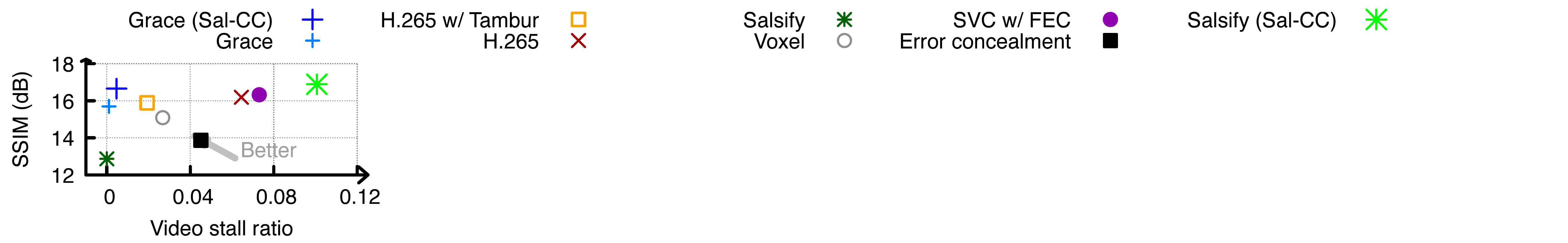}
         \tightcaption{One-way delay = 50ms}
         \label{fig:appgcc-50-25}
     \end{subfigure}
     \hfill
     \begin{subfigure}[t]{0.245\linewidth}
         \centering
         \includegraphics[width=\linewidth]{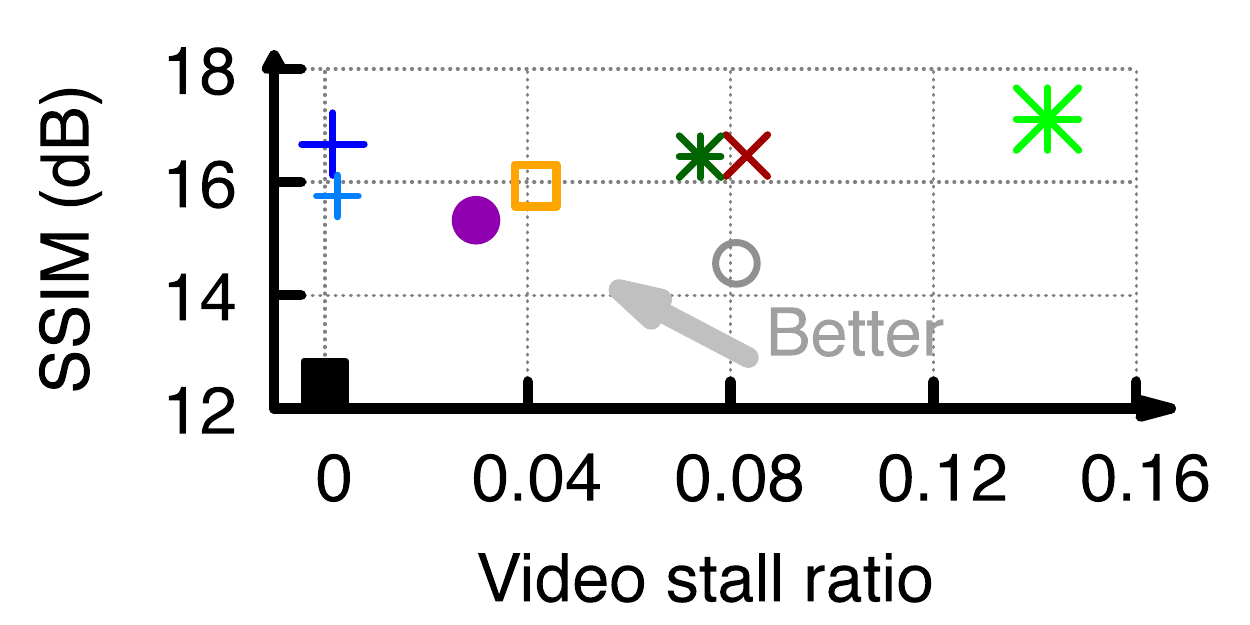}
         \tightcaption{One-way delay = 75ms}
         \label{fig:appgcc-75-25}
     \end{subfigure}
     \hfill
     \begin{subfigure}[t]{0.245\linewidth}
         \centering
         \includegraphics[width=\linewidth]{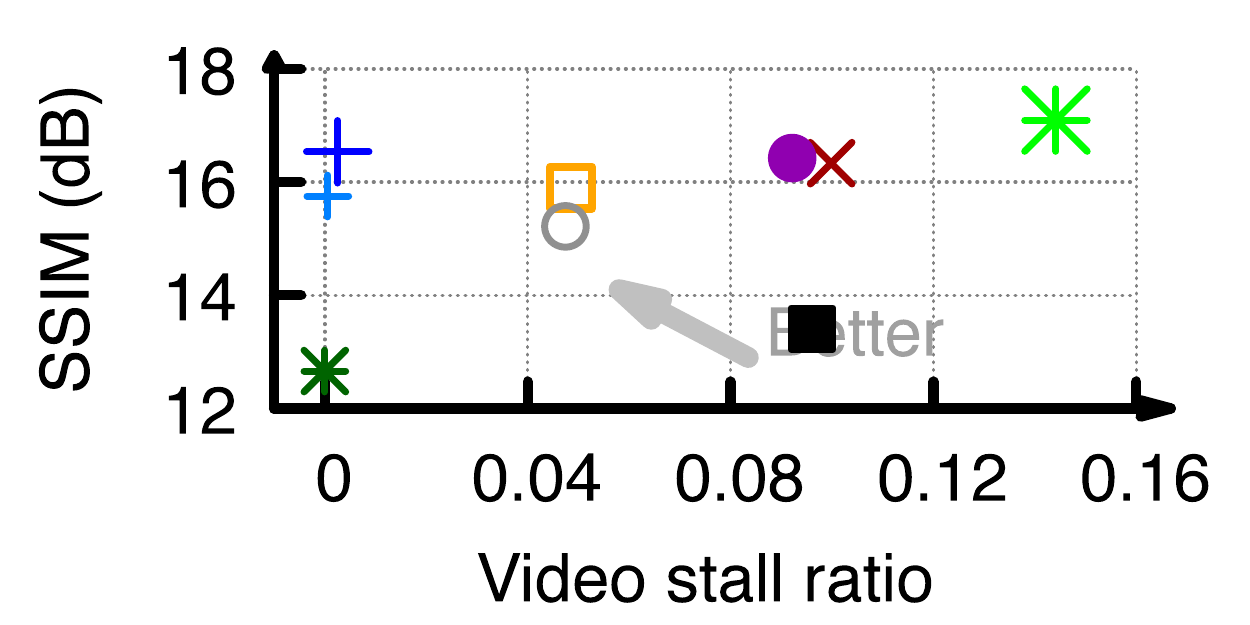}
         \tightcaption{One-way delay = 100ms}
         \label{fig:appgcc-100-25}
     \end{subfigure}
     \hfill
      \begin{subfigure}[t]{0.245\linewidth}
         \centering
         \includegraphics[width=\linewidth]{figs/appgccstall-queue25-delay100-new.pdf}
         \tightcaption{One-way delay = 150ms}
         \label{fig:appgcc-150-25}
     \end{subfigure}
     \hfill
     \vspace{0.3cm}
     \tightcaption{End-to-end simulation result under different one-way delay. Network queue length = 25 packets}
     \label{fig:appgcc-delay-simulation}
\end{figure*}

\name can also work with the congestion control algorithm proposed in Salsify (Sal-CC)~\cite{salsify}, which is more aggressive than GCC. Sal-CC has a higher average sending rate, while paying the cost of potentially having more packet losses. 
Figure~\ref{fig:appgcc-delay-simulation} show that changing from GCC to Sal-CC increases the average SSIM of 0.7-1.1dB for \name with a negligible increase in video stall ratio. 
In contrast, the video stall ratio for Salsify codec will increase a lot when using Sal-CC, because Salsify codec needs to keep skipping frames for more than one RTT when packet loss happens, which leads to frequent video stalls.

\tightsubsection{Working with super resolution}\label{app:sr}

\begin{figure}[t]
    \centering
    \begin{minipage}[t]{\linewidth}
        \centering
        \includegraphics[width=0.8\linewidth]{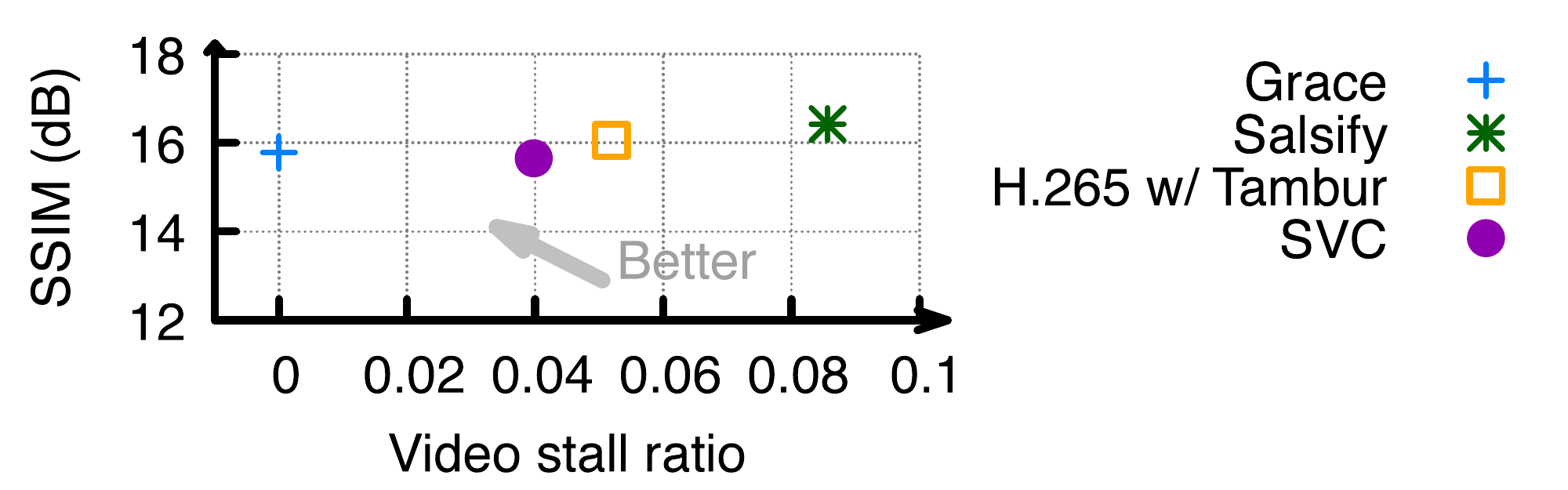}
        \tightcaption{The quality of \name and baselines after super-resolution}
        \label{fig:sr-experiment}
    \end{minipage}
    \hfill 
\end{figure}

In line with the discussion in \S\ref{subsec:related}, Super-Resolution (SR) can supplement the receiver-side video quality. We employed SwinIR~\cite{liang2021swinir}, a leading SR model, in our simulation to confirm that \name, like baselines, can also leverage SR benefits. Our experiments demonstrated that SR boosts receiver-side quality for all codecs, irrespective of the specific codec employed. For more details, refer to Appendix~\ref{app:sr}.

Figure~\ref{fig:sr-experiment} shows the tradeoff between quality and video stall ratio when using SR to enhance the quality at the receiver side. 
We run the simulation using LTE traces with a 100ms one-way delay and a 25-packet queue and then use a state-of-the-art SR model, SwinIR~\cite{liang2021swinir}, to improve the quality of the decoded videos.
When using SR, \name can still have on-par SSIM with Salsify codec and H.265 w/ Tambur: the SSIMs are 15.8 dB, 16.4 dB, and 16.0 dB respectively.
The SSIM of SVC (15.4 dB) is still lower than \name even with super-resolution. This is because packet loss can make higher layers of SVC undecodable, resulting in lower quality.
This shows SR technique is complementary to our work, as it can improve the quality for any codecs at the receiver side.

\subsection{Encoding/decoding time on CPU}\label{app:openvino}
We use OpenVINO library to run \name on a 32-core Intel(R) Xeon(R) Silver 4210R CPU. Table~\ref{tab:coding-time-openvino} shows the encoding/decoding time of a 720p/480p frame respectively.
It can encode/decode a 720p frame at 28.5~fps and 24.4~fps respectively.
\begin{table}[t!]
\begin{centering}
\begin{tabular}{c|cc|cc}
\toprule
                      & \multicolumn{2}{c|}{Encoding (ms)} & \multicolumn{2}{c}{Decoding (ms)} \\ \cline{2-5} 
                      & 720p             & 480p             & 720p             & 480p            \\ \hline
\namelite   & 35.1             & 17.2             & 40.9             & 21.6            \\
\bottomrule
\end{tabular}
\end{centering}
\vspace{3pt}
\tightcaption{Encoding/decoding time per frame for \namelite on Intel CPU}
\label{tab:coding-time-openvino}
\end{table}

\begin{table}[t!]
\begin{centering}
\begin{tabular}{c|c|c|c}
\toprule
           & \begin{tabular}[c]{@{}c@{}}
           SSIM\\ (dB)\end{tabular} & \begin{tabular}[c]{@{}c@{}}\% of non \\ rendered frames\end{tabular} & \begin{tabular}[c]{@{}c@{}}Video stall\\ ratio\end{tabular} \\ \hline
\name      & 15.53                                               & 0.21                                                                 & 0.0011                                                      \\ \hline
\namelite & 15.01                                               & 0.22                                                                 & 0.0012     \\ \hline                                                 
\namedecoder & 13.91                                               & 0.24                                                                 & 0.0014    \\ \hline                                    
\namepretrain & 12.53                                               & 0.33                                                                 & 0.0023  \\
\bottomrule
\end{tabular}
\vspace{3pt}
\tightcaption{End-to-end simulation shows \namelite has the same benefits in video realtimeness/smoothness compared to \name with marginal quality drop. Although \namedecoder and \namepretrain have similar video realtimeness/smoothness as \name, they suffer from low video quality.}
\label{tab:cpu-simulation}
\end{centering}
\end{table}

\subsection{Simulation results and visualization examples for \namelite, \namepretrain and \namedecoder}
\label{app:abl-simulation}

\begin{figure}[t!]
    \centering
    \includegraphics[width=0.9\linewidth]{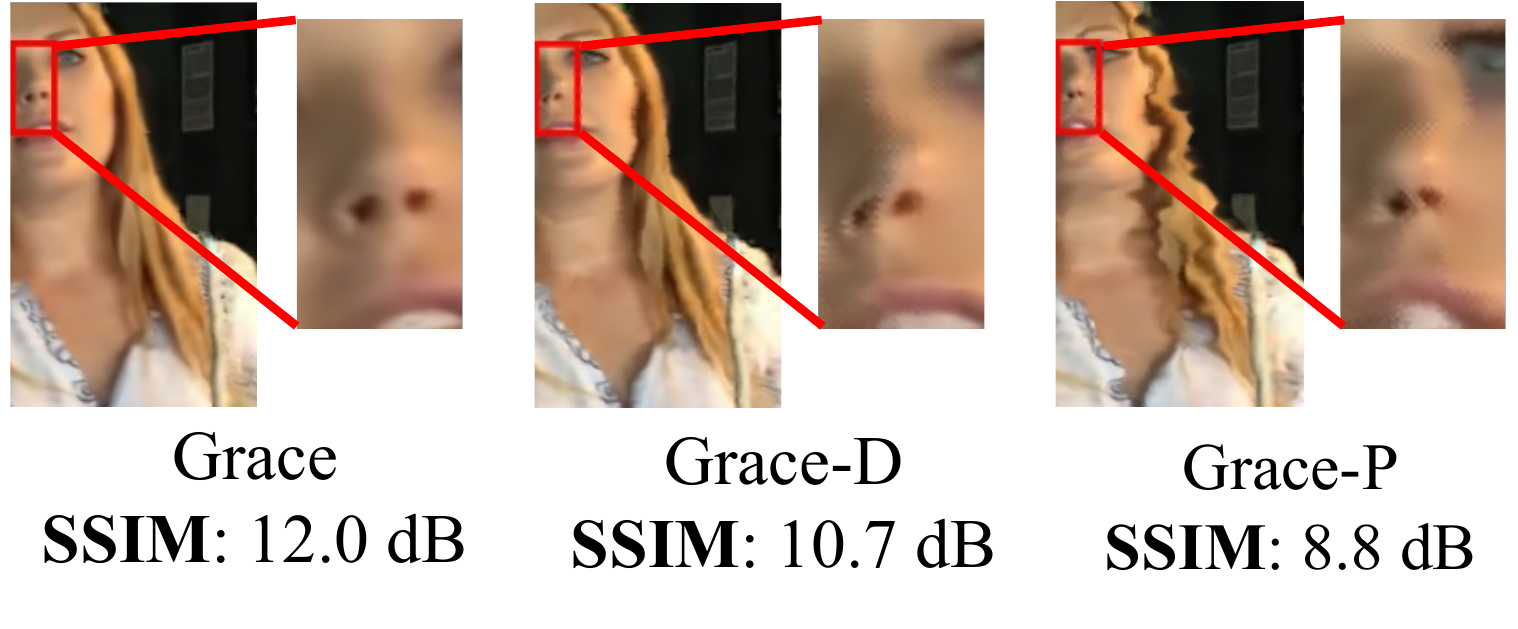}
    \tightcaption{Comparing reconstructed image when the same packet loss is applied to the pre-trained NVC (\namepretrain), a variant with only decoder fine-tuned with loss (\namedecoder), and \name (both encoder and decoder jointly fine-tuned).
    }
    \label{fig:late-visual-example}
    \vspace{-2pt}
\end{figure}

Table~\ref{tab:cpu-simulation} shows the end-to-end simulation results comparing \name, \namelite, \namedecoder, and \namepretrain.
We use the LTE traces, and set the one-way-delay to 100~ms and the network queue length to 25 packets.
\namelite has both similar quality and realtimeness/smoothness as \name.
Without jointly training the encoder and decoder with loss, \namepretrain and \namedecoder fail to achieve similar quality as \name.

Figure~\ref{fig:late-visual-example} visualizes the reconstructed frame of \name, \namepretrain, and \namedecoder when the same 50\% packet loss is applied to the encoded tensor of the same image, 
demonstrating that by jointly training both the encoder and decoder under various packet losses, \name delivers the best reconstruction quality without any prominent artifacts, and achieves a high SSIM.

\end{document}